\documentclass[aps,pre,showpacs,twocolumn]{revtex4-1}

\usepackage{amsmath,amssymb,amsfonts,hyperref}
\usepackage{graphicx,epsf}
\usepackage{xspace}
\usepackage{textcomp}
\usepackage{color}

\newif\ifhyper
\hypertrue
\ifhyper
\hypersetup{
  citecolor = {green},
  urlcolor = {blue} 
} 

\newcommand{\p}{\partial} 
\newcommand{\vx}{\vec{x}}

\newcommand{\vv}{\vec{v}}
\newcommand{\vu}{\vec{u}}
\newcommand{\vf}{\vec{f}\,}
\newcommand{\vJ}{\vec{J}}

\newcommand{\bq}{{\bf q}}
\newcommand{\bp}{{\bf p}}
\newcommand{\bx}{{\bf x}}
\newcommand{\vq}{{\vec{q}}}
\newcommand{\vp}{{\vec{p}}}
\newcommand{\td}{{\text{\tiny $D$}}}

\newcommand{\tf}{\tilde{f}}

\newcommand{\anz}{{ansatz}\xspace}

\begin{document}

\title{Fully developed  isotropic turbulence: nonperturbative renormalization 
group formalism and fixed point solution}

\author{L\'eonie Canet$^{1}$, Bertrand Delamotte$^{2}$, and  Nicol\'as 
Wschebor$^{2,3}$}
\affiliation{
$^1$LPMMC, Universit\'e Joseph Fourier Grenoble-Alpes, CNRS UMR 5493, 38042 
Grenoble Cedex, France\\
$^2$LPTMC, CNRS UMR 7600, Universit\'e Pierre et Marie Curie, 75252 Paris Cedex 
05, France \\
$^3$Instituto de F\'isica, Facultad de Ingenier\'ia, Universidad de la 
Rep\'ublica, J.H.y Reissig 565, 11000 Montevideo, Uruguay}

\begin{abstract}
We investigate the regime of 
fully developed homogeneous and isotropic turbulence
 of the Navier-Stokes (NS) equation in the presence of a stochastic forcing,
  using the nonperturbative (functional) renormalization group (NPRG). 
  Within a simple approximation based on symmetries,
 we obtain the fixed point solution  of the NPRG flow equations that corresponds 
to fully developed turbulence both in $d=2$ and $d=3$ dimensions. Deviations to 
the  dimensional scalings (Kolmogorov in $d=3$ or Kraichnan-Batchelor in $d=2$) 
are found for the two-point functions. To further analyze these deviations, we 
derive exact flow equations in the large wave-number limit, and show that
 the fixed point  {\it does not} entail the usual scale invariance, thereby 
identifying the mechanism for the emergence of intermittency
  within the NPRG framework.  The purpose of this work is to provide a 
detailed basis for NPRG studies of NS turbulence, the determination  of  the 
ensuing intermittency exponents is left for future work.

\end{abstract}

\maketitle

\section{Introduction}

The statistical theory of  turbulence is more than seventy years old and, 
despite intensive efforts,
 it remains unsatisfactory. In $d>2$, two length scales play a dominant role in 
the phenomenology of 
fully developed homogeneous and isotropic turbulence: the microscopic 
(Kolmogorov) scale $\eta$ where energy is dissipated by molecular viscosity
and the macroscopic integral scale $L$ where energy is injected in the system. 
These two scales  delineate the inertial range, 
 where energy is conserved and transferred towards the small scales in an energy 
cascade. Within the inertial range, the equal-time velocity correlation 
functions exhibit universal scaling,
that is, they behave as power laws with exponents independent of the precise 
mechanisms of  
energy injection and dissipation.  
 These observations
 lead to the celebrated K41 scaling theory, proposed by Kolmogorov in 1941 
\cite{Kolmogorov41a,Kolmogorov41b,Kolmogorov41c}.
 The energy flux constancy relation was derived, which yields the exact 
``four-fifth law'' for the three-velocity correlator.
   K41 also predicts power-law behaviors for all the correlation functions. 
Although the experimentally measured energy spectrum and low-order structure functions
 are well described by K41 theory, systematic deviations 
from K41 scalings were observed for higher-order correlation functions 
\cite{Frisch95,Oboukhov62}. 
 Calculating these exponents beyond K41 theory remains
 a great challenge in the study of fully developed turbulence.
  In $d=2$, two inertial ranges were predicted to coexist by Kraichnan 
\cite{kraichnan67a} as a consequence of the conservation of both energy and 
enstrophy (squared vorticity). In two-dimensional turbulence, part of the energy is 
transferred from the integral scale to the larger scales in an inverse energy 
cascade until it  is eventually dissipated at the boundaries of the system, 
while enstrophy flows towards the smaller scales in the direct cascade until it 
is dissipated at the molecular scale \cite{kraichnan67a,batchelor69}.
 In the direct cascade, the exponents of the structure functions are also 
believed to deviate from dimensional scalings \cite{Frisch95,Lesieur90} and 
their calculation remains a challenging issue.

This situation appears frustrating 
if compared  to that of critical phenomena occurring at equilibrium, which share 
many common
features with turbulence (e.g. scaling, chaos, universality) \cite{Eyink94} and 
where renormalization group (RG)
has led in most cases to a clear understanding of the physics at play and to 
accurate determinations
of the critical exponents \cite{zinnjustin89}. An essential difference is that, 
in standard equilibrium critical phenomena, a finite set of anomalous dimensions 
suffices to describe the scaling behavior of all the correlation functions, 
which is no longer true for turbulence. The correlation functions do exhibit 
power laws, but each with its specific exponent, which generates multiscaling, 
or multifractality, and constitutes one of the imprints of turbulence \cite{Frisch95}. 
This phenomenon, generically referred to as intermittency,  is investigated  in 
this paper
 using nonperturbative (functional) renormalization group (NPRG).

 Prior to giving an overview of existing RG approaches to describe 
fully developed homogeneous and isotropic turbulence, let us introduce the 
 relevant microscopic model, which is the Navier-Stokes equation with forcing:
\begin{equation}
 \partial_t v_\alpha+ v_\beta \partial_\beta v_\alpha=-\frac 1\rho 
\partial_\alpha p +\nu \nabla^2 v_\alpha+f_\alpha
\label{ns}
\end{equation}
where the velocity field $\vv$, the pressure field $p$, and the stochastic 
forcing $\vec f$ depend on the 
space-time coordinates $(t,\vx)$, and with  $\nu$  the kinematic viscosity and 
$\rho$ the density of the fluid.  
 Since we aim at studying the turbulent steady state,  the presence of the 
stirring force $\vf$ is essential to balance
 the dissipative nature of the  (unforced) NS equation which otherwise leads to 
the decay of  the velocity fields.
 We consider in the following incompressible flows, satisfying 
\begin{equation}
 \partial_\alpha v_\alpha = 0.
\label{compress}
\end{equation}
We focus  on the properties of the turbulent fluid within the inertial range of 
wavenumbers $p$ corresponding to $L^{-1} \ll p \ll \eta^{-1}$.
 In this  regime, the  steady-state correlation functions are expected to be 
universal in the sense that they do not depend on the precise form of the 
macroscopic forcing (as long as its Fourier transform is peaked around 
wave-numbers of the order  $L^{-1}$).
 This universality allows one, instead of choosing a deterministic  forcing, to 
take averages on various smooth forcings with an essentially arbitrary probability 
distribution, as long as the typical scale of the forcing remains the prescribed 
integral scale.
  One can hence conveniently  choose the simplest of probability distributions: 
a Gaussian one with
zero mean and variance
\begin{equation}
 \langle f_\alpha(t,\vx)f_\beta(t',\vx\,')\rangle=2 \delta(t-t')N_{L^{-1},\alpha 
\beta}(|\vx-\vx\,'|).
\end{equation}
This correlator is local in time, which is crucial to preserve Galilean 
invariance,
 and it is centered, in Fourier space, on the inverse of the integral scale $L$.
 The precise profile of $N_{L^{-1},\alpha\beta}(x)$,
 which should not affect universal properties in the inertial regime,
  will be specified in the following.\\

As in critical phenomena, scaling is observed in turbulence 
when the microscopic scale (the Kolmogorov scale $\eta$) is sent to zero and
the macroscopic one (the integral scale $L$)  to infinity. In this limit,
the expansion parameter, the Reynolds number, diverges. Field theoretic
techniques such as renormalization group (RG) are designed to handle the large 
scale fluctuations
developing in strongly correlated systems and we briefly review in the following 
some of the former attempts
in turbulence (for reviews, see \cite{Smith98,Adzhemyan99,Zhou10}).

The difficulty when applying RG in turbulence is not so much that the Reynolds 
number
diverges when the ultra-violet (UV) scale $\eta^{-1}$ is sent to infinity.  
Naively, one could think that the Reynolds number is the expansion parameter in 
a perturbative treatment, and its divergence would render odd  a perturbative 
analysis based on it.
  However, it is well known \cite{zinnjustin89} that once the RG is employed, 
the proper expansion parameter is not the bare Reynolds number but a 
renormalized parameter obtained from its RG evolution in the long-distance 
limit, such that the perturbative analysis can work. The real difficulty is to 
find a situation    
 where the renormalized
expansion parameter is small. For standard critical systems, this is achieved 
around the upper critical dimension $d_c$
and a double expansion in the coupling constant and $\epsilon=d_c-d$ renders  
the perturbative expansion well defined. 
As for turbulence, there is no upper critical dimension but a formal (second) 
expansion
parameter $\epsilon$ can be defined through the forcing profile 
$N_{L^{-1},\alpha \beta}(p)\propto p^{4-d-2\epsilon}$
where $p$ is the wave-number \cite{DeDominicis79,Fournier83,Yakhot86,Canuto96}. 
Typically, as explained above, $N_{L^{-1},\alpha \beta}(p)$ is not a power-law
in Fourier space, but is instead sharply peaked around the infra-red (IR) scale 
$L^{-1}$. One can show that the physical situation is   recovered only in a
precise limit, {\it eg.} when $\epsilon\to2$  (in $d=3$) or $\epsilon\to3$ (in the direct 
cascade in $d=2$). On the other hand, for $\epsilon=0$, the theory is exactly 
renormalizable
and a fixed point of order $\epsilon$ is found in any dimension 
\cite{Adzhemyan03,Adzhemyan08} (the $d=2$ case 
being particular 
\cite{Olla91,Honkonen96,Antonov97,Honkonen98,Honkonen02,Adzhemyan05b,Mayo05}). 
The challenge
for the perturbatively renormalized theory is therefore to extend the results 
obtained for $\epsilon\to0$ to $\epsilon=2$ (or 3)
which is far from trivial: The
difficulties encountered are very severe and have, up to now, hindered  real 
progress,  at least for the calculation
of multiscaling behavior in the NS problem \cite{Adzhemyan03,Adzhemyan08}.
This seems to be related to the appearance of operators with negative 
critical dimension at
finite $\epsilon$.

Let us emphasize that intermittency also occurs in the simpler Kraichnan's 
passive scalar model \cite{Kraichnan94}
  where a scalar field is advected by a prescribed 
Gaussian random field decorrelated in time and self-similar in space. 
In contrast with  Navier-Stokes turbulence,  the anomalous  exponents for the 
Kraichnan model have been determined 
under several controlled approximations including an $\epsilon$-expansion (with 
$\epsilon$ related to 
the power spectrum of the noise) or a $1/d$-expansion, and numerical simulations
  \cite{Gawedzki95,Chertkov95,Shraiman95,Chertkov96,Bernard96,Fairhall96,Gat97,
Frisch98,Adzhemyan98b,Mazzino01}
   (see \cite{Falkovich01,Antonov06} for reviews).
 In this model, composite operators with negative
critical dimensions,  called  ``dangerous''
 operators, were identified and the anomalous exponents could be calculated from
their critical dimensions \cite{Adzhemyan98b,Adzhemyan99b}.  The 
calculations were performed at three
loops \cite{Adzhemyan01a,Adzhemyan01b} and yielded reasonable results, under mild assumptions on
 the series behavior.
Unfortunately, the same
methodology does not seem to work satisfactorily for the Navier-Stokes
equation. In this case no operator with negative critical dimension
 could be identified at infinitesimal $\epsilon$ and thus, the $\epsilon$ 
expansion does not seem
to provide a reliable framework to compute anomalous exponents 
\cite{Adzhemyan03}.

Let us notice that another perturbative approach, almost ignored in the 
subsequent
literature,  does not rely on an $\epsilon$-expansion but on a self-consistent 
determination of the quadratic part
of the action around which perturbation theory is performed \cite{Giles01}. This 
approach, after elimination of what is named the ``sweeping
effect'' (the sweeping of the smaller scales by the larger) leads to the 
existence of an UV attractive 
fixed point from which,  performing an Operator Product Expansion (OPE), can be computed the multiscaling 
exponents. They
turn out to be quite accurate at least for the $n<10$ first equal-time 
correlation functions of the velocity differences.

A rather different field-theoretic approach, not based on RG, has  been 
developed by L'vov, Procaccia and collaborators.
To get rid of the sweeping effect that leads to severe IR singularities in 
renormalized perturbation theory, 
these authors use ``quasi-Lagrangian'' 
variables instead of the Eulerian velocities \cite{Belinicher87}. They show that 
the correlation functions of the differences
of these variables are finite order by order in perturbation theory in both 
limits where the UV and IR scales are removed.
As a consequence, Kolmogorov scaling holds at all finite orders of the 
perturbation theory. The only way out of this hindrance is 
to resum infinite classes of Feynman diagrams. The authors indeed show that 
these resummations  produce new singularities in terms
of the integral scale which is therefore the proper renormalization 
scale 
\cite{Lvov94a,Lvov94b,Lvov95a,Lvov95b,Lvov96a,Lvov96b,Lvov96c,Lvov96d,Lvov98a,
Belinicher98a,Belinicher98b}. They
 are then able to compute approximately the small order multiscaling exponents 
in terms of the first one \cite{Lvov00}.

An alternative RG approach that has been very
successful in the study of critical systems either at or out of equilibrium, is 
the nonperturbative 
renormalization group (NPRG), which is a modern version of the RG 
{\it \`a la Wilson} 
\cite{wetterich93c,ellwanger93b,ellwanger93c,tetradis94,morris94a,Berges02}.
In addition to avoiding  many problems encountered perturbatively, such as 
the need of explicit 
resummation of IR singularities or the asymptotic nature of the renormalized 
series \cite{zinnjustin89}, this approach 
has led not only to very accurate 
\cite{Berges02,tissier00,tissier00b,canet03a,canet03b,Benitez09,Essafi11,
Benitez12} but also fully nonperturbative
\cite{tarjus03a,canet04b,canet05a,tissier06a,tissier08,Gredat14} results in many 
systems.
 Of course, approximations are also unavoidable in the NPRG context and 
 they are not always easy to justify  
 when they are not controlled by a small parameter.
A very inspiring case 
is the Kardar-Parisi-Zhang (KPZ) equation describing the stochastic  growth of 
interfaces  \cite{kardar86}, which is  equivalent 
to Burgers equation in the context of fluids, and which shows fully 
nonperturbative behavior in the rough phase \cite{halpin95}. Contrary to
the standard perturbative RG which fails to all orders in perturbation to find 
the relevant fixed point and the associated scaling behavior \cite{wiese98}, 
the NPRG approach captures the strong coupling physics of the KPZ equation at 
and above one dimension  \cite{Canet10,Canet11a,Kloss12,Kloss14a,Kloss14b}. 
One of the aims
 of this article is to show how this method can be implemented to study  
Navier-Stokes turbulence. 

Let us now turn to other works on turbulence  using similar 
functional RG methods 
\cite{Tomassini97,leDoussal13,Monasterio12,Mathey14,Pagani15}.
 During the publication process of this manuscript, Kraichnan's 
model has  been studied using NPRG 
 methods in \cite{Pagani15}. The leading corrections to the exponents of the 
structure functions have been computed within this framework and they reproduce 
the known results, which provides an independent validation of the method.

As for the NS problem, the settings developed in Refs.~\cite{Tomassini97} and 
 \cite{Monasterio12} are closely related to the one presented here.  
    The relevant fixed point for turbulence  was already found in $d=3$ in 
Ref.~\cite{Tomassini97}. However, 
  important elements concerning the symmetries,  and 
multiscaling, were not identified in this early work, and we here bring them out.
 In particular, the regularization scheme chosen in Ref.~\cite{Tomassini97} prevented 
   from addressing the $d=2$ case. In contrast, the formalisms of Ref.~\cite{Monasterio12} and of the present work
    cure this problem and enable one to study both bidimensional and tridimensional turbulence
 within a unified framework. The main difference between the work of Ref.~\cite{Monasterio12} and ours is that 
 the former focuses on power-law forcing whereas we consider a forcing localized at a definite
 external or integral scale. The use of power-law forcing
 is essential in most perturbative treatments because only the limit
 of long-range enough forcing is well-controlled perturbatively \cite{Adzhemyan98b}.
  An extrapolation (depending on the dimension) is then required 
 to recover the behavior corresponding to a forcing dominated by the
integral scale, which constitutes an essential difficulty in most perturbative analyses.
In contrast, we  show that in the nonperturbative framework, this strategy
 is unnecessary. A fixed point that describes turbulence with integral-scale forcing is approached
 without the need of  power-law forcing. In fact, in \cite{Monasterio12}, both a power-law and a peaked
 component of the forcing are considered. Interestingly,  the authors
 observe a transition in the $\epsilon$-dependence of observables at a threshold value of $\epsilon$,
   beyond which  the power-law forcing plays a subdominant role 
 compared to the localized forcing. 
 A similar transition between a long-range (LR) regime with $\epsilon$-dependent
exponent and a
short-range (SR) regime with $\epsilon$-independent exponent, occurring at a
critical
value $\epsilon(d)$ of $\epsilon$, also exists in the KPZ
model in the presence of
 both a microscopic delta-correlated noise (SR) and a power-law noise (LR). In this model, the transition
from one
regime to the other is naturally explained by the presence of {\it two}
fixed points whose
stability and basin of attraction depend on $\epsilon$ \cite{Kloss14a}. At large
$\epsilon$ where the LR noise
is relevant, the long-distance behavior of the model is governed by one
fixed point
(the LR fixed point), characterized by  critical exponents depending
  on $\epsilon$. As $\epsilon$ is decreased, LR moves and
eventually
crosses the usual KPZ fixed point (the SR fixed
point) at $\epsilon_c(d)$. Below $\epsilon_c(d)$, the stability of
SR and LR are
interchanged and SR becomes fully attractive: the LR power-law part of the
noise  no longer plays a role.
We conjecture here that the same scenario  occurs in
the NS case: In the presence of both a power-law forcing and
a forcing centered at the integral scale,
{\it two} fixed points exist which collide and exchange their stability at
$\epsilon=\epsilon_c(d)$ (with $\epsilon_c(d=3)=3/2)$. Under this
hypothesis, the explanation
of the "saturation" of the exponent for $\epsilon>\epsilon_c(d)$ boils
down to the change of stability
of the two fixed points when they collide. An interesting outcome of our NPRG
approach is to show that  the fixed-point corresponding to a "physical"
 forcing applied at the integral scale only, can be described without
the need to introduce  a power-law component of the forcing.

Our analysis is based on the general strategy of the NPRG in its modern 
implementation \cite{Berges02},
 adapted for classical non-equilibrium systems \cite{Canet04a,Canet11b}. The  
NPRG formalism to study the NS equation
 is set up and presented in Sec. \ref{SEC-NPRG}.
As common with field-theoretic methods, the symmetries play a crucial role.
 The accuracy of the results obtained with the NPRG approach depends 
 on the order of the approximation implemented, and 
 to preserve all the symmetries of the initial problem along the RG flow is of 
particular importance to ensure that it
 takes place in the appropriate functional (in fields, momenta and frequencies) 
space. 
  The Navier-Stokes field theory
  admits, besides the well-known Galilean 
 invariance and its time-gauged  (also named time-dependent, or extended) 
version 
\cite{DeDominicis79,teodorovich89,Adzhemyan99,Adzhemyan94,Antonov96,berera07},  
another gauge symmetry, presented in Ref.~\cite{Canet15a}.
 Both these gauge symmetries are briefly reviewed in Sec. \ref{SEC-SYM}.

 We then follow two complementary routes.
 The first one is closely related to the works of Refs. 
\cite{Canet10,Canet11a,Kloss12,Kloss14a,Kloss14b} on the KPZ equation,
 and also to \cite{Tomassini97,Monasterio12} in the NS context.
  An ansatz for the 
scale-dependent generating functional $\Gamma_\kappa$ of the
(one-particle-irreducible) correlation and response functions is proposed and 
its evolution is followed
 between the Kolmogorov microscale and the macroscopic scale. 
 The choice of the ansatz is strongly constrained by 
 the gauge symmetries of the NS field theory, which hence play a fundamental 
role. 
 We  begin in Sec. \ref{SEC-LO} with building the appropriate ansatz (at Leading 
Order (LO) approximation), which exactly encodes these symmetries, and
  derive the corresponding NPRG flow equations. We show in Sec. \ref{SEC-LOnum} 
that the RG flow is generically (without fine-tuning any parameter) attracted
 towards a fixed point, which corresponds to stationary fully developed 
turbulence  generated by integral-scale forcing, both in $d=2$ and $d=3$. 

 The scaling properties of the turbulent steady state are analyzed within this 
approximation. Let us recall that, in $d=3$, Kolmogorov K41 theory predicts a 
$p^{-5/3}$ decay of the energy spectrum in the inertial range, associated with 
the direct cascade of energy.
 In $d=2$, part of the energy is transferred towards the large scale in the inverse 
cascade \cite{kraichnan67a}, with a $p^{-5/3}$ spectrum, whereas enstrophy flows 
towards the small scale in the direct cascade, yielding a steeper $p^{-3}$ 
energy spectrum, according to the Kraichnan-Batchelor (KB) theory  
\cite{kraichnan67a,batchelor69}.
 Within the LO approximation, we find that the energy spectrum (and the 
second-order structure function) computed at the fixed point  follow
   K41 predictions in $d=3$ and KB ones in $d=2$. This 
was already observed by Tomassini in  $d=3$ \cite{Tomassini97}, and by the authors of \cite{Monasterio12}
 in the regime dominated by the localized forcing. 
This observation  is compatible with 
experiments and numerical simulations, in so far as they find  very small -- if 
any -- deviations from the dimensional scaling for these quantities 
\cite{Frisch95}. 
  One hence needs to very precisely study the large wave-number behavior to 
determine whether  there exist such corrections.
  However, the LO approximation is not appropriate for this. Indeed, 
 even though this approximation is well controlled and reliable for quantities 
defined at wave-numbers smaller than or comparable to the inverse integral 
scale, such as the existence of the fixed point, it is  not justified  at 
wave-numbers much larger than the inverse integral scale and its predictions in 
this regime should be  taken with care. 
 
 In Sec. \ref{SEC-LP}, we undertake a  second complementary strategy, 
specifically focusing on the large wave-number regime of the NPRG flow 
equations, and deeply rooted in the symmetries.  This second approach has never been undertaken in previous works.
 We show that  a set of {\it exact} and {\it closed} flow
  equations for the two-point functions can be derived in the large (compared to 
the running scale of the flow) wave-number   regime, by using  the 
Ward identities ensuing from the gauge symmetries of the theory.
From these equations, 
 we prove (without approximations) that the exact fixed point {\it does not} 
entail the usual scale invariance in the large wave-number regime, and we 
expound the mechanism of emergence of multiscaling within the NPRG framework. It 
originates in this formalism in a violation of the property 
 of decoupling of the wave-number scales (the two fundamental UV and IR scales 
play a role all along the flow), which is not encountered  in ordinary critical 
phenomena.  The consequence is that the behavior at large-wave numbers
   of the correlation functions (the exponent of the power law) is not fixed in 
term of the scaling dimension of the velocity field and  hence may deviate  from 
dimensional predictions (K41 in $d=3$ and KB in $d=2$).   The non-decoupling 
property is related to   the absence of a regular limit when the integral scale 
(the typical length scale of energy injection) tends to infinity.
 It means that the salient scale for intermittency is the integral scale and not 
the UV one. The same observation underlies the OPE  approaches in the 
perturbative context \cite{Adzhemyan08}. However, the precise link between the 
absence of operators of negative dimensions in the OPE and the non-decoupling 
property is not straightforward,
 and deserves further investigations.
 We emphasize that 
 these exact equations in the large wave-number sector
 complement the flow equations obtained from the LO approximation, which is 
valid in the small wave-number regime. Their numerical solution, that we leave 
for a future 
 publication, should allow us to obtain from first principles
 (although approximately) the intermittency 
exponents for the two-point  functions.

\section{NPRG formalism for Navier-Stokes turbulence}
\label{SEC-NPRG}

\subsection{Navier-Stokes field theory}

The NS equation (\ref{ns}) in the presence of the stochastic forcing $\vf$ 
 formally resembles a Langevin equation. One can resort to the 
standard Martin-Siggia-Rose-Janssen-de Dominicis procedure 
\cite{martin73,Janssen76,DeDominicis76} to derive the associated field theory. 
Following Ref. \cite{Canet15a},  we introduce
  Martin-Siggia-Rose response fields $\bar v_\alpha$ and $\bar p$  to enforce 
both the equation of motion  (\ref{ns}) and the incompressibility constraint 
(\ref{compress}). Note that in this derivation, the pressure field is kept 
(instead of being eliminated as the solution,  expressed in terms of the 
velocity and of the forcing, of a Poisson equation), because the pressure sector 
turns out to be very simple to handle since it is  not renormalized 
\cite{Canet15a}. Once the response fields are introduced, the stochastic forcing 
can be integrated out and  one obtains the generating functional \cite{Canet15a}
\begin{align}
 {\cal Z}[\vJ,\bar{\vJ},K,\bar{K}] &= \int \mathcal{D}\vv \,\mathcal{D}p 
\,\mathcal{D}\bar{\vv} \,\mathcal{D}\bar p\,
  \, e^{-({\cal S}_0[\vv,\bar{\vv},p,\bar p] +\Delta {\cal 
S}_{0,L^{-1}}[\vv,\bar{\vv}])} \nonumber\\
 &\times  e^{\int_{\bx}\{ \vJ\cdot \vv+\bar{\vJ}\cdot \bar{\vv}+K p+\bar K\bar 
p\}}
\label{Z}
 \end{align}
where $\vJ$,  $K$, $\bar{\vJ}$ and $\bar K$ are sources for the velocity,  
pressure and response fields \footnote{
 Note that no Grassmann fields have been introduced, because the It$\bar{\rm o}$'s 
discretization has been chosen, in which the Jacobian involved in the MSRJD 
procedure is simply a constant (independent of the fields), absorbed in the 
normalization, see Ref.\ \cite{Canet15a} for the detailed derivation of the 
generating functional.
 The consequences of this choice and of causality are expounded in Refs. \ 
\cite{Benitez12,Canet11b}, and taken into account throughout this work.},
 and where the NS action, splitted in a local and a nonlocal contribution for 
convenience, is given by
\begin{align}
 {\cal S}_0[\vv,\bar{\vv},p,\bar p]&= \int_{\bx}\Big\{\bar p 
(\bx)\,\partial_\alpha v_\alpha(\bx)+\bar v_\alpha(\bx)\Big[ \partial_t 
v_\alpha(\bx)
 \nonumber\\
 & -\nu \nabla^2 v_\alpha(\bx)+v_\beta (\bx)\partial_\beta v_\alpha(\bx)+\frac 
1\rho \partial_\alpha p (\bx) \Big]\Big\}\nonumber\\
 \Delta {\cal S}_{0,L^{-1}}[\vv,\bar{\vv}]&=-\int_{t,\vec x,\vec x'}\bar 
v_\alpha(t,\vec x) N_{L^{-1},\alpha \beta}(|\vec x-\vec x'|)\bar v_\beta(t,\vec 
x').
\label{NSaction}
\end{align}
where ${\bx}\equiv(t,\vx)$ and $\int_\bx \equiv \int d^d \vx dt$.
 Let us now discuss the choice of the forcing profile. Without loss of 
generality, in order to preserve rotational invariance along the flow, it can be 
written in Fourier space as
\begin{equation} 
N_{\kappa,{\alpha \beta}}(\vq) \equiv \delta_{\alpha\beta}\, 
N_\kappa(\vq)+q_\alpha q_\beta \hat N_\kappa(\vq),
\end{equation}
where the inverse integral scale $L^{-1}$ is denoted $\kappa$ in anticipation 
since  it will be running in the following. 
The  Fourier convention, used throughout this work, is
\begin{align}
 f(\bq)  \! &= \! \!\int_{\bx} \! \! \! f({\bx})\! \; e^{-i \vq \cdot\vx + 
i\omega t} ,\nonumber\\
 f(\bx) \! &=  \!\! \int_{\bq} \! \! \! f (\bq)\, e^{i \vq \cdot\vx - i\omega 
t},
\end{align}
where  $\bq\equiv (\omega,\vq)$ and $\int_\bq \equiv\int \frac{d^d 
\vq}{(2\pi)^d}\frac{d\omega}{2\pi}$. In practice, due to the incompressibility 
condition, the term proportional to $q_\alpha q_\beta$ plays no role and
 can be omitted.
 We further parametrize the function $N_\kappa$ as
\begin{equation}
N_\kappa(\vq) = D_\kappa \left({|\vq|}/{\kappa}\right)^2 \hat 
n\left({|\vq|}/{\kappa}\right) \label{dkdef}
\end{equation} 
where  $D_\kappa$ is a  scale-dependent coefficient, discussed  in 
Sec.~\ref{dimensionlessflow}.
 $N_\kappa(\vq)$ vanishes at $\vq=0$ in order not to imprint a global motion to 
the fluid. The
  stirring force profile $\hat n$, peaked at the inverse integral scale 
$\kappa$, can be  typically shaped as
\begin{equation}
\hat n(x) = e^{-x^2}.\label{hatn}
\end{equation}
Let us report that ten different forcing profiles have been studied in Ref. 
\cite{Tomassini97}, which shows that the influence of the precise form of the 
stirring  is negligible, or inexistent (as expected from universality)  in the 
sense that  the  properties
 of the turbulent flow in the stationary regime do not change. We can hence 
restrict our analysis to the specific profile (\ref{hatn}).  It also corresponds
 to the local component of the forcing chosen in \cite{Monasterio12}.

\subsection{NPRG formalism}

The general NPRG formalism for  nonequilibrium systems in classical physics is 
presented  in details in Refs.\  \cite{Canet04a,Canet11b}. 
In the spirit of Wilson's RG ideas, it consists in building  a sequence of 
scale-dependent  effective models   
such that fluctuations are smoothly averaged as the (wave-number) scale $k$ is 
lowered 
from the  microscopic UV scale $k=\eta^{-1}$ (inverse Kolmogorov scale), where 
no fluctuations are yet included, to the macroscopic IR scale $k=0$ (infinite 
volume), where they
 are all summed over \cite{Berges02,Delamotte05}.
 The procedure is formally the same as in equilibrium \cite{Berges02}, 
 but with the presence of response fields, and additional requirements
 stemming from  It$\bar{\rm o}$'s discretization and  causality issues 
\cite{Canet11b,Benitez12b}. 

In this work, we identify  the RG wave-number scale $k$ with the inverse of the 
integral scale $\kappa$, {\it i.e.}  $k\equiv \kappa$. The integral scale is 
therefore running and  eventually sent to infinity when $k\to 0$ (similarly to Refs. \cite{Tomassini97,Monasterio12}). For other 
purposes,  the integral (injection) scale and the inverse volume scale can be 
kept independent, and the RG scale chosen as one of them while keeping the other 
fixed.  For instance, the study of the RG flow at a fixed integral scale $L$ in 
the infinite volume limit  (RG scale tends to zero) would be relevant to access 
the properties of the inverse cascade in bidimensional turbulence. This 
important issue will be investigated in a future work. 

To achieve the separation of fluctuation modes within the NPRG procedure, a 
wave-number and scale-dependent quadratic (regulator) term  $\Delta {\cal 
S}_\kappa$ is  added to the original action ${\cal S}_0 +\Delta {\cal 
S}_{0,L^{-1}}$. On the one hand, we let the inverse integral scale run in 
the nonlocal quadratic term $\Delta{\cal S}_{0,L^{-1}}\to \Delta{\cal 
S}_{0,\kappa}$  (as in \cite{Tomassini97} and \cite{Monasterio12}). On the other hand, we include an 
additional scale dependent quadratic term  to obtain the following regulator 
\begin{align}
 \Delta {\cal S}_\kappa[\vv,\bar \vv] &=-\int_{t,\vec x,\vec x'}\bar 
v_\alpha(t,\vec x) N_{\kappa,\alpha \beta}(|\vec x-\vec x'|)\bar v_\beta(t,\vec 
x')\nonumber \\
&+\int_{t,\vec x,\vec x'}\bar v_\alpha(t,\vec x) R_{\kappa,{\alpha \beta}}(|\vec 
x-\vec x'|) v_\beta(t,\vec x').
\label{deltaSk}
\end{align}
 The additional $R_\kappa$ term, proportional to the velocity, can be 
interpreted as an Eckman friction term. Its presence is fundamental in $d=2$ to 
damp energy transfer towards larger and larger scales. It introduces an 
effective energy dissipation at the boundary of the effective volume  
$\kappa^{-d}$. Its effect is hence to  suppress fluctuations with wavenumbers 
smaller than $\kappa$. Varying the scale $\kappa$ of $R_\kappa$ is  conceptually 
 equivalent to varying the volume of the system, it is hence analogous in a way 
to studying finite-size scaling. 
As previously, by using the incompressibility of the flow, the  function 
$R_\kappa$ can be chosen diagonal. We write it in Fourier space as
\begin{equation}
 R_{\kappa,{\alpha \beta}}(\vq) =\delta_{\alpha\beta}\, R_\kappa(\vq)= 
\delta_{\alpha\beta}\, \nu_\kappa \,\vq\,^2 \, \hat r(q^2/\kappa^2) 
\label{rkdef}
\end{equation}
 with $q=|\vq|$ and where $\nu_\kappa$ is the scale-dependent viscosity, 
discussed in Sec. \ref{dimensionlessflow}.
The cutoff function $\hat r(x)$  ensures the selection of 
fluctuation modes: $\hat r(x)$ is required to almost vanish for $x\gtrsim 1 $ 
such that 
the fluctuation modes $v_\alpha(q \gtrsim \kappa)$ and $\bar v_\alpha(q \gtrsim 
\kappa)$ are unaffected by 
 the $R_\kappa$ term in $\Delta {\cal S}_\kappa$,
and to be large  when $x\lesssim 1 $ such that the other modes 
($v_\alpha(q\lesssim \kappa)$ and $\bar v_\alpha(q\lesssim \kappa)$) are 
essentially frozen.
 One can show that the form (\ref{deltaSk}) of  regulator term  preserves all 
the symmetries and 
 causality properties  of the problem  as done in a very similar case in  
\cite{Canet11a}.
 We work here with the following cutoff function
\begin{equation}
\hat r(x)=\frac{a}{e^x -1} 
\label{eq:expReg}
\end{equation}
where $a$ is a free parameter, which can be varied to assess the accuracy of the 
approximation scheme \footnote{Whereas physical quantities do not depend on the 
shape of $R_\kappa$ in the exact theory, they acquire a (spurious) dependence on 
$a$ when approximations are performed. This parameter can therefore be utilized 
to optimize results \cite{canet03a,canet03b}}.
Let us emphasize that the addition of the regulator term $R_\kappa$ is essential 
to properly implement the RG procedure and to correctly regularize the flow, 
both in the UV and in the IR, {as already realized in \cite{Monasterio12}}. This constitutes a fundamental difference 
  with the work of Ref. \cite{Tomassini97}, where the 
term $R_\kappa$ is missing, and only the forcing term $N_\kappa$ acts to  select 
the fluctuation modes. Although the procedure of
  \cite{Tomassini97} qualitatively leads to the correct behavior in $d=3$, it 
clearly prevents from studying the $d=2$ case because the flow equations  are IR 
divergent in this dimension without the $R_{\kappa}$ term. 
Conversely, the NPRG flow equations derived in  \cite{Monasterio12} and in 
the present work are properly  regulated in any dimensions.

In the presence of the regulator term $\Delta {\cal S}_\kappa$, the generating 
functional (\ref{Z}) becomes
 scale dependent
\begin{align}
 {\cal Z}_\kappa[\vJ,\bar{\vJ},K,\bar K] &= \int \mathcal{D}\vv \,\mathcal{D}p 
\,\mathcal{D}\bar{\vv} \,\mathcal{D}\bar p\,
  \, e^{-({\cal S}_0[\vv,\bar{\vv},p,\bar p] +\Delta {\cal 
S}_\kappa[\vv,\bar{\vv}])} \nonumber\\
 &\times  e^{\int_{\bx}\{ \vJ\cdot \vv+\bar{\vJ}\cdot \bar{\vv}+K p+\bar K\bar 
p\}}.
\label{Zk}
 \end{align}
  Field expectation values in the presence of the external sources $\vJ$, 
$\bar{\vJ}$, $K$, and $\bar K$ 
 are obtained as functional derivatives of  ${\cal W}_\kappa = \log {\cal 
Z}_\kappa$ as 
\begin{equation}
  u_\alpha({\bx}) = \langle v_\alpha({\bf x}) \rangle = \frac{\delta {\cal 
W}_{\kappa}}{\delta J_\alpha({\bf x})}  \, \, , \, \,
  \bar u_\alpha({\bf x}) = \langle \bar v_\alpha({\bf x}) \rangle = \frac{\delta 
{\cal W}_{\kappa}}{\delta \bar J_\alpha({\bf x})} \nonumber
\end{equation} 
 and similarly for the pressure fields, for which  for simplicity the same 
notation can be kept for the fields and their average values  
\begin{equation}
  p({\bf x}) \equiv \langle p({\bf x}) \rangle = \frac{\delta {\cal 
W}_{\kappa}}{\delta K({\bf x})}  \, \, , \, \,
  \bar p({\bf x}) \equiv \langle \bar p({\bf x}) \rangle = \frac{\delta {\cal 
W}_{\kappa}}{\delta \bar K({\bf x})} .\nonumber
\end{equation} 
The effective average action $\Gamma_\kappa[\vu,\bar\vu,p,\bar p]$ is defined as the 
Legendre transform of  ${\cal W}_\kappa$
(up to  terms proportional to $R_{\kappa,\alpha\beta}$ or 
$N_{\kappa,\alpha\beta}$) \cite{Berges02,Canet11b}:
\begin{align}
\Gamma_\kappa[\vu,\bar\vu,p,\bar p]& +{\cal W}_\kappa[\vJ,\bar{\vJ},K,\bar K] = 
\int_{\vx} \! j_i \varphi_i \nonumber\\
 &-\int_{t,\vx,\vx'} \Big\{ \bar u_\alpha \,R_{\kappa,\alpha \beta} \,u_\beta 
- \bar u_\alpha \,N_{\kappa, \alpha \beta} \,\bar u_\beta \Big\}
\label{legendre}
\end{align}
where  $\varphi_i$, $i=1,\dots,4$, stand for the fields $u_\alpha,\bar 
u_\alpha,p$ and $\bar p$, respectively,  and  $j_i$
  for the sources $J_\alpha,\bar J_\alpha,K$ and $\bar K$, respectively. 
 From $\Gamma_\kappa$, one can derive 2-point correlation and response 
functions, which can be gathered in a $4\times4$ matrix as
\begin{equation}
[\,\Gamma_\kappa^{(2)}\,]_{i_1 i_2}({\bf x}_1,{\bf x}_2, \{\varphi_i\}) = 
\frac{\delta^2 \Gamma_\kappa[\{\varphi_i\}]}{\delta\varphi_{i_1}({\bf 
x}_1)\delta\varphi_{i_2}({\bf x}_2)}
\end{equation}
and more generally $n$-point correlation functions that are also written 
  in a $4\times4$ matrix form  as
\begin{equation}
\Gamma_{\kappa,i_3,...,i_n}^{(n)}({\bf x}_1,...,{\bf x}_n, \{\varphi_i\}) =
\frac{\delta^{n-2} \Gamma_\kappa^{(2)}({\bf x}_1,{\bf x}_2, \{\varphi_i\})}
{\delta\varphi_{i_3}({\bf x}_3)...\delta\varphi_{i_n}({\bf x}_n)} \;.
\end{equation}
The exact flow for $\Gamma_{\kappa}[\vu,\bar\vu,p,\bar p]$ is given  by 
Wetterich 
equation, which reads in Fourier space \cite{Wetterich93,Berges02}
\begin{equation}
\partial_\kappa \Gamma_\kappa = \frac{1}{2}\, {\rm Tr}\! \int_{\bf q}\! 
\partial_\kappa {\cal R}_\kappa(\bq) \cdot G_\kappa (\bq) ,
\label{dkgam}
\end{equation}
where  ${\cal R}_\kappa(\bq)$ is the Fourier transform of the $4\times4$ matrix 
${\Delta {\cal S}^{(2)}_\kappa}$.
One can infer from definition (\ref{deltaSk}) that its sole non-vanishing 
elements are $[{\cal R}_\kappa]_{2 2} = -2 N_{\kappa,\alpha\beta}$
 and  $[{\cal R}_\kappa]_{1 2} =[{\cal R}_\kappa]_{2 1}= 
R_{\kappa,\alpha\beta}$. The matrix
\begin{equation}
 G_\kappa\equiv \left[\Gamma_\kappa^{(2)}+{\cal R}_\kappa\right]^{-1}
 \label{eq:propag}
\end{equation}
is the full, that is, field-dependent, renormalized at scale $\kappa$ propagator 
of the theory.
When the RG scale $\kappa$ is lowered from the UV scale $\eta^{-1}$ to zero, 
$\Gamma_\kappa$ interpolates 
between the microscopic model $\Gamma_{\kappa=\eta^{-1}}={\cal S}_0$  and the 
full effective action $\Gamma_{\kappa=0}$
 that encompasses all the macroscopic properties of the system (for a detailed 
discussion in nonequilibrium processes, see Ref.~\cite{Canet11b}).
Differentiating Eq.~(\ref{dkgam}) twice with respect to the fields
and evaluating the resulting identity in a uniform and stationary field 
configuration $\varphi_i(\bx)=\varphi_i$ (since the model is analyzed in its 
long time and large distance regime where it is translationally invariant in 
space and time)
one obtains the flow equation for the 2-point functions:
\begin{eqnarray}
\partial_\kappa \Gamma^{(2)}_{\kappa,ij}(\bp)\! &=& \! {\rm Tr}\! \int_{\bq} 
\partial_\kappa {\cal R}_\kappa(\bq) \cdot G_\kappa(\bq) \cdot
\!\bigg(\!\!-\!\frac{1}{2}\, \Gamma^{(4)}_{\kappa,ij}(\bp,-\bp,\bq) \nonumber\\
&& \hspace{-2.2cm} +\Gamma^{(3)}_{\kappa,i}(\bp,\bq) \cdot G_\kappa(\bp+\bq) 
\cdot
\Gamma^{(3)}_{\kappa,j}(-\bp,\bp+\bq) \bigg) \cdot G_\kappa(\bq)
\label{dkgam2}
\end{eqnarray}
where the background field $\varphi_i$ dependencies are implicit,
as well as the last arguments of the $\Gamma_\kappa^{(n)}$ which are determined 
by
frequency and wave-vector conservation \cite{Canet11b}.

Of course Eq.~(\ref{dkgam})  cannot be solved exactly and  one has to resort to 
an appropriate   approximation scheme, 
adapted to the physics of the model under study,  and in particular to its 
symmetries, that are reviewed in the next section.

\section{Symmetries and related Ward identities}
\label{SEC-SYM}
In this section, we briefly review the three gauge symmetries of the NS action 
expounded in Ref.~\cite{Canet15a} and the  ensuing  non-renormalization theorems 
and general Ward identities derived in this Reference.

\subsection{Symmetries}
\label{SECpressure}

The NS action  ${\cal S} \equiv {\cal S}_0+\Delta{\cal S}_{\kappa}$ given by 
(\ref{NSaction})  admit three gauge symmetries:
\begin{itemize}
 \item  (i) invariance under gauged shifts of the pressure fields,
 \item (ii) time-gauged Galilean symmetry,
 \item (iii) invariance under a time-gauged shift of the response fields.
\end{itemize}
The symmetry (i) is the invariance of ${\cal S}$ under the local  shifts 
$p(t,\vx)\to p(t,\vx)+\epsilon(t,\vx)$ or 
$\bar p(t,\vec x)\to \bar p(t,\vec x)+\bar \epsilon(t,\vec x)$, which implies 
that the equations of motion for $p$ and $\bar p$ are exactly given by  the 
minimization of the  bare action ${\cal S}_0$. The infinitesimal time-gauged 
Galilean symmetry~(ii), also referred to as time-dependent 
\cite{Adzhemyan94,Adzhemyan99,Antonov96}, or extended \cite{berera07,berera09} 
Galilean symmetry, consists in the following  field transformation 
\begin{align} 
 \delta v_\alpha(\bx)&=-\dot{\epsilon}_\alpha(t)+\epsilon_\beta(t) 
\partial_\beta v_\alpha(\bx)\nonumber\\
 \delta \bar v_\alpha(\bx)&=\epsilon_\beta(t) \partial_\beta \bar 
v_\alpha(\bx)\nonumber\\
 \delta p(\bx)&=\epsilon_\beta(t) \partial_\beta p(\bx)\nonumber\\
 \delta \bar p(\bx)&=\epsilon_\beta(t) \partial_\beta \bar p(\bx)
\label{defG}
\end{align}
where $\dot{\epsilon}_\alpha = \p_t \epsilon_\alpha$. When 
$\vec\epsilon(t)\equiv \vec \epsilon$ is an arbitrary constant vector,   the 
transformation corresponds to a translation in space, and  when 
$\vec\epsilon(t)\equiv \vec \epsilon\;t$ it corresponds to the usual 
(non-gauged) Galilean transformation.
Lastly, the infinitesimal time-gauged shift symmetry~(iii) consists in the field 
transformation 
\begin{align}
 \delta \bar v_\alpha(\bx)&=\bar \epsilon_\alpha(t) \nonumber\\
 \delta \bar p(\bx)&= v_\beta(\bx) \bar \epsilon_\beta(t).\label{shiftjauge}
\end{align}
For each of these transformations,  the different terms of the NS action ${\cal 
S}$
 are either invariant, or have a linear variation in the fields.  The 
corresponding non-invariant terms play the role of gauge fixing controllable 
terms.
By explicitly performing  these transformations as changes of variables in the 
functional integral (\ref{Zk}),
 and exploiting that they must leave it unaltered, one  deduces general Ward 
identities. These identities
 are derived in Ref.~\cite{Canet15a} for the original (microscopic) NS field 
theory. The procedure
 can be directly transposed to the scale dependent effective average action 
$\Gamma_\kappa$, by simply replacing the original quantities ({\it e.g.} ${\cal 
Z}$,  $\Gamma$) by the running ones  ({\it e.g.} ${\cal Z}_\kappa$, 
$\Gamma_\kappa$). Note also that since  $\Delta {\cal S}_\kappa$  is substracted 
in the (modified)  definition of the Legendre transform (\ref{legendre}) in 
contrast to Ref.~\cite{Canet15a},  the corresponding terms (proportional to 
$N_{\kappa,\alpha\beta}$ or $R_{\kappa,\alpha\beta}$) are removed from the Ward 
identities.
These identities are recapitulated below.

\subsection{Ward identities}

The Ward identities ensuing from the gauged-shift symmetries~(i) simply read 
\begin{equation}
 \frac{\delta \Gamma_\kappa}{\delta p(\bx)}=\frac{\delta {\cal S}_0}{\delta 
p(\bx)}\quad \hbox{and}\quad\frac{\delta \Gamma_\kappa}{\delta \bar 
p(\bx)}=\frac{\delta {\cal S}_0}{\delta \bar p(\bx)}
\end{equation}
which means that the dependence in $p(\bx)$ and $\bar p(\bx)$ of both  the 
effective action  $\Gamma_\kappa$ and  the bare one ${\cal S}_0$ are identical. 
One thus concludes that the whole pressure sector is not renormalized.
Of course, connected correlation functions of the pressure do have corrections coming
 from fluctuations. This is a simplifying feature of the 1-PI effective action,
  which  keeps exactly the same pressure dependence as the bare action.

The NS action is invariant under the time-gauged Galilean transformation 
(\ref{defG}), but for the term proportional to the Lagrangian time derivative 
$ D_t v_\alpha(\bx)\equiv  \partial_t v_\alpha(\bx)+ v_\beta(\bx) \partial_\beta 
v_\alpha(\bx)$,
which variation is
\begin{equation}
  \delta \int_{\bx} \bar v_\alpha(\bx) D_t  v_\alpha(\bx) \equiv \delta {\cal 
S}=
 -\int_{\bx}\ddot{\epsilon}_\alpha(t) \bar v_\alpha(\bx).
\end{equation}
Hence, requiring that the change of variables  (\ref{defG}) leaves the functional 
integral (\ref{Zk}) unaltered, one obtains the following  Ward identity 
\begin{align}
\label{wardgalilee}
&  \int_{\vx} \Big\{\big(\delta_{\alpha\beta}\p_t +  \partial_\beta 
u_\alpha(\bx)\big)\frac{\delta \Gamma_\kappa}{\delta u_\alpha(\bx)}   \nonumber
+ \partial_\beta \bar u_\alpha(\bx) \frac{\delta \Gamma_\kappa}{\delta \bar 
u_\alpha(\bx)} \nonumber\\
 &+\partial_\beta p(\bx) \frac{\delta \Gamma_\kappa}{\delta p(\bx)}  
 +\partial_\beta \bar p(\bx) \frac{\delta \Gamma_\kappa}{\delta \bar p(\bx)} 
\Big\}= -\int_{\vx} \p_t^2 \bar u_\beta(\bx),
\end{align}
which implies that the variations of both the effective action and  the bare one 
under time-gauged Galilean transformations are identical.
This entails that, apart from the term $\int_{\bx}\bar u_\alpha(\bx) D_t 
u_\alpha(\bx)$ which is not renormalized and remains equal to its bare 
expression, $\Gamma_\kappa$ is invariant under these transformations.

As for the time-gauged shift symmetry~(iii), the variation of the NS action 
under (\ref{shiftjauge}) is
 \begin{equation}
 \delta {\cal S} = -\int_{\bx}  \dot{\bar\epsilon}_\beta(t) v_\beta(\bx),
\end{equation}
and the related  Ward identity reads
 \begin{equation}
\label{wardshift}
\int_{\vx} \Big\{\frac{\delta \Gamma_\kappa}{\delta \bar u_\alpha(\bx)}
 + u_\alpha(\bx) \frac{\delta \Gamma_\kappa}{\delta \bar p(\bx)} \Big\}= 
\int_{\vx}  \p_t u_\alpha(\bx),
\end{equation}
meaning that, apart from the term $\int_{\bx} \bar u_\alpha \partial_t u_\alpha$ 
which is not renormalized, the effective action $\Gamma_\kappa$ is invariant 
under  time-gauged shift transformations.\\

 Furthermore, as shown in Ref. \cite{Canet15a}, a functional Ward identity 
associated with a fully gauged (both in time and space) version of the shift 
symmetry~(\ref{shiftjauge}) can be obtained
  in the presence of a local source term bilinear in the velocity field. 
 This functional identity entails an infinite set of {\it exact} and {\it local} 
relations between correlation functions,
 which includes in particular the K\'arm\'an-Howarth relation \cite{Karman38}.
 From this fundamental relation,   the exact Kolmogorov  law for the 
third-order structure function in $d=3$ can be derived, assuming the existence
 of a dissipative anomaly (finite mean dissipation rate in the inviscid limit).
  The gauged shift symmetry hence plays a 
crucial role since it directly roots  in  symmetries the four-fifth law.
 From the functional Ward idendity  can also be deduced
 another exact identity  for a fourth-order pressure-velocity 
correlation function
 recently derived  in Ref.~\cite{Falkovich10}, and further generalized in 
\cite{Canet15a}.

\subsection{General structure of the effective action $\Gamma_\kappa$}

One can infer from the previous Ward identities the general form of the 
effective action $\Gamma_\kappa$:
\begin{align}
\Gamma_\kappa[\vu,\bar \vu,p,\bar p] &= \int_{\bx}\Big\{ \bar u_\alpha 
\Big(\partial_t u_\alpha+ \lambda u_\beta \partial_\beta u_\alpha 
+\frac{\partial_\alpha p}{\rho}\Big)\nonumber\\
& +\bar p \partial_\alpha u_\alpha\Big\} +\tilde \Gamma_\kappa[\vu,\bar \vu]
 \label{anzGk}
\end{align}
where the explicit terms are not renormalized and thus keep their bare forms,
  and the functional  $\tilde \Gamma_\kappa$ is invariant under time-gauged 
Galilean and  shift transformations.
 The coefficient $\lambda$ is introduced in front of the nonlinear term  for 
later power counting purposes. 
Of course, $\lambda$ can always be set equal to one in appropriate units. Note 
that including this coefficient in the original NS action (\ref{NSaction})
  induces some slight modifications of the related Ward identities  by trivial 
factors $\lambda$. Yet,
 it still leads  to  the general form (\ref{anzGk}) of the effective action 
where $\lambda$ is not renormalized.

\section{Conservation laws and energy spectrum} 
\label{SEC-energy}

In this section, we study the different contributions to the energy and to the 
enstrophy
 and we show that the conservation of energy in $d=3$, and of both energy and 
enstrophy in $d=2$, yields constraints which fix the values of  the anomalous 
dimensions.
 We also derive the general expressions for  the energy spectrum and the second 
order structure function.
  These observables  are explicitly computed in Sec.~\ref{SEC-LOnum} at LO 
approximation.
  
 All these quantities can be expressed in terms of  connected two-point 
correlation and response functions: $G^{u  u}_{\alpha\beta}=\langle v_\alpha  
v_\beta\rangle_c$ and   $G^{u \bar u}_{\alpha\beta}=\langle v_\alpha \bar 
v_\beta\rangle_c$. These functions are elements of the propagator matrix  
$G_\kappa$, which is defined as  
  the inverse of the matrix $\Gamma_\kappa^{(2)}$.
 The general structure of the propagator matrix  $G_\kappa$ is determined  in 
Appendix A,  its components in the velocity sector are reported below  
Eq.~(\ref{propGk}). 
  
  Let us  clarify  notation.  Because of  rotational and parity invariance,
   any generic two-(space)index function (in Fourier space) 
$F_{\alpha\beta}(\omega,\vp)$ can be decomposed  into a longitudinal and a 
transverse part 
\begin{equation}
 F_{\alpha\beta}(\omega,\vec p)=P_{\alpha\beta}^\perp(\vec p) 
F_{\perp}(\omega,\vec p\,^2)+P_{\alpha\beta}^\parallel(\vec p) 
F_{\parallel}(\omega,\vec p\,^2)
 \label{proj-decomp}
\end{equation}
where  the transverse and longitudinal projectors are defined by
\begin{equation}
 P_{\alpha\beta}^\perp(\vec p)=\delta_{\alpha\beta}-\frac{p_\alpha p_\beta}{\vec 
p\,^2}, \hspace{.5cm}\mathrm{and}\hspace{.5cm} P_{\alpha\beta}^\parallel(\vec 
p)=\frac{p_\alpha p_\beta}{\vec p\,^2}.
\end{equation}
As shown in Appendix A,  the incompressibility condition entails that
the components of the propagator in the velocity sector are purely transverse 
(that is, all the longitudinal parts vanish) and are given by
\begin{align}
G^{\bar u\bar u}_{\alpha\beta}(\omega, \vec q)&= 0\nonumber\\
 G^{u \bar u}_{\alpha\beta}(\omega, \vec q)&= P_{\alpha\beta}^\perp(\vec q) 
\frac{1}{\Gamma^{(1,1)}_\perp(-\omega, \vec q) + R_\kappa(\vec q)}\nonumber\\
G^{u u}_{\alpha\beta}(\omega, \vec q)&= -P_{\alpha\beta}^\perp(\vec q) 
\frac{\Gamma^{(0,2)}_\perp(\omega, \vec q) -2 N_\kappa(\vec 
q)}{\left|\Gamma^{(1,1)}_\perp(\omega, \vec q)+R_\kappa(\vec q)\right|^2}.
\label{propGk}
\end{align}

For the following discussion, we also need to introduce 
 renormalized and dimensionless quantities, denoted with a hat symbol. The 
wave-vectors and frequencies are respectively measured in units of $\kappa$ and 
$\kappa^2 \nu_\kappa$, where $\nu_\kappa$ is the running viscosity. We thus 
define  {\it e.g.} $\hat p = p/\kappa$ and $\hat \omega = 
\omega/(\kappa^2\nu_\kappa)$.   Let us consider the expression  
({\ref{deltaSk}}) of the term $\Delta{\cal S}_\kappa$. The dimension of the 
cutoff term  $R_{\kappa,\alpha\beta}$ is given by
 \begin{equation}
  \left[  \int_{\vec x'} R_{\kappa,\alpha \beta}(\vec x-\vec x')\right] = \left[ 
\kappa^2 \nu_\kappa\right] \label{nukdef}
 \end{equation}
 and the dimension of the forcing term  $N_{\kappa,\alpha\beta}$  can be 
inferred from  definition (\ref{dkdef})
\begin{equation}
 \left[  \int_{\vec x'} N_{\kappa,\alpha \beta}(\vec x-\vec x')\right] = \left[ 
D_\kappa\right] \label{Dkdim}.
  \end{equation}
  We  associate two running anomalous dimensions $\eta_\kappa^\nu$ and 
$\eta_\kappa^\td$ with these
    running coefficients as
   \begin{equation}
   \eta_\kappa^\nu = -\p_s \ln \nu_\kappa \quad\quad \hbox{and}\quad\quad 
\eta_\kappa^\td=-\p_s \ln D_\kappa
   \end{equation}
where $\p_s \equiv \kappa \p_\kappa$.
According to Eqs.~(\ref{nukdef}) and (\ref{Dkdim}), and since $\Gamma_\kappa$ 
and $\Delta {\cal S}_\kappa$ have the same dimension,  one deduces from Eq. 
(\ref{anzGk}) the dimensions of the fields: $[u] =  \left[\kappa^{d-2}D_\kappa 
\nu_\kappa^{-1}\right]^{1/2}$ and 
 $[\bar u] = \left[\kappa^{d+2}\nu_\kappa D_\kappa^{-1}\right]^{1/2}$. We also 
introduce the dimensionless coupling  $\hat\lambda_\kappa$ as
 \begin{equation}
  \lambda = \left(\kappa^{-d+4}\nu_\kappa^3 D_\kappa^{-1}\right)^{1/2} 
\hat\lambda_\kappa.
  \end{equation}
 Since $\lambda$ is not renormalized, the flow equation for $\hat 
\lambda_\kappa$ is purely dimensional and reads
 \begin{equation}
 \partial_s \hat \lambda_\kappa = \hat\lambda_\kappa\left(\frac d 2 -2 +\frac 3 
2 \eta_\kappa^\nu  - \frac 1 2 \eta_\kappa^\td\right).
\label{dklambda}
\end{equation} 
 Of course, this is a direct consequence of Galilean symmetry and this equation (or similar forms)
 has already been obtained in other RG approaches \cite{Adzhemyan08,Tomassini97}. 
It follows from this equation that  any non-Gaussian fixed point, with $\hat\lambda_\kappa\neq 0$, 
is characterized by a {\it single independent anomalous dimension}, for instance 
$\eta_*^\td$, with
\begin{equation}
 \eta_*^\nu = 4/3 +(\eta_*^\td-d)/3.
\label{valetanu}
\end{equation}
We now show that the value of $\eta_\kappa^\td$ is actually fixed by 
conservation laws.

\subsection{Energy conservation}
\label{SUBSECeta}

 As recalled in the introduction, energy must be permanently injected at the 
integral scale $\kappa^{-1}$
  to maintain a turbulent
 flow since the  NS equation is dissipative. Hence, reaching the stationary 
regime of fully developed turbulence requires that the mean rate of  injected 
power by unit mass $\langle \epsilon_{\rm inj} \rangle$ of the fluid compensates 
the mean rate of dissipated power by unit mass.
Of course, if the full dynamics was studied starting
from well-defined initial conditions, the conservation
of energy would be automatically satisfied at all times
and in particular in the long-time limit where the system
reaches stationarity. Here, we  directly study the
steady state (assuming translational invariance in
time).  Indeed, the inclusion of the transcient regime is challenging and would be interesting to address  in the future.
  Hence, we need  to impose 
  stationarity as an external constraint, in the form of a  balance between
the injected and dissipated energy. This was already realized in \cite{Tomassini97}. In fact, this is reminiscent of the additional assumption (existence of a dissipative  anomaly)
 needed to derive the four-fifth law from the K\'arm\'an-Howarth relation.

There are two sources of energy 
dissipation in the NPRG setting, the dissipation by molecular viscosity at the 
microscopic Kolmogorov scale $\eta$ and the effective dissipation at the 
boundaries induced by the cutoff term  proportional to $R_\kappa$ in 
(\ref{deltaSk}). Denoting $\langle \epsilon_{\rm dis}^{1/\eta} \rangle$ and 
$\langle \epsilon_{\rm dis}^\kappa \rangle$ their respective rate per unit mass, 
 the energy balance equation  in the steady state reads
\begin{equation}
\langle \epsilon_{\rm inj} \rangle = \langle \epsilon_{\rm dis}^{1/\eta} 
\rangle+\langle \epsilon_{\rm dis}^\kappa \rangle.
\label{energybudget}
\end{equation} 
Let us express these different contributions. As explained in Appendix C, 
 the average injected power per unit mass can be expressed as
 \begin{align}
\langle \epsilon_{\rm inj} \rangle&= \big\langle f_\alpha(t,\vx) v_\alpha(t,\vx) 
\big\rangle\nonumber\\
  & =\lim_{\delta t\to 0^+}\int_{\vec x'} N_{\kappa,\alpha\beta}(|\vec x-\vec 
x'|)\, G^{u\bar u}_{\alpha\beta}(t+\delta t,\vec x;t,\vec x')\nonumber\\
&=(d-1) \,D_\kappa \kappa^d  \lim_{\delta t\to 0^+}
\int_{\hat \omega,\hat \vq} \hat N(\hat \vq) e^{-i\hat \omega \hat {\delta 
t}}\hat G^{u\bar u}_\perp(\hat \omega,\hat \vq).
\label{eps-inj}
\end{align} 
The average effective dissipated power at the scale $\kappa$ can be written as
\begin{align}
\langle \epsilon_{\rm dis}^{\kappa} \rangle &= \Big\langle \int_{\vx'} 
v_\alpha(t,\vx) R_{\kappa,\alpha\beta}(|\vx-\vx'|)  
v_\beta(t,\vx')\Big\rangle\nonumber\\
      & = \int_{\omega,\vq} R_{\kappa,\alpha\beta}(\vq) 
\,G^{uu}_{\alpha\beta}(\omega,\vq)\nonumber\\
    & = (d-1)\, D_\kappa \kappa^d \int_{\hat \omega,\hat \vq} \hat q^2 \,\hat 
r(\hat \vq) \hat G^{u  u}_\perp(\hat \omega,\hat \vq).
\label{eps-disk}
\end{align}
Let us emphasize that, in the present analysis, once the scale $\kappa$ of 
$R_\kappa$ and the scale $k$ of $N_k$ have been equated, the  effective energy 
dissipation (at the volume scale) and the  energy injection (at the integral 
one) scale in the same way, as $D_\kappa \kappa^d$.
 The average  dissipated power at the Kolmogorov scale per unit mass is 
expressed as
\begin{align}
 \langle \epsilon_{\rm dis}^{1/\eta} \rangle &= \big\langle \nu  \p_j v_i(t,\vx) 
\p_j v_i(t,\vx)\big\rangle\nonumber\\
      & =\nu (d-1)\int_{\omega,\vq} q^2 
\,G^{uu}_\perp(\omega,\vq)\label{eps-diseta}.
\end{align}
The behavior of this integral can be analyzed using the canonical behavior of 
$G_\perp^{uu}$ (intermittency corrections, if any, can be neglected in this 
argument, see Sec. \ref{SEC-LOnum}). This canonical behavior  can be deduced 
assuming (standard) scale invariance, see Eq.~(\ref{scalingform}) and Sec. 
\ref{SECnonDCP}, and is given in the inertial regime by
\begin{equation}
G^{uu}_\perp(\omega,\vq) = q^{-\eta^\td-4+2\eta^\nu}\, 
g\left({\omega}/{q^{2-\eta^\nu}}\right), \label{canoGuu}
\end{equation}
where $g$ is a scaling function. 
One then obtains 
\begin{align}
 \langle \epsilon_{\rm dis}^{1/\eta} \rangle &=\frac{\nu 
(d-1)}{2^{d-1}\pi^{d/2}\Gamma(d/2)} \int_0^{1/\eta} dq \,q^{\frac 2 3 
(d-\eta^\td)+\frac 1 3}\nonumber\\
 &\times  \int_0^{\infty} \frac{dx}{\pi} g(x) 
\end{align}
where Eq.~(\ref{valetanu}) is used. 
If $\eta^\td \lesssim d+2$, the integral on $q$ is UV divergent when $\eta\to 
0$, which means that it is dominated by the UV scale $\eta^{-1}$. It follows 
that 
if $\eta^\td \lesssim d+2$,  $\langle \epsilon_{\rm dis}^{1/\eta} \rangle$ is 
independent of $\kappa$, and remains finite in the RG (infinite volume) limit
 $\kappa \to 0$. Hence, one deduces that to satisfy the energy budget equation 
(\ref{energybudget}), the average injected power $\langle \epsilon_{\rm inj} 
\rangle$ given by Eq. (\ref{eps-inj}) minus the average effective dissipated power 
 $\langle \epsilon_{\rm dis}^{\kappa} \rangle$ given by Eq. (\ref{eps-disk}) 
should also remain  finite in the limit $\kappa \to 0$.  Accordingly, either the two terms are both finite,
 or if one of them diverges in this limit, the other term must also diverge to compensate it.
This implies $\kappa^d 
D_\kappa \ge \mathcal{O}(1)$ (as the integrals in   (\ref{eps-inj}) and 
(\ref{eps-disk}) are finite), that is $\eta^\td  \ge d$.

\subsection{Case $d=3$}

In $d=3$, the energy is expected to be dominantly dissipated  at the microscopic 
 Kolmogorov scale $\eta$.
 This corresponds to the case $\eta^\td \lesssim d+2$, with an average rate of 
dissipated power $\langle \epsilon_{\rm dis}^{1/\eta} \rangle$  independent of 
$\kappa$.  In the limit $\kappa \to 0$, the effective dissipation at the volume 
scale (fixed by $R_\kappa$) and  the injected energy are of the same order ($\propto \kappa^d 
D_\kappa$). 
However, as explained before, the integral scale and the volume scale 
could be kept independent. In  $d=3$, the volume scale could be safely  removed 
by sending $R_\kappa\to 0$ while keeping $N_\kappa\neq 0$  (as in 
Ref.~\cite{Tomassini97}) and the flow equations would remain well-behaved. 
Otherwise stated, by taking the volume scale much larger than the integral 
scale, the dissipated energy at the volume scale can be rendered negligible 
compared to the injected energy in $d=3$. Accordingly,
the injected energy cannot be compensated by the
energy dissipated at the volume scale alone. This  implies that it must behave  as 
  the energy dissipated at the Kolmogorov scale.
As a consequence, the three terms in (\ref{energybudget}) must scale identically  when $\kappa \to 0$
(that is, they must be independent of $\kappa$ when $\kappa \to 0$).

 We choose to impose this condition all along the RG flow (as should be done 
without the regulator $R_\kappa$).
 \footnote{This choice is justified because we only study  the $\kappa\to 0$ 
limit where the dimensionless flow 
 approaches a universal fixed point, where the precise energy dissipation 
mechanism at large $\kappa$ is washed out.}. To keep a constant injected power 
while taking the RG (infinite volume) limit $\kappa\to 0$ thus requires to fix
 $\eta_\kappa^\td = d$ for all $\kappa$, which means that $N_\kappa(\vq)$ must 
scale as $\kappa^{-d}$.
This yields in $d=3$ that at the fixed point, $\eta_*^\nu = 4/3$ according to 
Eq. (\ref{valetanu}). Hence the values of the two running 
 anomalous dimensions
 at the fixed point are determined. For simplicity, since we are merely 
interested in the fixed point properties,   we  fix  
\begin{equation}
 \eta_\kappa^\td = 3 \hspace{0.5cm}\hbox{and}\hspace{0.5cm} \eta_\kappa^\nu = 
4/3 \hspace{0.5cm}\hbox{in}\hspace{0.5cm} d=3
\label{expod3}
\end{equation}
for all $\kappa$, which has no influence on the fixed point properties.

\subsection{Case  $d=2$}

In $d=2$, the situation is different. The energy is transferred both towards the 
small  scales (direct cascade) and the large scales (inverse cascade) and is 
thus dissipated both at the microscopic Kolmogorov scale and at the boundaries 
of the system.  This is manifest on the NPRG equations (\ref{dkfnu},\ref{dkfd}) that are no longer
regular when sending $R_\kappa\to 0$ while keeping $N_\kappa\neq 0$. Accordingly,
the argument used in $d=3$ (exploiting that the dissipation at the volume scale can be rendered negligible)
  is no longer valid in $d=2$ and  only the previously shown inequality $\eta^\td  \ge d$ holds (one cannot conclude that
 $\eta_\kappa^\td = d$ in this dimension).
In order to fix $\eta^\td$ for $d=2$, another conservation law is necessary.

 One can exploit the enstrophy conservation 
to fix $\eta^\td$ in a similar way that the conservation of energy was used in the
$d=3$ case. The enstrophy flux is towards the small scale, in the 
direct cascade, and  
 the dissipation of vorticity is dominated by the microscopic scale, and is thus 
independent  of $\kappa$.
 The vorticity is defined as  $\vec \omega(t,\vx) = \vec \nabla \times 
\vv(t,\vx)$ which reduces
 in $d=2$ to the (pseudo-)scalar $\omega(t,\vx) = \epsilon_{i j} \p_i 
v_j(t,\vx)$.
 The mean dissipation rate of vorticity can be expressed  as \cite{Gawedzki99}
\begin{align}
\big\langle \omega_{\rm dis} \big\rangle &= \ \big\langle \nu (\vec 
\nabla\omega(t,\vx))^2 \big\rangle \nonumber\\
  &=  \nu \int_{\omega,\vq} \,q^{\,4}\, G^{u 
u}_\perp(\omega,\vq).\label{eps-vort}
\end{align}
Using the canonical behavior (\ref{canoGuu}) of $G^{u u}_\perp$,  one can indeed 
check that for  $\eta_\td <7$ in $d=2$ the   integral on $q$ is UV divergent 
without a cutoff, and is hence  dominated by the UV scale, and thus independent 
of $\kappa$.
 The conservation of enstrophy then requires that the mean injection rate of 
vorticity should also be independent of $\kappa$.
This rate can be expressed as \cite{Gawedzki99}
\begin{align}
  \big\langle \omega_{\rm inj} \big\rangle &= \big\langle (\vec\nabla \times 
\vec f\;)(t,\vx)\cdot \vec \omega(t,\vx)  \big\rangle \nonumber\\
    &=  \lim_{\delta t\to 0^+}\int_{\omega,\vq} q^2\, N(\vq)\, G_\perp^{u\bar 
u}(\omega,\vq) e^{-i\omega \delta t}\nonumber\\
  &=  D_\kappa \kappa^{4}   \lim_{\delta t\to 0^+} \int_{\hat\omega, \hat\vq} 
\hat q^2\, \hat N(\vq)\, \hat G_\perp^{u\bar u}(\hat \omega,\hat \vq) 
e^{-i\hat\omega \hat{\delta t}}.
\end{align}
One hence concludes that to keep a constant rate of injection of vorticity in 
the RG limit $\kappa\to 0$ requires
 to fix $\eta_\kappa^\td=4$ for all $\kappa$.  This identity satisfies the 
constraint $\eta_\kappa^\td \ge d$
 stemming from the energy conservation. Thus both enstrophy and energy are 
conserved in $d=2$. 
 \footnote{Note that in this limit, the mean rate of energy dissipation at the 
microscopic scale Eq. (\ref{eps-diseta}) is only logarithmically divergent, 
which corroborates the fact that energy is not dominantly dissipated at this 
scale in $d=2$.}

 This choice yields   at the fixed point $\eta_*^\nu = 2$, according to Eq. 
(\ref{valetanu}). 
The values of the two running anomalous dimensions  at the fixed point are also 
determined in $d=2$.
As in $d=3$, and since we only consider the stationary regime,
  we simply fix these values all along the flow 
\begin{equation}
 \eta_\kappa^\td = 4 \hspace{0.5cm}\hbox{and}\hspace{0.5cm} \eta_\kappa^\nu = 2 
\hspace{0.5cm}\hbox{in}\hspace{0.5cm} d=2.
\label{expod2}
\end{equation}

\subsection{Energy spectrum}

The energy spectrum in dimension $d$ is usually defined as \cite{Frisch95}
\begin{equation}
 E^{(d)}(\vp) = \displaystyle\frac{2 \pi^{d/2}}{\Gamma(d/2)}\,p^{d-1}\, {\cal 
E}^{(d)}(\vp)
 \end{equation}
where ${\cal E}^{(d)}(\vp)$ is the Fourier transform of the equal-time 
velocity-velocity correlation function $\langle \vec v(t,\vec x)\cdot \vec 
v(t,\vec 0)\rangle$. 
The velocity-velocity correlation function (at arbitrary times) can be expressed 
as the inverse Fourier transform of
$G_{\alpha\beta}^{u u}(\omega,\vq)$, that is
\begin{align}
&\left\langle  v_\alpha(t,\vec x) v_\beta(0,\vec 0)\right\rangle = \displaystyle 
\int_{\bq}  \,e^{-i(\omega t - \vec q \cdot \vec x)}\,  G_{\alpha\beta}^{u 
u}(\omega,\vq)\nonumber\\
&=\displaystyle -\int_{\bq}  \,e^{-i(\omega t - \vec q \cdot \vec x)}\, 
P_{\alpha\beta}^\perp(\vec q) \,\frac{\Gamma_\perp^{(0,2)}(\omega,\vec q)-2 
N_\kappa(\vec q)}{\Big|\Gamma_\perp^{(1,1)}(\omega,\vec q)+ R_\kappa(\vec 
q)\Big|^2}.
\end{align}
One hence obtains
\begin{align}
  {\cal E}^{(d)}(\vp) &=  \displaystyle \int d^d \vec x \,e^{-i \vec p \cdot 
\vec x}\, \left\langle  v_\alpha(t,\vec x) v_\alpha(t,\vec 0)\right\rangle 
\nonumber \\
 &=- (d-1)\displaystyle \int_0^\infty \frac{d\omega}{\pi} 
\frac{\Gamma_\perp^{(0,2)}(\omega,\vec p)-2 N_\kappa(\vec 
p)}{\Big|\Gamma_\perp^{(1,1)}(\omega,\vec p)+ R_\kappa(\vec p)\Big|^2}.
\end{align}
We focus on the inertial regime where $|\vec p| \gg \kappa$,  both in $d=2$ and 
$d=3$. In this limit the functions $N_\kappa(\vec p)$
and $R_\kappa(\vec p)$ tend to zero rapidly yielding
\begin{equation}
  {\cal E}^{(d)}(\vp) \stackrel{|\vec p| \gg \kappa}{\simeq}- (d-1)\displaystyle 
\int_0^\infty \frac{d\omega}{\pi} \frac{\Gamma_\perp^{(0,2)}(\omega,\vec 
p)}{\Big|\Gamma_\perp^{(1,1)}(\omega,\vec p)\Big|^2}.\label{spectrum}
\end{equation}

\subsection{Second-order structure function}

The longitudinal structure function of order $n$ is defined as the average of 
the $n^{\rm th}$ power of the equal time longitudinal velocity increment
\begin{equation}
 S^{(n)}(\ell) = \Big\langle \big[(\vec v(t,\vec \ell) -\vec v (t,\vec 0))\cdot 
\hat{\ell} \big]^n  \Big\rangle ,
\end{equation}
where $\hat{\ell} = \vec\ell/|\vec\ell|$. Exploiting translation invariance, the 
second-order structure function hence reads
\begin{equation}
 S^{(2)}(\ell) =-2 \hat\ell_i \hat\ell_j \Big\langle v_j(t,\vec \ell) v_i 
(t,\vec 0) - v_j(t,\vec 0) v_i (t,\vec 0)\Big\rangle
\end{equation}
and in terms of  $G^{u u}_{\alpha\beta}$, one has
\begin{equation}
 \Big\langle v_\alpha(t,\vec \ell) v_\beta (t,\vec 0) - v_\alpha(t,\vec 0) 
v_\beta (t,\vec 0)\Big\rangle = \displaystyle \int_{\bq} \, G^{u 
u}_{\alpha\beta}(\omega,\vec q)\,\Big[e^{i \vec q\cdot \vec \ell} -1\Big].
\end{equation}
It then follows that
\begin{align}
 S&^{(2)}(\ell) = -2  \displaystyle \int_{\bq} \, G^{u u}_\perp(\omega,\vec 
q)\,\Big[e^{i \vec q\cdot \vec \ell} -1\Big]\,\Big[1-\frac{(\hat \ell\cdot \vec 
q)^2}{q^2} \Big]\nonumber\\
  & = \displaystyle\gamma_d\,\int_0^\infty \frac{d\omega}{\pi}\int_0^\infty dq  
q^{d-1} \frac{\Gamma_\perp^{(0,2)}(\omega,\vec q)-2 N_\kappa(\vec 
q)}{\Big|\Gamma_\perp^{(1,1)}(\omega,\vec q)+ R_\kappa(\vec q)\Big|^2}\, I_d(q 
\ell)\label{integS2}
\end{align}
with
\begin{align}
\gamma_d &\equiv \frac{4 
\pi^{(d-1)/2}}{(2\pi)^d\Gamma\left(\frac{d-1}{2}\right)}\nonumber\\
 I_d(v) &\equiv \int_0^\pi d\theta \sin^{d}\theta \Big[e^{i v \cos\theta} 
-1\Big] \\
  &=\int_{-1}^1 du (1-u^2)^{(d-1)/2} \Big[e^{i v u} -1\Big].
\end{align}
In dimensions $d=2$ and $d=3$, the integrals $I_d(v)$ are given by
\begin{align}
 I_3(v) &= \displaystyle \frac{4}{v^3}\Big[\sin v -v \cos v 
-\frac{v^3}{3}\Big]\nonumber\\
I_2(v) &= \pi\Big[-\frac 1 2 +\frac 1 v J_1(v)\Big]
\end{align}
where $J_1(v)$ denotes a  type $J$ Bessel function of  the first kind.

Again, let us focus on the inertial regime corresponding to $\ell \kappa \ll 1$. 
Since the integral in (\ref{integS2})
is dominated by values of $q$ such that $q \ell$ is of order one, it is 
dominated by values of  $q \gg \kappa$ in the inertial range. Accordingly,
one can neglect in this regime the functions $N_\kappa(\vec q)$ and 
$R_\kappa(\vec q)$ that tend to zero rapidly  and one obtains
\begin{equation}
 S^{(2)}(\ell) \stackrel{\ell \kappa \ll 1}{\simeq}
 \displaystyle\gamma_d\,\int_0^\infty \frac{d\omega}{\pi}\int_0^{\infty} dq \, 
q^{d-1} \frac{\Gamma_\perp^{(0,2)}(\omega,\vec 
q)}{\Big|\Gamma_\perp^{(1,1)}(\omega,\vec q)\Big|^2}\, I_d(q \ell). \label{s2}
\end{equation}

\section{NPRG flow equations at the Leading Order approximation}
\label{SEC-LO}
In this section, we devise a simple approximation to study the stationary regime 
of the stirred NS equation,
 designed to provide a reliable description of the  large distance properties of 
the system (that is, small wave-number sector), and is therefore appropriate to 
investigate the existence of a fixed point of the NPRG flow.
This approximation is similar to the approximation implemented in \cite{Tomassini97,Monasterio12},
 although presented differently, with emphasis on its condition of validity and limitations.

\subsection{Approximation scheme}
\label{approxLO}

We here explain the principles and justifications of the approximation scheme. 
This technical subsection
 may be skipped in a first lecture.
The NPRG flow of the effective action $\Gamma_\kappa$ is given by the exact 
equation (\ref{dkgam}).
However, as this equation cannot be solved exactly, one has to devise some 
approximation.
 The standard approximation schemes within the NPRG framework are the derivative 
expansion \cite{Berges02} and the BMW approximation scheme \cite{Benitez12}.

  The derivative expansion consists in an expansion of $\Gamma_\kappa$ in powers 
of gradients and time derivatives.
 It is tailored to provide an accurate description of the long time and large 
distance properties of the theory (zero external wave-vector and frequency 
sector), which encompass phase diagrams and critical exponents. 
 However, if one needs the wave-vector and/or frequency dependencies of the 
vertex functions, one has to resort to a more sophisticated approximation scheme 
such as the BMW one.
 It consists in a closure at a given order $n$ of the  hierarchy
 of  flow equations for the $p$-point vertex functions $\Gamma_\kappa^{(p)}$ 
with $p\le n$ by approximating the $\Gamma_\kappa^{(n+1)}$ and 
$\Gamma_\kappa^{(n+2)}$ vertices. For instance, for $n=2$, the set of flow 
equations (\ref{dkgam2}) for the 2-point functions
 involves the vertices $\Gamma_\kappa^{(3)}$ and  $\Gamma_\kappa^{(4)}$, that 
are approximated in the BMW scheme in such a way that they can be expressed in 
terms of derivatives of $\Gamma_\kappa^{(2)}$ thus yielding a  closed set of 
equations.

In this work, our aim is to compute the wave-vector dependent 2-point functions, 
in order to calculate the energy spectrum (\ref{spectrum}) and the second order 
structure function (\ref{s2}) of the stationary NS incompressible flow. Hence 
the derivative expansion is not appropriate and we resort to the BMW scheme.
 However, the standard implementation of this  scheme, well-established for 
equilibrium problems \cite{Benitez12}, 
 is hindered here by the symmetries. The reason is that, on the one hand, the 
BMW approximation requires an expansion of the vertex functions 
$\Gamma_\kappa^{(3)}$ and  $\Gamma_\kappa^{(4)}$ in the internal wave-vector and 
frequency, 
  but on the other hand,  these dependencies are very much constrained through 
the Ward identities, ensuing in particular from the time-gauged Galilean 
symmetry, such that both are very difficult to conciliate {\it a priori}.

 In fact, this obstacle has been successfully circumvented in another closely 
related nonequilibrium classical problem, the Kardar-Parisi-Zhang (KPZ) 
equation, which is a stochastic (Langevin) equation describing interface growth 
and roughening \cite{halpin95}.
  The KPZ equation turns out to share with the NS one  
  a very similar invariance under time-gauged Galilean transformations, and also 
under time-gauged shifts (although of the field itself instead of  the  response 
field) \cite{Canet11a}. This similarity is easily conceivable as the KPZ 
equation  maps onto the Burgers equation.
  Thus the approximation scheme devised in the context of the KPZ equation 
\cite{Canet10,Canet11a,Kloss12} can be quite simply transposed to the NS 
equation.
 
For the KPZ equation, the solution consists in constructing an \anz for 
$\Gamma_\kappa$ explicitly preserving the symmetries
 of the action, in particular the time-gauged Galilean symmetry. At second order 
(SO) of this scheme,  the effective action $\Gamma_\kappa$
 is truncated  at quadratic order in the response field, while retaining  for 
the 2-point functions an arbitrary dependence 
 in wave-vectors and frequencies (and also in the field itself through arbitrary 
powers of the covariant time derivative). 
 The 2-point correlation and response functions for the KPZ problem were 
calculated within the SO approximation in $d=1$ \cite{Canet11a}.
  The existence of scaling forms for these functions  could be proved 
analytically, and the related scaling functions in one dimension 
  turned out to reproduce
  with an impressive agreement the exact results established in 
\cite{Praehofer04}, including the finest detail of their tails.
The SO approximation  can therefore be roughly considered as a BMW-like scheme 
(for the field but not for the response field)
 rendered compatible with the KPZ symmetries. 

From a numerical viewpoint, the SO approximation is  demanding
 in dimensions larger than one.
 Two simpler approximations were thus proposed in 
\cite{Canet10,Canet11a,Kloss12}. They both consist in a simplification of the 
frequency sector, by either completely neglecting the frequency dependence of 
the 3-point vertices (leading order (LO) approximation) or approximating it 
(next-to-leading (NLO) approximation).  
The predictions obtained at NLO for some universal amplitude ratios of the KPZ 
problem in \cite{Kloss12} were very accurately confirmed in $d=2$ in recent 
large-scale simulations \cite{halpin13}. As for the LO approximation, which only 
retains the bare frequency dependence, but preserves the full wave-vector 
dependence of the 2-point functions,  
  it clearly  suffices in the KPZ problem to obtain the full phase diagram, 
including the strong-coupling rough phase. The related estimates for the 
critical exponents  are in good agreement with the numerical ones in dimensions 
$d=2$ and $d=3$ \cite{Canet10,Kloss12}.  Let us underline that the values of the 
KPZ critical exponents  are greatly improved at LO compared to those obtained 
within the Derivative Expansion \cite{canet05b}, which is  probably to be 
imputed to the derivative nature of the bare vertex.  

As we merely consider, in the current work,  the wave-vector dependence of the 
2-point functions of the NS problem, we choose to implement the  LO 
approximation for the NS equation, which is achieved in the next section. For 
the NS effective action $\tilde \Gamma_\kappa$ in Eq.~(\ref{anzGk}), it consists 
 in 
\begin{itemize}
 \item (i) performing a field expansion in $\bar \vu$ at order two
\item (ii) keeping only the bare wave-vector and frequency dependencies of all 
$n$-point  functions with $n\ge 3$
\item (iii) preserving an arbitrary wave-vector dependence of the 2-point 
functions while restricting to their bare frequency dependence.
\end{itemize}
In this approach, all the symmetries of the theory are automatically encoded by 
writing the proposed ansatz for $\tilde \Gamma_\kappa$ in terms of Galilean 
scalars only. Notice that within the LO approximation, point (ii) implies that 
all $n$-point vertex functions with $n\ge 4$ vanish 
 and point (iii) that $\tilde \Gamma_\kappa$ is also truncated at order two in 
$\vu$ since the dependence in the $D_t$ covariant derivative is neglected. The 
LO ansatz is given by Eq.~(\ref{anzhGk}). 

Let us comment on the validity/accuracy of the LO approximation.
Point (ii) implies that this approximation is valid only for wave-vectors 
typically smaller than $\kappa$ and frequencies smaller than $\kappa^{1/z}$ 
because keeping the bare wave-vector dependence of the $n$-point functions for 
$n\ge 3$ is equivalent to keeping their leading terms in a  wave-vector 
expansion. Notice that once this expansion is performed, the frequency sector is 
entirely fixed by the symmetries gauged in time, see Sec.~\ref{SEC-LP}. This 
expansion is certainly valid for the internal wave-vector $\vq$ in 
Eq.~(\ref{dkgam2}) since it is suppressed for $|\vq|\gg \kappa$ by the $\p_\kappa 
{\cal R}_\kappa$ term but, a priori, is only justified for small external 
wave-vector, that is $|\vp|\lesssim \kappa$  in Eq.~(\ref{dkgam2}). In fact, in 
most systems, including the KPZ growth, 
 the flow of the 2-point functions actually stops 
 when the external wave-vector becomes larger than the RG scale $\kappa$, a 
phenomenon called {\it decoupling}.
Thus, the determination of the momentum and frequency dependencies remains 
accurate within the LO approximation because (a) when $\kappa \gtrsim \vert \vp \vert $ 
(and $\kappa \gtrsim \nu^{1/z} $) the LO approximation is controlled and (b) when 
$\kappa$ becomes smaller than $\vert \vp \vert $, the flow almost stops  and 
although the LO approximation is not valid in this region of wave-numbers, this 
has negligible  impact on the two-point functions.

 However, as stressed in the following, the  NS problem is very peculiar because 
the RG flow of the 2-point functions does not
  satisfy the decoupling property of the large wave-vector sector. Thus  a 
complementary scheme is necessary to determine the wave-vector dependence of 
$\Gamma_\kappa^{(2)}(\vp)$ at finite $\vp$ when $\kappa\to 0$. The LO 
approximation analyzed below is therefore only valid in the limit $|\vp|\to 0$ 
when $\kappa\to 0$,  which is sufficient to determine the fixed point structure 
and critical exponents, but not to investigate multiscaling. On the other hand, 
as shown in Sec.~\ref{SEC-LP}, exact RG flow equations can be derived 
 in the large $\vp$ sector, relying on the very constraining  gauge symmetries 
  of the NS field theory, which can take over from the LO equations when 
$\kappa$ becomes smaller than $|\vp|$.
  In fact, a source for the multiscaling behavior of the $n$-point functions precisely 
  emerges in the NPRG framework from the non-decoupling of the large $\vp$ sector  
and the associated  nontrivial behavior in both $|\vp|$ and $\nu$ (see 
Sec.~\ref{SEC-LP}).

\subsection{LO \anz}

As explained in the previous section, 
 one can construct an \anz for $\tilde \Gamma_\kappa$ in (\ref{anzGk}) which 
explicitly preserves the time-gauged Galilean symmetry  by using as building 
blocks  Galilean scalars (see  Ref.~\cite{Canet11a} for detail). The shift 
gauged symmetry can then also be simply 
 enforced.
 At LO,  the functional $\tilde \Gamma_\kappa$ is truncated at quadratic order 
in the response field $\bar \vu$
and moreover, the (non-bare) frequency dependence ({\it i.e.} a dependence in 
the covariant time derivative $D_t$ of the running functions 
$f^{\nu,\td}_{\kappa,\alpha\beta}$ below) is neglected. Hence the LO \anz simply 
reads
\begin{align}
\tilde \Gamma_\kappa[\vu,\bar \vu] &= \int_{t,\vec x,\vec x'}\Big\{\bar 
u_\alpha(t,\vec x) f^{\nu}_{\kappa,\alpha \beta}(\vec x-\vec x') u_\beta(t,\vec 
x') \nonumber\\
& -\bar u_\alpha(t,\vec x) f^{\td}_{\kappa,\alpha \beta}(\vec x-\vec x') \bar 
u_\beta(t,\vec x')\Big\}.
 \label{anzhGk}
\end{align}
 To further specify the running functions $f_{\kappa,\alpha \beta}^{\nu,\td}$ , 
let us introduce the notation for the vertex functions
\begin{align}
 &\Gamma_{\alpha_1,\dots,\alpha_m,\beta_1,\dots,\beta_n}^{(m,n)}(\bx_1,\dots,
\bx_m,\bar \bx_1,\dots,\bar \bx_n)\nonumber\\
 &=\frac{\delta^{n+m}\Gamma_ \kappa}{\delta u_{\alpha_1}(\bx_1)\dots\delta 
u_{\alpha_m}(\bx_m)\delta \bar u_{\beta_1}(\bar \bx_1)
 \dots\delta \bar u_{\beta_n}(\bar \bx_n)}
\end{align}
where from now on, the explicit index $\kappa$  is dropped for the vertex 
functions $\Gamma^{(m,n)}\equiv \Gamma_\kappa^{(m,n)}$ and  for the running 
functions $f_{\alpha \beta}^{\nu,\td}\equiv f_{\kappa,\alpha \beta}^{\nu,\td}$. 
 The  Fourier transforms of the $\Gamma^{(m,n)}$ only depend on $n+m-1$ 
wave-vectors and frequencies because of translation invariance.

The dependence of the running functions $f_{\alpha \beta}^{\nu}$ and $f_{\alpha 
\beta}^{\td}$ in $\vec x-\vec x'$ is in fact through gradients as can be 
inferred
 from the Ward identities (\ref{Wshift02}) and (\ref{Wgal11}). The latters imply 
that these two functions vanish at zero wave-vector
\begin{equation}
 f^{\nu}_{\alpha\beta}(\vec p=\vec 0)=f^{\td}_{\alpha\beta}(\vec p=\vec 0)=0.
\label{constr}
\end{equation}
The initial conditions of the flow at scale $\kappa=\eta^{-1}$ for the two 
running functions  are
\begin{align}
\left.f^{\td}_{\alpha \beta}(\vec x-\vec x')\right|_{\kappa=\eta^{-1}}&=0 
\nonumber\\
 \left. f^{\nu}_{\alpha \beta}(\vec x-\vec x')\right|_{\kappa=\eta^{-1}}&=-\nu 
\delta_{\alpha \beta}\nabla_x^2(\delta^{(d)}(\vec x-\vec x'))
\end{align}
to recover the original NS action (\ref{NSaction}) at the microscopic scale.

The calculation of the 2-point functions from the LO \anz is straightforward. At 
vanishing fields and in Fourier space, one obtains
\begin{align}
\tilde\Gamma_{\alpha\beta}^{(2,0)}(\omega,\vec p)&=0\nonumber\\
\tilde\Gamma_{\alpha\beta}^{(1,1)}(\omega,\vec p)&=f^{\nu}_{\alpha\beta}(\vec 
p)\nonumber\\
\tilde\Gamma_{\alpha\beta}^{(0,2)}(\omega,\vec p)& =-2f^{\td}_{\alpha\beta}(\vec 
p).
\label{gam2LO}
\end{align}
Within the LO approximation, all vertex functions of order $m+n\ge 4$ vanish, 
and the only non-zero 3-point vertex function is the bare one, which reads in 
Fourier space
\begin{equation}
 \Gamma_{\alpha\beta\gamma}^{(2,1)}(\omega_1,\vec p_1,\omega_2,\vec p_2) = -i 
\lambda (p_2^\alpha \delta_{\beta\gamma} + p_1^\beta \delta_{\alpha\gamma}).
\end{equation}

\subsection{Derivation of the LO flow equations}

In this section, we derive the LO flow equations of the two running functions 
$f_{\alpha \beta}^{\nu}$ and $f_{\alpha \beta}^{\td}$.
In fact, for  incompressible flows, only the transverse sector plays a role, 
which means that only their transverse components, denoted  $f_\perp^\nu$ and 
$f_\perp^\td$, are eventually needed. 
The flow equations of  $f_\perp^{\nu}$ and $f_\perp^{\td}$
 are proportional to the projection in the transverse sector of the flow 
equations of the 2-point functions $\Gamma_{\alpha\beta}^{(1,1)}(0,\vp)$ and 
$\Gamma_{\alpha\beta}^{(0,2)}(0,\vp)$, which are given in matrix form by 
Eq.~(\ref{dkgam2}) evaluated at zero external frequency.
 At LO, the matrices $\Gamma^{(4)}_{\kappa,ij}$ entering these equations are 
zero and  only one 3-point vertex, the bare one $\Gamma^{(2,1)}$,  contributes 
in the matrices $\Gamma^{(3)}_{\kappa,i}$.
  The transverse components of the propagator in the $\vu$,$\bar\vu$ sector 
(\ref{propGk}) are given within the LO approximation, {\it i.e.} with the \anz 
(\ref{gam2LO}), by
\begin{align}
G^{\bar u\bar u}_\perp(\omega,\vp) &=0\nonumber\\
G^{u\bar u}_\perp(\omega,\vp) &= \frac{1}{-i\omega+\tf^{\nu}_\perp(\vec 
p)}\nonumber\\
G^{u u}_\perp(\omega,\vp) &= \frac{2 \tf_\perp^{\td}(\vec 
p)}{\omega^2+(\tf_\perp^{\nu}(\vec p))^2}
\end{align}
where
\begin{align} 
\tf_\perp^{\nu}(\vec p)&\equiv f_\perp^{\nu}(\vec p)+ R_\kappa(\vq)\nonumber\\
 \tf_\perp^{\td}(\vec p)&\equiv f_\perp^{\td}(\vec p)+  N_\kappa(\vq).
\end{align}
One has to compute the trace of the matrix product (\ref{dkgam2}), and project 
the result onto the transverse sector.
In the obtained expression, since the frequency dependence remains the bare one 
at LO,  the integral over the internal frequency $\omega$ can be 
 analytically carried out. These calculations are detailed in Appendix B. The 
resulting flow equations for the two running functions $f_\perp^\nu$ and 
$f_\perp^\td$  are  given by
\begin{widetext}
\begin{align}
 \partial_s f_\perp^{\nu}(\vec p)&= \frac{\lambda^2}{(d-1)}\int_{\vec 
q}\Bigg\{\frac{\partial_s R_\kappa(\vec q) \,\tf_\perp^\td(\vec p+\vec q)}{ 
\tf_\perp^{\nu}(\vec p +\vec q) \big(\tf_\perp^{\nu}(\vec q) + 
\tf_\perp^{\nu}(\vec p+\vec q)\big)^2}\Big[\Big(-\vec p\,^2+\frac{(\vec p \cdot 
(\vec p+\vec q))^2}{(\vec p+\vec q)^2}\Big)(d-1)-2 \vec p\cdot \vec 
q\Big(1-\frac{(\vec p\cdot \vec q)^2}{\vec q\,^2 \vec p\,^2}\Big) 
\Big]\nonumber\\
&+ \frac{1}{\tf_\perp^{\nu}(\vec q) \big( \tf_\perp^{\nu}(\vec q) +  
\tf_\perp^{\nu}(\vec p+\vec q)\big)} \Big[ \partial_s R_\kappa(\vec q) \frac{ 
\tf_\perp^\td(\vec q)\big(2  \tf_\perp^{\nu}(\vec q) + \tf_\perp^{\nu}(\vec 
p+\vec q)\big)}{\tf_\perp^{\nu}(\vec q) \big(  \tf_\perp^{\nu}(\vec q) +  
\tf_\perp^{\nu}(\vec p+\vec q)\big)}- \partial_s N_\kappa(\vec 
q)\Big]\nonumber\\
&\times \Big[\Big(-\vec p\,^2+\frac{(\vec p \cdot \vec q)^2}{\vec 
q\,^2}\Big)(d-1)+2 \frac{\vec p\cdot (\vec p+\vec q)}{(\vec q+\vec 
p)^2}\Big(\vec q\,^2-\frac{(\vec p\cdot \vec q)^2}{\vec p\,^2}\Big) \Big]\Bigg\}
\label{dkfnu}\\
 \partial_s f_\perp^\td(\vec p)&
 =-\frac{\lambda^2}{2(d-1)}\int_{\vec q}  \Bigg\{
\frac{2 \tf_\perp^\td(\vec q+\vec p)}{\tf_\perp^{\nu}(\vec p+\vec q) 
\tf_\perp^{\nu}(\vec q)\big( \tf_\perp^{\nu}(\vec q) +  \tf_\perp^{\nu}(\vec 
p+\vec q)\big)}\Big[ \partial_s R_\kappa(\vec q) \frac{\tf_\perp^\td(\vec 
q)\big(2  \tf_\perp^{\nu}(\vec q) + \tf_\perp^{\nu}(\vec p+\vec q)\big) 
}{\tf_\perp^{\nu}(\vec q)\big(  \tf_\perp^{\nu}(\vec q) + \tf_\perp^{\nu}(\vec 
p+\vec q)\big) }-  \partial_s N_\kappa(\vec q)\Big]\nonumber\\
&\times\Big[\Big(2\vec p\,^2+\frac{(\vec p \cdot (\vec p+\vec q))^2}{(\vec 
p+\vec q)^2}-\frac{(\vec p \cdot \vec q)^2}{\vec q\,^2}\Big)(d-1)+2 
\frac{1}{\vec q\,^2 (\vec p+\vec q)^2}\Big(\vec q\,^2-\frac{(\vec p\cdot \vec 
q)^2}{\vec p\,^2}\Big) 
 \Big(\vec p\,^2 \vec p\cdot \vec q+2 (\vec p\cdot \vec q)^2-\vec p\,^2 \vec 
q\,^2\Big)\Big]\Bigg\}
 \label{dkfd}
\end{align}
\end{widetext}
where $\p_s\equiv \kappa \p_\kappa$.

\subsection{Dimensionless LO flow equations}
\label{dimensionlessflow}

As we seek the fixed point solution of the  flow equations, we work with 
dimensionless and renormalized quantities, as defined in Sec. \ref{SEC-energy}.
Let us  determine the dimensions of the terms entering the LO \anz 
(\ref{anzGk},\ref{anzhGk}). 
  According to the  definitions (\ref{nukdef}) and (\ref{Dkdim}), the two 
functions $\hat{r}(q^2/\kappa^2)$ and $\hat{n}(q^2/\kappa^2)$ introduced
   in Eqs.~(\ref{dkdef}) and (\ref{rkdef}) are  dimensionless, and the flow of 
the regulator terms in Eqs.~(\ref{dkfnu}) and (\ref{dkfd}) are  given by
   \begin{align}
\partial_s R_\kappa(\vq) &= \nu_\kappa q^2 \big(-\eta_\kappa^\nu \hat r(\hat 
q^2) -2 \hat q^2 \partial_{\hat q^2}\hat r(\hat q^2)\big) \nonumber\\
\partial_s N_\kappa(\vq) &= D_\kappa \hat q^2 \big(-(\eta_\kappa^\td+2) \hat 
n(\hat q^2) -2 \hat q^2 \partial_{\hat q^2}\hat n(\hat q^2)\big).
\end{align}
 
 Then, according to Eq.~(\ref{anzhGk}), the dimensions of the running functions 
$f_{\alpha\beta}^\nu$ and  $f_{\alpha\beta}^\td$ are the same as the ones of 
$R_{\kappa, \alpha\beta}$ and $N_{\kappa, \alpha\beta}$  in Eqs.~(\ref{nukdef}) 
and (\ref{Dkdim}) and we thus define 
 the dimensionless functions $\hat h^\nu$ and $\hat h^\td$ as
\begin{equation}
 f_\perp^\nu(\vp) = \nu_\kappa \,\kappa^2\,\hat p^2\, \hat h^\nu(\hat \vp) 
\hspace{0.4cm}\hbox{and} \hspace{0.4cm} f_\perp^\td(\vp) = D_\kappa \, \hat 
p^2\, \hat h^\td(\hat \vp).
 \end{equation}
 Their flow equations are  given by
\begin{align}
 \partial_s \hat h^\nu(\hat \vp) &=  \eta_\kappa^\nu  \hat h^\nu(\hat \vp) +\hat 
\vp \partial_{\hat \vp} \hat h^\nu (\hat \vp) + \nu_\kappa^{-1} \frac{\partial_s 
f_\perp^\nu(\vp)}{p^2} \nonumber \\
\partial_s \hat h^\td(\hat \vp) &=  (\eta_\kappa^\td+2) \hat h^\td(\hat \vp) 
+\hat \vp \partial_{\hat \vp} \hat h^\td (\hat \vp) + D_\kappa^{-1} 
\frac{\partial_s f_\perp^\td(\vp)}{\hat p^2}
\label{flowh}
\end{align}
with the substitutions for dimensionless quantities  in the flow equations 
(\ref{dkfnu}) and (\ref{dkfd})
 for $\partial_s f_\perp^\nu(\vp)$ and $\partial_s f_\perp^\td(\vp)$.

\section{Fixed point solutions}
\label{SEC-LOnum}

One has to integrate the two flow equations (\ref{flowh})  to determine the 
scale evolution of the two running functions $h^\nu(\hat \vp)$ and $h^\td(\hat 
\vp)$. For simplicity,  the  two running exponents $\eta_\kappa^\nu$ and  
$\eta_\kappa^\td$ are fixed to the values (\ref{expod3}) in $d=3$ and 
(\ref{expod2}) in $d=2$ at any $\kappa$, that is all along the flow (note that 
$\p_s \hat \lambda_\kappa=0$ for all $\kappa$ once the exponents are fixed).
  We checked that it does not affect the critical properties at the fixed point.
 These flow equations are integrated numerically from the initial conditions 
$\hat h^\td(\vp) = 0$ and $\hat h^\nu(\vp) = 1$,
 with different values of $\hat\lambda$ and different  values of the $a$ 
parameter in Eq. (\ref{eq:expReg}). 
  The detail of the numerical procedure is summarized in Appendix D.
 We observe that the flow always reaches a (fully attractive) fixed point  
without
 fine-tuning any parameter, and independently of the initial conditions,  both 
in  $d=3$ and in $d=2$.
 Hence the corresponding stationary regime is universal.

\subsection{Fixed point functions $\hat h^\nu$ and $\hat h^\td$}

 \begin{figure}[t]
\epsfxsize=9.cm
\hspace{-.5cm}
\vspace{0.8cm}
\epsfbox{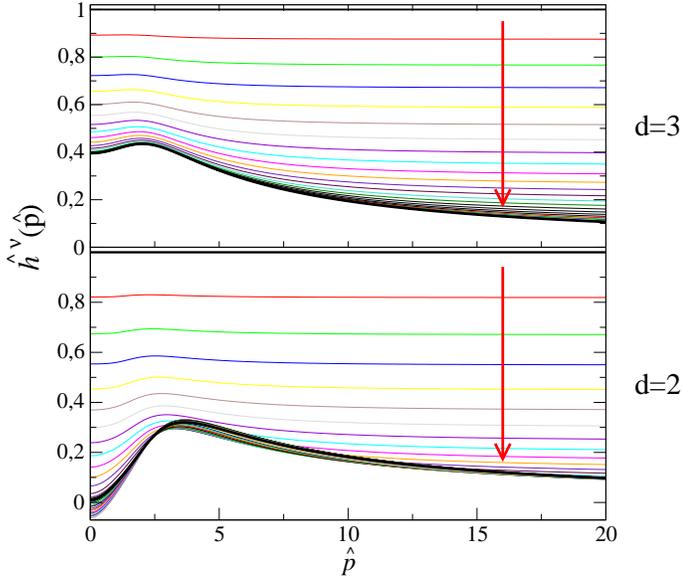}
\caption{(Color online) RG evolution of the dimensionless running function $\hat 
h^\nu(\hat p)$ in dimensions $d=3$ 
(upper panel) and $d=2$ (lower panel) starting from the initial condition  $\hat 
h^\nu(\hat p)=1$ at the microscopic 
scale $\kappa=\eta^{-1}$. The red arrow corresponds to decreasing RG scales 
$\kappa$ and the black thick line to the fixed point function.}
\label{fig1}
 \end{figure}

\begin{figure}[t]
\epsfxsize=8.75cm

\vspace{0.2cm}
\epsfbox{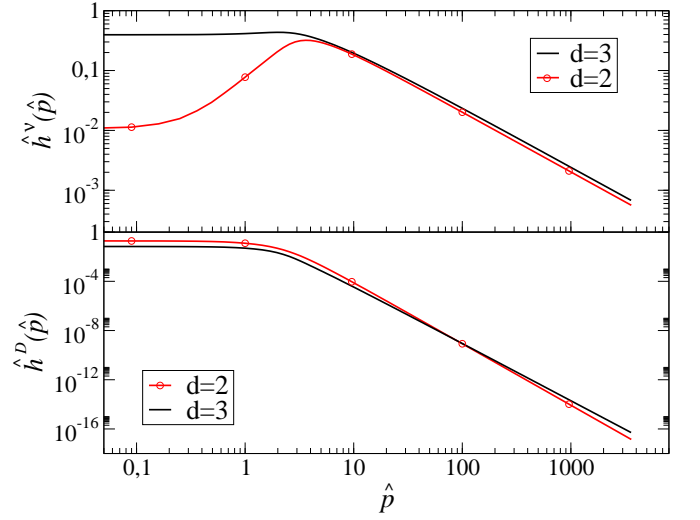}
\caption{(Color online) Fixed point functions $\hat h^\td(\hat p)$ (lower panel) 
and $\hat h^\nu(\hat p)$ (upper panel) in dimensions $d=3$ (black curves) and 
$d=2$ (red curves). Both horizontal and vertical axes are in logarithmic 
scales.}
\label{fig2}
 \end{figure}

 Along the flow, the two functions  $\hat h^{\nu,\td}$ are smoothly deformed to 
acquire a fixed form, which is illustrated
  on the example of the function $\hat h^\nu$ in $d=2$ and $d=3$ in 
Fig.~\ref{fig1}. 
 The fixed point profile of the two functions in both dimensions is displayed
   in logarithmic scales in Fig.~\ref{fig2}. This figure shows that both 
functions decay algebraically
    at large wave-number. We determined the corresponding decay exponents and 
observed that they deviate from the expected (dimensional) scalings.
We indeed found
  \begin{equation}
   \hat h^\nu(\hat p) \sim \hat p^{-\eta^\nu+\alpha} \quad\hbox{and} \quad \hat 
h^\td(\hat p) \sim \hat p^{-(\eta^\td+2)+\beta}  \label{asympLO}
  \end{equation}
where $\alpha$ and $\beta$ are the deviations from the dimensional (K41 or KB) 
scalings, with the estimated values  $\alpha\simeq \beta \simeq 0.33$ in $d=3$
 and $\alpha\simeq \beta \simeq 1.00$ in $d=2$.
 The values found  are in agreement  with the values reported in $d=3$
  in Ref. \cite{Tomassini97}. They are also very close to the values that can be estimated from the Figures 3 and 7 of Ref. \cite{Monasterio12}, which confirms that the large $\epsilon$ regime in this work corresponds in fact to the 
  fixed-point with localized forcing,
 although this was not noticed by the authors. It was shown in particular in  \cite{Tomassini97} that 
these exponents are independent of the choice of the stirring profile. In fact, 
one can  prove  by inspection of the regime $p\gg \kappa$ of the LO flow 
equations (\ref{dkfnu}) and (\ref{dkfd}) that the exponent $\alpha$ is exactly 
$1/3$ in $d=3$ and $1$ in $d=2$. Moreover, we  found numerically that 
$\beta\simeq1/3$ in $d=3$ and $\beta\simeq 1$ in $d=2$, if not  exactly at least 
 very precisely \footnote{Notice that the dependence of these values on the 
regulator  ({\it via} the $a$ parameter in Eq.~(\ref{eq:expReg})) could also be 
studied. However, it is meaningless within the LO approximation since this 
approximation is not appropriate to  study with precision the large wave-number 
sector, as already pointed out in Sec.~\ref{approxLO}}. 
 Let us now probe  the effect of these deviations on physical observables. 

\subsection{Energy spectrum}

\begin{figure}[ht]
\epsfxsize=8.75cm

\vspace{1cm}
\hspace{-.2cm}
\epsfbox{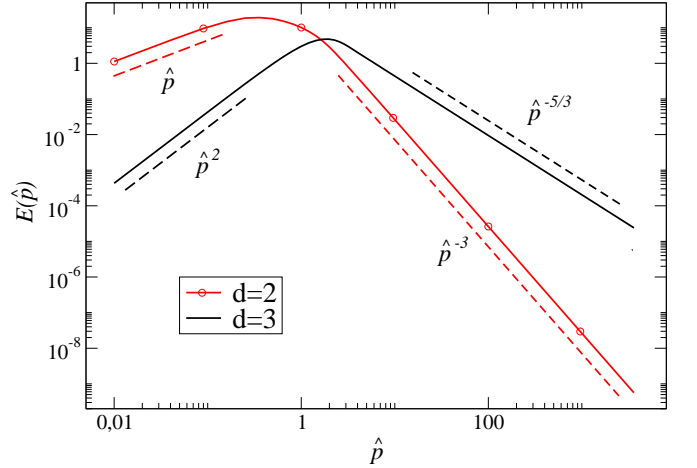}
\caption{(Color online) Energy spectrum (multiplied by the appropriate power 
$\kappa^{-5/3}$ in $d=3$ and $\kappa^{-3}$ in $d=2$) of the turbulent flow in 
dimensions $d=3$  and $d=2$ as a function of the dimensionless wave-number
 $\hat p = |\vp|/\kappa$. The dashed lines are guidelines for the eyes.}
\label{fig3}
\end{figure}

Within the LO approximation, the 2-point functions are given by
\begin{align}
\Gamma^{(0,2)}_\perp(\omega,\vec q) &=-2 D_\kappa \hat q^2 \hat h^\td(\hat 
\vq)\nonumber\\
\Gamma^{(1,1)}_\perp(\omega,\vec q) & = \nu_\kappa\kappa^2 (i\hat \omega+\hat 
q^2 \hat h^\nu(\hat q)).\label{gamMini}
\end{align}
Inserting these expressions into Eq.~(\ref{spectrum}) and performing the 
integral over the frequency, one obtains for the energy spectrum 
\begin{equation}
{E}^{(d)}(\vp)  {\simeq} \displaystyle\frac{2 \pi^{d/2}}{\Gamma(d/2)}\,(d-1)\, 
p^{d-1}\,\frac{D_\kappa}{\nu_\kappa \kappa^{2}}\,\frac{\hat h^\td(\hat p)}{\hat 
h^\nu(\hat p) }.
\end{equation}
The corresponding energy spectra in $d=3$ and $d=2$ are displayed on 
Fig.~\ref{fig3}. At small wave-vector, the two functions  $\hat h^{\nu,\td}(\hat 
p)$ tend to a finite constant. It follows that the energy spectra  grow as 
\begin{equation}
 {E}^{(d)}(\vp) \sim \displaystyle  p^{d-1}\,\kappa^{-\eta^\td+\eta^\nu-2}
\end{equation}
with the power law $p^{d-1}$ reflecting equipartition of energy. At large 
wave-vector, the two functions $\hat h^{\nu,\td}(\hat p)$
 follow the asymptotics (\ref{asympLO})
and the energy spectra in both dimensions thus decay algebraically as
\begin{align}
{E}^{(d)}(\vp) &\sim \displaystyle p^{d-3-\eta^\td+\eta^\nu} 
\left(\frac{p}{\kappa}\right)^{\beta-\alpha} \nonumber\\
               &=\left\{ \begin{array}{l l}\displaystyle p^{-5/3} 
\left(\frac{p}{\kappa}\right)^{\beta-\alpha} & \hbox{\hspace{0.5cm}in $d=3$}\\
               \displaystyle p^{-3} \left(\frac{p}{\kappa}\right)^{\beta-\alpha} 
& \hbox{\hspace{0.5cm}in $d=2$} 
               \end{array} \right.
\label{decay2D}
\end{align}
using Eqs. (\ref{expod3}) and (\ref{expod2}).

In $d=3$, we hence recover the Kolmogorov scaling when $\alpha=\beta$ which 
describes the decay of the energy spectrum when all the 
 energy is dominantly  transferred towards the small scale in  the direct 
cascade. 

Let us compute at LO approximation the rate of dissipated power at the 
microscopic scale in $d=3$ and check that the presence of deviations from 
dimensional scalings does not alter the analysis  of Sec. \ref{SUBSECeta}.
The average  dissipated power per unit mass (\ref{eps-diseta}) is expressed as
\begin{equation}
 \langle \epsilon_{\rm dis}^{1/\eta} \rangle
      =-\nu(d-1)\int_{\omega,\vq} q^2 
\,\frac{\Gamma_\kappa^{(0,2)}(\omega,\vq)-2N_\kappa(\vq)}{\big|\Gamma_\kappa^{(1
,1)}(\omega,\vq)+R_\kappa(\vq)\big|^2}.
\end{equation}
For $\vq$ in the inertial range, $N_\kappa$ and $R_\kappa$ are negligible, and 
within the LO approximation one obtains
\begin{align}
 \langle \epsilon_{\rm dis}^{1/\eta} \rangle &= \nu \frac{\kappa^d 
D_\kappa}{\nu_\kappa} 
\frac{d-1}{2^{d-1}\pi^{d/2}\Gamma(d/2)}\int_0^{1/(\kappa\eta)} d\hat q\; \hat 
q^{d+1}\,\frac{\hat h^\td(\hat q)}{\hat h^\nu(\hat q)}.
\end{align}
This integral is UV divergent in $d>2$ without an UV cutoff, and  hence 
dominated by the UV scale $\eta^{-1}$. Indeed,
 for $\kappa \ll \eta^{-1}$, the two functions $\hat h^\td$ and $\hat h^\nu$ 
follow the asymptotics (\ref{asympLO}) and thus in $d=3$
\begin{align}
 \langle \epsilon_{\rm dis}^{1/\eta}\rangle &\propto 
\kappa^{d-\eta^\td+\eta^\nu} \int_0^{1/(\kappa\eta)} d \hat q\;\hat q^{1/3 
+\beta-\alpha} \nonumber\\ &\propto { \eta^{-4/3-\beta+\alpha}} 
\kappa^{\alpha-\beta},
\end{align}
 and is independent of $\kappa$ for $\alpha\simeq \beta$.
This corroborates the analysis of Sec. \ref{SUBSECeta}. Thus, the flux of energy 
is  constant 
 in the inertial range between the injection scale $\kappa^{-1}$ and the 
dissipation scale $\eta$, as expected in three-dimensional turbulence.\\

 In $d=2$, we find according to Eq. (\ref{decay2D}) that the decay of the energy 
spectrum in the direct cascade
 is  steeper (slope $-3$) in $d=2$ than in $d=3$. This is expected since
  part of the energy is transferred in the inverse cascade  in $d=2$. We recover 
the $p^{-3}$ decay in the direct cascade  predicted  by KB theory with no (or 
very small) corrections when $\alpha\simeq \beta$. One can check explicitly that 
the mean rate of dissipated vorticity (\ref{eps-vort}) is indeed dominated by 
the UV scale $\eta^{-1}$  (behaving as $\eta^{-2}$) and thus independent of 
$\kappa$, confirming the analysis of Sec. \ref{SUBSECeta}. We remind that we do 
not have access here to the inverse cascade since the integral scale is merged 
with the (inverse) RG scale and sent to infinity with it,  that is the energy is 
effectively injected at the boundaries of the system. To investigate the inverse 
cascade, the integral scale (energy injection scale) should be kept fixed while 
the volume scale, identified in that case with the RG scale, should diverge in 
the limit $\kappa \to 0$. This important study is left for future work. \\

Both these results seem to indicate that, although the 2-point functions do 
display substancial deviations to their dimensional scalings, these deviations 
cancel out (or almost) for the energy spectrum, which is in agreement with 
experimental and numerical results which find very small (if any) corrections 
\cite{Frisch95}. However,  as explained previously, the LO approximation is not 
appropriate to study this regime, since it is based on an expansion of all 
wave-vectors, including the external ones, in powers of wave-vectors over 
$\kappa$. 
 Definite conclusions are postponed to Sec.~\ref{SEC-LP}.

\subsection{Second-order structure function}

Within the LO approximation, one obtains for the second order structure function 
(\ref{s2}) inserting the
 expressions (\ref{gamMini})
\begin{equation}
  S^{(2)}(\ell) {\simeq} - 
\displaystyle\gamma_d\,\frac{D_\kappa}{\nu_\kappa}\kappa^{d-2}\,\int_0^{\infty} 
d\hat q \, \hat q^{d-1}\frac{\hat h^\td(\hat q)}{\hat h^\nu(\hat q)}\,I_d(\kappa 
\hat q \ell).
  \end{equation} 
Let us determine the behavior of this quantity within the inertial regime, which 
corresponds to  the limit $\kappa \ell \ll 1$. Performing the change of 
variables  $x= \kappa \hat q \ell$, one obtains
\begin{align}
  S^{(2)}(\ell) &\simeq - 
\displaystyle\gamma_d\,\frac{D_\kappa}{\nu_\kappa}\kappa^{d-2}\,(\kappa 
\ell)^{-d} \int_0^{\infty} dx \, x^{d-1}\frac{\hat h^\td(\frac{x}{\kappa \ell} 
)}{\hat h^\nu(\frac{x}{\kappa \ell})}\,I_d(x) \nonumber \\
   &\sim  - 
\displaystyle\gamma_d\,\kappa^{\alpha-\beta}\ell^{
2-d+\eta^\td-\eta^\nu+\alpha-\beta}\nonumber\\
  & \times \int_0^{\infty} dx \, x^{d-3-\eta^\td+\eta^\nu+\beta-\alpha} 
\,I_d(x).
\end{align}
where in the second equality the asymptotics (\ref{asympLO}) are used. The 
integral over $x$ is both IR and UV finite.
Hence, in $d=3$,  
\begin{equation}
 S^{(2)}(\ell)  \sim - 
\displaystyle\gamma_d\,\kappa^{\alpha-\beta}\ell^{2/3+\alpha-\beta} 
\int_0^{\infty} dx \, x^{-5/3+\beta-\alpha}\,I_d(x),
 \end{equation}
  the Kolmogorov scaling is again recovered (or receives a very small 
correction) for $\alpha\simeq\beta$. 
In $d=2$,
\begin{equation}
 S^{(2)}(\ell)  \sim - 
\displaystyle\gamma_d\,\kappa^{\alpha-\beta}\ell^{2+\alpha-\beta} 
\int_0^{\infty} dx \, x^{-3+\beta-\alpha}\,I_d(x),
 \end{equation}
and no deviation (or very small effect) from the KB scaling is found when 
$\alpha\simeq\beta$. Both these results are again not in contradiction with 
experimental and numerical results \cite{Frisch95}.

As previously, one should be cautious with these results  
 because the regime of  wave-numbers much larger than $\kappa$ is not  
controlled in the LO approximation. 
Let us also point out that finding standard scaling for $S^{(2)}$ does not 
entail that the higher-order structure functions $S^{(n)}$, $n>3$ do not exhibit 
intermittency either,  because this result for $S^{(2)}$ relies on    
compensations  which are not likely to be generic. This is further discussed in 
Sec.~\ref{SEC-LP}.

\subsection{Large wave-number sector and limit of the LO approximation}
\label{SECdecLO}

In this section, we analyze the large wave-number limit of the flow equations 
(\ref{flowh}).
  This analysis unveils  that the nonlinear parts of these equations given by 
Eqs.~(\ref{dkfnu}) and (\ref{dkfd})
 do not become negligible at the fixed point compared to the linear 
(dimensional) parts for large external wave-vectors $\vp$. This means  
  that the large wave-number sector does not decouple from the flow when $\kappa 
\ll |\vp|$, which is a very unusual property.
 As a consequence, the existence of the fixed point
 does not  lead to the usual scale invariance, as encountered in ordinary 
critical phenomena. 
  Indeed, the non-decoupling entails that the large wave-vector sector is not 
determined by the small wave-vector one, that is the large $|\vp|$ behavior -- 
the exponents of the algebraic tails of the correlation functions -- 
  is not fixed in terms of the anomalous dimensions ($\eta^\nu$ and $\eta^\td$) 
and can develop  deviations from them.

 These deviations cannot be reliably computed at LO because this approximation 
is fully justified only when {\it all} the wave-numbers are small. As already 
explained, whereas this expansion is always valid
 for the internal wave-vector because of the presence of the regulator term 
$\p_\kappa {\cal R}_\kappa$ which effectively cuts off the
 contributions  $|\vq|~\gg~\kappa$, it is justified only for  small external 
wave-vectors $\vp$. 
  Hence,  one has to devise an alternative approximation to properly
   describe the large wave-number sector. This is achieved in Sec. \ref{SEC-LP}. 
In fact,
  we  prove in this section that the non-decoupling is a real feature of the 
exact NPRG flow of the NS problem and not an artifact of the LO approximation. 
Therefore, it is instructive to first understand how this non-decoupling 
  works on the example of the LO flow equations. 

To this aim, we  now study  the large $\vp$ sector of the flow equations 
(\ref{flowh}). We observed that the dimensionless running functions $\hat 
h^{\nu,\td}$ reach a fixed point for $\kappa \ll \eta^{-1}$,
which means by definition $\partial_s \hat h^\nu(\hat p)=\partial_s \hat 
h^\td(\hat p)=0$. We concentrate in the following on this regime.
Let us assume for a moment that,  as it generally occurs,  the nonlinear terms 
of the flow equations (\ref{flowh}), denoted ${\cal L}^\nu \equiv \partial_s 
f_\perp^\nu(\vp)/(\nu_\kappa p^2)$ and ${\cal L}^\td\equiv \partial_s 
f_\perp^\td(\vp)/(D_\kappa \hat p^2)$ and which explicit expressions are given 
by (\ref{dkfnu}) and (\ref{dkfd}), become negligible  in the large
wave-number limit $|\vp| \gg \kappa$ compared to the linear terms, that is, they 
decouple.
 One then deduces that the general  solutions at the fixed point of the 
remaining homogeneous (linear) parts of the flow equations   are the scaling 
forms
  \begin{equation}
  \hat h_*^\nu(\hat p) = \hat p^{-\eta^\nu} \zeta^\nu\quad\hbox{and}  \quad \hat 
h_*^\td(\hat p) = \hat p^{-(\eta^\td+2)} \zeta^\td \label{canscaling}
  \end{equation}
 where $\zeta^{\nu,\td}$ are  constants.  
Had we considered wave-vector {\it and} frequency dependent running functions, 
we would have obtained, {\it e.g.} $\hat h^{\nu}(\hat \omega,\hat p) = \hat 
p^{-\eta^\nu} \zeta(\hat\omega/\hat p^{2-\eta^\nu})$. 
 We now show that this leads to a contradiction, that is, these scaling 
solutions are incompatible with the assumed decoupling of the large wave-number 
sector.  For this, the leading contribution in $\hat p$
  of  each nonlinear term ${\cal L}^\nu$ and ${\cal L}^\td$ can be analytically 
determined substituting the two functions $\hat h^{\nu,\td}$ with the scaling 
solutions (\ref{canscaling}).
 One obtains that the dominant contributions of ${\cal L}^\nu$ and ${\cal 
L}^\td$ are respectively    $\hat p^{-\eta^\nu}$ and  $\hat p^{-(\eta^\td+2)}$. 
Hence, they are not negligible (sub-dominant) compared to the linear terms, but 
of the same order. One concludes that in LO approximation the nonlinear parts of 
the flow equations do not decouple at the fixed point, which invalidates the 
general scaling solutions (\ref{canscaling}).

This non-decoupling property entails that the existence of the fixed point does 
not  generate  scale invariance  as usual (in critical phenomena). 
As a matter of fact,
  we found in the previous section that the two functions $\hat h^{\nu,\td}$ do 
decay algebraically, but not with the dimensional  exponents
 (\ref{canscaling}). Instead, they exhibit the deviations $\alpha$ and $\beta$ 
to these scalings following the asymptotics Eqs.~(\ref{asympLO}). 
  Of course, the realm of the non-decoupling property for the exact NS flow 
equations cannot be asserted at the LO level and the  obtained values of these 
deviations are not to be  trusted. However, 
  we now specifically address the large wave-number regime, and prove that  
non-decoupling indeed occurs in the exact NPRG flow for NS.

\section{Exact NPRG flow equations in the large wave-number regime}
\label{SEC-LP}

The LO approximation is justified and expected to be accurate in the small 
wave-number regime
 $|\vp| \lesssim \kappa$. In this section, we devise an alternative and 
complementary approximation for the large wave-number regime, $|\vp| \gtrsim 
\kappa$ which 
   becomes exact in the limit $|\vp| \gg \kappa$, or equivalently in the limit 
$\kappa \to 0$ for any fixed external 
   wave-vector and frequency. 
  The flow equations for the 2-point  functions can indeed be exactly closed in 
this limit using Ward identities
 \footnote{As a remark, let us notice that the exact closure of the flow 
equations in the large wave-number sector constitutes an essential property of 
the BMW scheme. However,
 it generally requires to keep arbitrary external fields since within this 
scheme, the 3- and 4- point vertex functions are  expressed as derivatives
 of 2-point functions with respect to the fields. Here, the closure  entirely 
relies on  the symmetries,
 and thus can be achieved even at zero external fields. In that sense, the BMW 
scheme appears simpler here than in standard situations ({\it e.g.} in 
equilibrium scalar theories).
 The resulting approximation is very similar to the full SO approximation for 
the KPZ problem briefly 
 presented in Sec. \ref{approxLO}, but with the difference that whereas the SO 
approximation for KPZ requires a truncation at quadratic order in the response 
field,  it is here exact in the large $|\vp|$ limit because of the symmetries.}.

\subsection{Derivation of the flow equations in the large wave-number limit}

\begin{figure}[t]
\epsfxsize=9.cm
\hspace{-.6cm}
\epsfbox{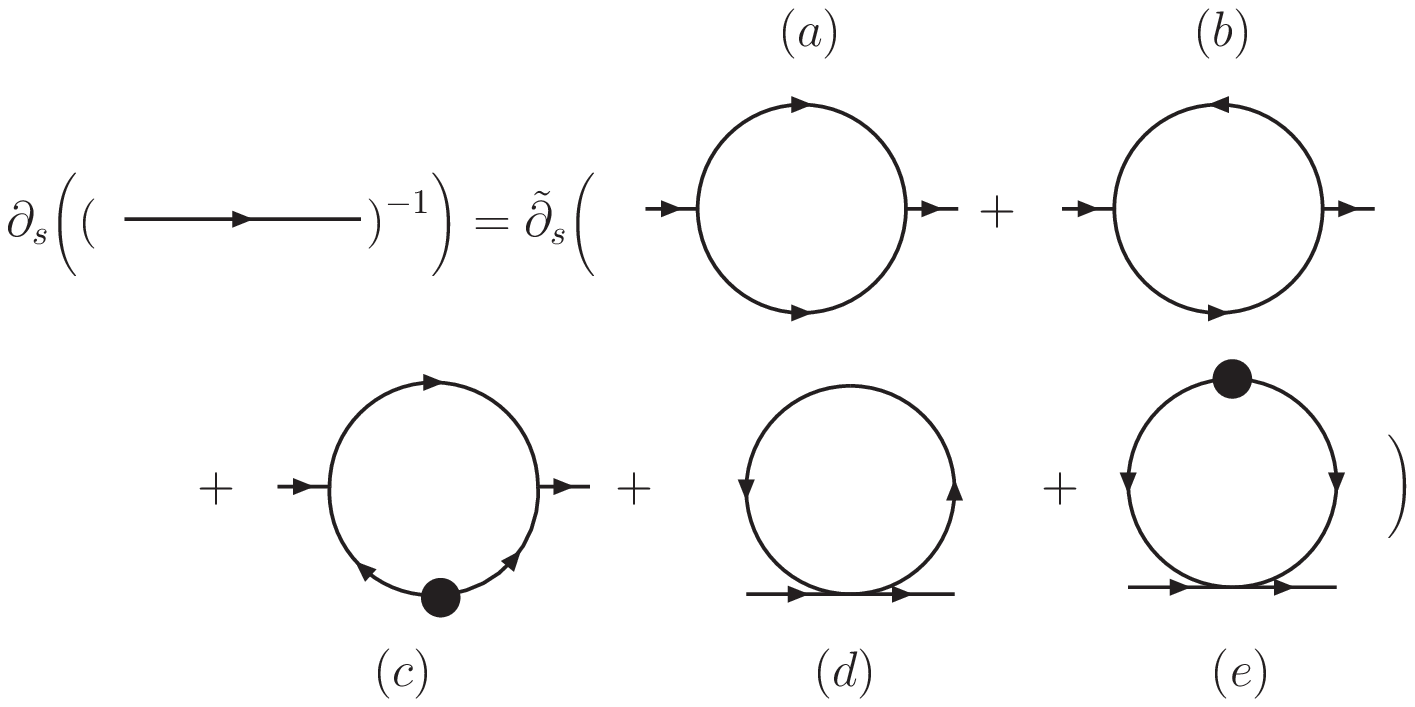}
\caption{Diagrammatic representation of the exact flow equation of 
$\Gamma^{(1,1)}_{\perp}(\nu, \vec p)$, with $\tilde \p_s \equiv \displaystyle 
\p_s R_\kappa \frac{\p}{\p R_\kappa} + \p_s N_\kappa \frac{\p}{\p N_\kappa}$. 
The combinatorial factors are not explicitly written, and
 diagrams involving $\Gamma_\kappa^{(n,0)}$ vertices are omitted since they are 
vanishing.}
\label{fig4}
 \end{figure}
 \begin{figure}[t]
\epsfxsize=9.7cm
\hspace{-1.1cm}
\epsfbox{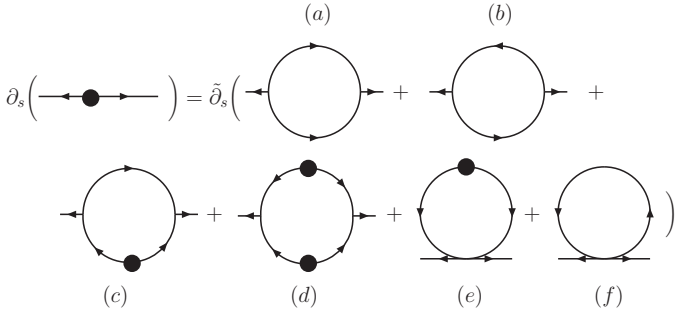}
\caption{Diagrammatic representation of the exact flow equation of 
$\Gamma^{(0,2)}_{\perp}(\nu, \vec p)$, with $\tilde \p_s \equiv \displaystyle 
\p_s R_\kappa \frac{\p}{\p R_\kappa} + \p_s N_\kappa \frac{\p}{\p N_\kappa}$.  
The combinatorial factors are not explicitly written, and
 diagrams involving $\Gamma_\kappa^{(n,0)}$ vertices are omitted since they are 
vanishing.}
\label{fig5}
 \end{figure}
The exact flow equations for the 2-point functions (\ref{dkgam2}) are 
represented diagrammatically in Figs.~\ref{fig4} and \ref{fig5}.   
 The diagrams involving  $\Gamma_\kappa^{(n,0)}$  vertices are not depicted 
since they are vanishing as a consequence of the general properties of
 NPRG within the Janssen-de Dominicis formalism and  It$\bar{\rm o}$'s discretisation 
\cite{Benitez12b,Canet11b}.
 
Let us show that the remaining diagrams in Figs.~\ref{fig4} and \ref{fig5} are 
either negligible, or closed (expressed in terms of 2-point functions) in the 
large $|\vp|$ limit. Indeed, the presence of the regulator term $\p_\kappa {\cal 
R}_\kappa$ (implicit in the $\tilde\p_s$ operator) effectively 
  cuts off the internal wave-vector $\vq$ integral to values of order $|\vq| 
\lesssim \kappa$ (and similar conclusions can be drawn when $\vp+\vq$ is cut off 
instead of $\vq$, see Appendix F). 
 In the limit of large external wave-number
 $|\vp| \gg \kappa$, the internal wave-vector $\vq$  is  negligible in all 
vertices compared to $\vp$ and can be set to zero.  Let us emphasize that this 
can be safely done since analyticity is ensured for all vertex functions  at any 
scale $\kappa\neq 0$ \footnote{This holds true in the presence of the regulator 
$R_\kappa$, and would not without it.}, and only $|\vp| \gg |\vq|$ is required, 
without assumptions on $\kappa$.
 This implies that, in this limit, the 3- and 4-point vertex functions are to be 
evaluated at one, respectively two, vanishing wave-vectors, and this implies
\begin{itemize}
 \item   if the zero wave-vector  is carried by a $\bar \vu$-leg, then the 
corresponding vertex is vanishing as a consequence of the time-gauged shift 
symmetry 
  (encoded in the general  Ward identity (\ref{WshiftNM})) and thus such 
diagrams are negligible in the large $\vp$ limit (diagrams (a), (b),  (d) in 
Fig.~\ref{fig4} and (a), (b), (f) in Fig.~\ref{fig5}). 
\item  if the zero wave-vector  is carried by a $\vu$-leg, then the 
corresponding vertex is exactly related  to lower order vertices by  time-gauged 
Galilean Ward identities and thus such diagrams are closed in the large $\vp$ 
limit.
 \end{itemize}
Let us make this last assertion explicit. There are two 3-point vertex 
functions, $\Gamma^{(2,1)}$ and $\Gamma^{(1,2)}$, involved in the (non-zero) 
diagrams (c) of Fig.~\ref{fig4} and (c) and (d) of Fig.~\ref{fig5}, which are to 
be evaluated at a zero wave-vector on a $\vu$-leg. As derived in Appendix E, 
they are related by a time-gauged Galilean Ward identity to 2-point vertex 
functions as
\begin{align}
 \Gamma^{(2,1)}_{\alpha\beta\gamma}(\nu,&\vec 0;\omega,\vec 
q)=-\frac{q^\alpha}{\nu}\Big(\Gamma^{(1,1)}_{\beta\gamma}(\nu+\omega,\vec 
q)-\Gamma^{(1,1)}_{\beta\gamma}(\omega,\vec q)\Big)\label{ward-21}\\
 \Gamma^{(1,2)}_{\alpha\beta\gamma}(\nu,&\vec 0;\omega,\vec 
q)=-\frac{q^\alpha}{\nu}\Big(\Gamma^{(0,2)}_{\beta\gamma}(\nu+\omega,\vec 
q)-\Gamma^{(0,2)}_{\beta\gamma}(\omega,\vec q)\Big).\label{ward-12}
\end{align}
Similarly, one can deduce from the Ward identities (\ref{WG22}), (\ref{WG31}), 
(\ref{ward-21}) and (\ref{ward-12}) that the two 4-point vertex functions, 
$\Gamma^{(2,2)}$ and $\Gamma^{(3,1)}$, involved in the remaining diagrams (e) of 
Figs.~\ref{fig4} and \ref{fig5}, to be evaluated at two vanishing wave-vectors, 
are related to 2-point vertex functions as
\begin{widetext}
\begin{align}
\Gamma^{(2,2)}_{\alpha\beta\gamma\delta}(\omega,\vec q = \vec 0, -\omega, -\vec 
q = \vec 0,\nu, \vec p) &= \frac{p^\alpha p^\beta}{\omega^2}\Bigg[ 
\Gamma^{(0,2)}_{\gamma\delta}(\omega+\nu,\vec p) -2 
\Gamma^{(0,2)}_{\gamma\delta}(\nu,\vec p) 
+\Gamma^{(0,2)}_{\gamma\delta}(-\omega+\nu,\vec p)  \Bigg] \label{ward-22} \\
\Gamma^{(3,1)}_{\alpha\beta\gamma\delta}(\omega,\vec q = \vec 0, -\omega, -\vec 
q = \vec 0,\nu, \vec p)& = \frac{p^\alpha p^\beta}{\omega^2}\Bigg[ 
\Gamma^{(1,1)}_{\gamma\delta}(\omega+\nu,\vec p) -2 
\Gamma^{(1,1)}_{\gamma\delta}(\nu,\vec p) 
+\Gamma^{(1,1)}_{\gamma\delta}(-\omega+\nu,\vec p)  \Bigg] \label{ward-31}.
\end{align}
\end{widetext}
Thus, the expressions of  all the non-vanishing diagrams contributing to the 
flows of the 2-point  functions can be exactly closed, {\it i.e.} expressed in 
terms of 2-point functions only, in the limit of large external wave-number. 

Let us emphasize that,  since the Galilean symmetry is gauged in time, {\it no} 
approximation is needed for the internal frequency once the internal wave-vector 
is neglected. 
This is a great advantage since an expansion on the internal frequency would not 
be justified as it is not cut off. Indeed, the regulator functions $N_\kappa$ 
and $R_\kappa$ only depend on momenta, but not on frequencies (which is required 
in order to maintain the various symmetries of the model along the flow). Hence, 
the internal momentum can safely be neglected when $|\vp|\gg \kappa$, and once 
this is done, the internal frequency dependence is entirely fixed by the 
symmetries and no  approximation is necessary in the frequency sector.

The non-zero  diagrams (c), (e) of Fig.~\ref{fig4} and (c), (d), (e) of 
Fig.~\ref{fig5} are explicitly calculated in Appendix F. Gathering their 
contributions,  the NPRG evolution of the transverse 2-point functions in the 
limit $|\vp| \gg \kappa$
is given by the exact flow equations 
\begin{widetext}
\begin{align}
\partial_s \Gamma^{(1,1)}_{\perp}(\nu, \vec p)&= \frac{(d-1)}{d} p^2 
\int_{\omega}   \Bigg\{ -\left[\frac{\Gamma^{(1,1)}_\perp(\omega+\nu, \vec p)- 
\Gamma^{(1,1)}_\perp(\nu, \vec p)}{\omega}\right]^2  G^{u\bar 
u}_{\perp}(-\omega-\nu,\vec p) \nonumber\\
& + \frac 1 {2 \omega^2}\Bigg[ \Gamma^{(1,1)}_{\perp}(\omega+\nu,\vec p) -2 
\Gamma^{(1,1)}_{\perp}(\nu, \vec p) +\Gamma^{(1,1)}_{\perp}(-\omega+\nu,\vec p)  
\Bigg] \Bigg\}
\times \tilde\partial_s \int_{\vec q} G^{ u u}_{\perp}(\omega,\vec q) 
\label{dtGam11}\\
\partial_s \Gamma^{(0,2)}_{\perp}(\nu, \vec p)&= \frac{(d-1)}{d} p^2 
\int_{\omega}   \Bigg\{
 \Bigg|\frac{\Gamma^{(1,1)}_\perp(\nu, \vec p)- \Gamma^{(1,1)}_\perp(\nu+\omega, 
\vec p)}{\omega} \Bigg|^2 \,    G^{ u u}_{\perp}(\omega+\nu,\vec p) \nonumber\\
&- 2 \left[\frac{\Gamma^{(0,2)}_\perp(\omega+\nu, \vec p)- 
\Gamma^{(0,2)}_\perp(\nu, \vec p)}
{\omega}\right] \times \Re \left\{\left[\frac{\Gamma^{(1,1)}_\perp(\omega+\nu, 
\vec p)- \Gamma^{(1,1)}_\perp(\nu, \vec p)}{\omega}\right]  G^{u \bar 
u}_{\perp}(-\omega-\nu,\vec p) \right\}\nonumber \\
&+ \frac 1 {2\omega^2}\Bigg[ \Gamma^{(0,2)}_\perp(\omega+\nu,\vec p) -2 
\Gamma^{(0,2)}_\perp(\nu,\vec p) +\Gamma^{(0,2)}_\perp(-\omega+\nu,\vec p)  
\Bigg] \Bigg\}  \times \tilde\partial_s \int_{\vec q} G^{ u 
u}_{\perp}(\omega,\vec q). 
\label{dtGam02}
\end{align}
\end{widetext}
We  study below the decoupling property of these equations and its consequences.

\subsection{Study of the (non-)decoupling}
\label{SECnonDCP}

In this section, we prove that the large wave-number sector does not decouple in 
the  NS flow equations Eqs.~(\ref{dtGam11}) and (\ref{dtGam02}). Our strategy is 
as previously to assume that such a decoupling does take place  and then to show 
that this leads to a contradiction.

We consider the inertial regime of wave-numbers $|\vp|$ much larger than the 
running inverse integral scale $\kappa$ and much smaller than the inverse  
microscopic Kolmogorov scale. The effective action $\Gamma_\kappa$ has thus  
already approached the  IR attractive fixed point, and it is convenient to 
rewrite Eqs.~(\ref{dtGam11}) and (\ref{dtGam02}) in terms of  dimensionless 
quantities:
\begin{align}
 \partial_s &\Gamma_\perp^{(1,1)}(\nu,p)=\kappa^2\nu_\kappa \Big\{(2-\eta^\nu) 
\hat\Gamma_\perp^{(1,1)}(\hat \nu,\hat p)
 -\hat p\partial_{\hat p}\hat\Gamma_\perp^{(1,1)}(\hat \nu,\hat p)\nonumber\\
& -(2-\eta^\nu) \hat \nu \partial_{\hat \nu}\hat\Gamma_\perp^{(1,1)}(\hat 
\nu,\hat p)+\partial_s \hat\Gamma_\perp^{(1,1)}(\hat \nu,\hat p)\Big\} 
\nonumber\\
 \partial_s &\Gamma_\perp^{(0,2)}(\nu,p)=D_\kappa\Big\{-\eta^\td 
\,\hat\Gamma_\perp^{(0,2)}(\hat \nu,\hat p)-\hat p\partial_{\hat 
p}\hat\Gamma_\perp^{(0,2)}(\hat \nu,\hat p)\nonumber\\
& -(2-\eta^\nu) \hat \nu \partial_{\hat \nu}\hat\Gamma_\perp^{(0,2)}(\hat 
\nu,\hat p)+\partial_s \hat\Gamma_\perp^{(0,2)}(\hat \nu,\hat p)\Big\} 
\label{dimensionlessflowexact}
\end{align}
where
\begin{align}
 \Gamma_\perp^{(1,1)}(\nu,p)&=\kappa^2\nu_\kappa \hat\Gamma_\perp^{(1,1)}(\hat 
\nu,\hat p) \nonumber \\
 \Gamma_\perp^{(0,2)}(\nu,p)&=D_\kappa \hat\Gamma_\perp^{(0,2)}(\hat \nu,\hat 
p). 
 \end{align}
 At the fixed point,  by definition
 $\partial_s \hat\Gamma_\perp^{(1,1)}(\hat \nu,\hat p)=\partial_s 
\hat\Gamma_\perp^{(0,2)}(\hat \nu,\hat p)=0$ and the running coefficients behave 
as $D_\kappa  \sim \kappa^{-\eta^\td}$ and
 $\nu_\kappa \sim \kappa^{-\eta^\nu}$.
We now  assume  that for $|\vp|\gg \kappa$, the right hand sides of  
Eqs.~(\ref{dtGam11}) and (\ref{dtGam02}) are negligible (decoupled), that is 
$\partial_s \Gamma_\perp^{(0,2)}\simeq \partial_s \Gamma_\perp^{(1,1)}\simeq 0$ 
in
 Eqs.~(\ref{dimensionlessflowexact}). The general solutions of the remnant 
homogeneous linear equations are the scaling forms 
\begin{align}
 \hat\Gamma_\perp^{(1,1)}(\hat \nu,\hat p)&=\hat p^{2-\eta^\nu} \hat 
\chi^{(1,1)}\big(\hat \nu/\hat p^{2-\eta^\nu}\big), \nonumber\\
 \hat\Gamma_\perp^{(0,2)}(\hat \nu,\hat p)&=\hat p^{-\eta^\td} \hat 
\chi^{(0,2)}\big(\hat \nu/\hat p^{2-\eta^\nu}\big),
\end{align}
or, equivalently, 
\begin{align}
 \Gamma_\perp^{(1,1)}(\nu, p)&= p^{2-\eta^\nu} \chi^{(1,1)}\big(\nu/ 
p^{2-\eta^\nu}\big), \nonumber\\
 \Gamma_\perp^{(0,2)}(\nu, p)&= p^{-\eta^\td}  \chi^{(0,2)}\big(\nu/ 
p^{2-\eta^\nu}\big).
\label{scalingform}
\end{align}
Both functions $\chi^{(i,j)}(z)$ and $\hat\chi^{(i,j)}(z)$ are equal up to some
 (non-universal)  normalisations (of the functions and of their arguments).

Let us now prove that this is  inconsistent. For this, we substitute the 
obtained solutions (\ref{scalingform}) in the right hand sides of 
Eqs.~(\ref{dtGam11}) and (\ref{dtGam02}) and show that they are not negligible 
compared to the  other  terms of Eqs.~(\ref{dimensionlessflowexact}).  
 To determine the behavior of the right hand sides of Eqs.~(\ref{dtGam11}) and 
(\ref{dtGam02}), we assign to  each quantity appropriate powers of $p$:
\begin{itemize}
\item the internal wave-vector $\vq$ is tailored to values $q\sim \kappa$ by the 
presence of the term
  $\p_s\mathcal{R}_\kappa$, that is, it is  of order one as $p\gg \kappa$. 
 \item the external frequency $\nu$ scales as $p^{2-\eta^\nu}$.
 \item the internal frequency $\omega$  satisfies $\omega \ll \nu \sim 
p^{2-\eta^\nu}$, as shown below.
 \end{itemize}
As a matter of fact,  the internal frequency
is not cut off by the regulator and it is not clear {\it a priori} which region 
of integration on $\omega$ dominates. There are essentially two scales, $\kappa$ 
and $p$. If one assumes that $\omega\sim\nu\sim p^{2-\eta^\nu}$ then 
 the resulting integral on $\omega$ behaves for small $\omega$ as $\int 
d\omega/\omega^2$ which is IR divergent.
  This means that the dominating internal frequencies are $\omega \ll \nu \sim 
p^{2-\eta^\nu}$. In this limit, the flow equations (\ref{dtGam11}) and 
(\ref{dtGam02}) acquire a simpler  form
\begin{align}
\partial_s \Gamma^{(1,1)}_{\perp}(\nu, \vec p)&=\frac{(d-1)}{d} p^2 I_0
\Big\{\frac 1 2 \partial_\nu^2\Gamma^{(1,1)}_\perp(\nu, \vec p)\nonumber\\
&-  \left( \partial_\nu\Gamma^{(1,1)}_\perp(\nu, \vec p)\right)^2 \;G^{u\bar 
u}_{\perp}(-\nu,\vec p)\Big\}\label{eqsnugrand}\\
\partial_s \Gamma^{(0,2)}_{\perp}(\nu, \vec p)&= \frac{(d-1)}{d} p^2 I_0
\Big\{\frac 1 2 \partial_\nu^2\Gamma^{(0,2)}_\perp(\nu, \vec p) \nonumber\\
&\hspace{-1cm}- 2 \partial_\nu\Gamma^{(0,2)}_\perp(\nu, \vec p)\times \Re \Big[ 
\partial_\nu\Gamma^{(1,1)}_\perp(\nu, \vec p) G^{u \bar u}_{\perp}(-\nu,\vec p) 
\Big]\nonumber\\
&\hspace{-1cm}+ \Big| \partial_\nu\Gamma^{(1,1)}_\perp(\nu, \vec p) \Big|^2 
G^{uu}_{\perp}(\nu,\vec p) \Big\}\label{eqsDgrand}
\end{align}
where
\begin{equation*}
I_0 =  \tilde\partial_s \int_{\omega,\vec q} G^{uu}_{\perp}(\omega,\vec q). 
\end{equation*}
Under this form,  the relevant scale for the internal frequency $\omega$ is 
manifestly $\kappa$ since the integral $I_0$ does not depend on the external 
scales $p$ or $\nu$.

 Substituting the scaling solutions (\ref{scalingform}) in the right-hand sides 
of Eqs.~(\ref{eqsnugrand})  and (\ref{eqsDgrand})
 one obtains that the equations for $\Gamma^{(1,1)}(\nu,p)$ and 
$\Gamma^{(0,2)}(\nu,p)$ behave as $p^{\eta^\nu}$ and  
$p^{-2-\eta^\td+2\eta^\nu}$, respectively.  Since the left-hand sides behave as 
$p^{2-\eta^\nu}$ and  $p^{-\eta^\td}$, respectively, this yields violations of 
the scaling
which are not marginal, but quite substancial: the right-hand sides are not 
sub-leading compared to the left-hand ones but dominating by a factor $p^{2/3}$ 
in $d=3$ and $p^2$ in $d=2$. This clearly is not consistent and proves that 
there is no decoupling of  the large  wave-number sector $|\vp|\gg \kappa$.\\

§

\subsection{Consequences of the non-decoupling and intermittency}

This non-decoupling property is {\it extremely peculiar}. It means that 
correlation functions remain sensitive to the integral scale even at $|\vp|\gg 
\kappa$, which  is completely different from what occurs in critical phenomena, 
where correlation functions have a  well-behaved infinite-volume limit.  The 
prominence of the integral scale in the onset of intermittency was already 
observed in the perturbative context  \cite{Adzhemyan08}. The origin of this 
difference can be intuitively understood.
 A dissipative system such as a fluid cannot sustain well-defined stationary 
correlation functions without injection of energy, and thus it remains  in some 
way sensitive
to the corresponding scale, the integral scale, even at much larger wave-length 
scales.
 It is therefore reasonable to infer that this violation of scaling is general 
in fully developed turbulence, and not restricted to 2-point correlation 
functions. This could explain the origin of intermittency: correlation functions 
are dominated by the existence of an IR
fixed point, leading to power-law behavior, but the absence of decoupling 
prevents the existence of usual scaling (determined by a finite set of anomalous 
dimensions, here $\eta^\nu$ and $\eta^\td$),  and opens the door to 
multiscaling.

Let us make one further step and  try to explain  why the lowest order structure 
functions display only very small corrections to the dimensional scaling 
exponents. Obviously
 the  four-fifth theorem forbids anomalous corrections for the $S^{(3)}$ 
structure function. 
 As for  $S^{(2)}$, one can justify very small but non-zero corrections  in the 
following way.
In the regime of large wave-numbers, the equations (\ref{eqsnugrand}) and 
(\ref{eqsDgrand}) become exact. 
 These equations read in terms of the connected 2-point correlation functions 
\begin{align}
\label{eqsnugrandG1}
\partial_s G^{u\bar u}_{\perp}(\nu, \vec p)&=\frac{(d-1)}{2d} p^2 I_0 
\partial_\nu^2 G^{u\bar u}_\perp(\nu,\vp),\\
\partial_s G^{uu}_{\perp}(\nu, \vec p)&= \frac{(d-1)}{2d} p^2 I_0 \partial_\nu^2 
G^{uu}_\perp(\nu,\vp).
\label{eqsnugrandG2}
\end{align}
Again, if the associated dimensionless  functions approach a fixed point and if 
decoupling is assumed,
 one can show that  right-hand sides  are enhanced with respect to the other 
terms  by a factor $p^{2/3}$ in $d=3$ and $p^2$ in $d=2$. 
 However, the functions that are
usually measured experimentally are not directly the functions $G^{u\bar 
u}_{\perp}(\nu, \vec p)$ or $G^{uu}_{\perp}(\nu, \vec p)$, but correlators at 
{\it equal times} such as
\begin{equation}
 \int \frac{d\omega}{2\pi}G^{uu}_{\perp}(\omega, \vec p),
\end{equation}
see {\it e.g.} Eq.~(\ref{s2}). As the function $G^{uu}_{\perp}(\omega, \vec p)$ 
is expected  to remain regular in frequency when $\kappa\to 0$, one can 
integrate Eq.~(\ref{eqsnugrandG2}) over the  frequency, which yields
\begin{equation}
 \partial_s \int \frac{d\omega}{2\pi}G^{uu}_{\perp}(\omega, \vec p)\ll 
\hbox{leading terms}.
\end{equation}
This means that the {\it leading} term that violates the decoupling is zero when 
integrated over frequencies.  Accordingly, the possible leading intermittency 
correction
to this quantity  comes from a sub-leading contribution. This could explain the 
smallness of the deviation 
  for the second order structure
function. On the contrary, higher-order $n$-point functions bear a more 
complicated frequency structure and such
  compensations are very unlikely to occur, and thus intermittency effects could 
be much larger for higher-order structure functions,
 as observed in experiments and numerical simulations.

\section{Conclusion}

In this paper, we  expounded the NPRG formalism  to investigate the regime of  
fully developed isotropic and homogeneous turbulence
 of the NS equation in the presence of a stochastic forcing. We then developed 
two complementary approaches to solve the NPRG flow equations.
  We first implemented a simple approximation,
  called the LO approximation, which consists in proposing an ansatz for the 
running effective action $\Gamma_\kappa$
  based on the NS symmetries. 
  By numerically integrating the corresponding  equations, 
 we found a fully attractive fixed point in dimension $d=2$ and $d=3$,
 governing the stationary regime of fully developed turbulence in the presence of an integral-scale forcing.
 This fixed point was already identified with approximations similar to ours in $d=3$ in Refs. \cite{Tomassini97,Monasterio12}
  and in $d=2$ in Ref. \cite{Monasterio12}. 
 The remarkable feature of this fixed point is the emergence of deviations to 
the dimensional scaling for the two-point functions.
  These deviations turn out to compensate very precisely for the energy spectrum 
and the second-order structure function, such that for instance for  the energy 
spectrum, the 
 Kolmogorov scaling $p^{-5/3}$ in $d=3$ and the Kraichnan-Batchelor one $p^{-3}$ 
in $d=2$ (in the direct cascade), are recovered.

 To further analyze the regime of large wave-number where these deviations lie, 
  we derived a set of closed 
 flow equations for the 2-point functions, which are exact in this regime. 
  We proved that the usual decoupling property of NPRG flows is violated, that 
is the large wave-number sector
 does not decouple from the flow equations, which in turn prevents the usual 
scale invariance. 
 More precisely,  on the one hand, the existence of the fixed point entails 
power-law behavior for the correlation functions.
 On the other hand, the non-decoupling of the large wave-numbers allows for 
violations of simple scaling to occur,  
     which means that the exponents
 can deviate from their dimensional values. This opens the door to multiscaling 
and intermittency. 
 We also suggested why these deviations remain small for the low-order structure 
functions (and the energy spectrum), but may be larger for higher-order ones.
The value of the corresponding intermittency exponents can be computed
 by integrating the exact  flow equations obtained in the large wave-number 
regime, which will be investigated in a future work.
  It would also be interesting to work out the link between the absence of 
operators of negative dimensions in the OPE in the perturbative context and the 
non-decoupling property unveiled in the NPRG framework.
 More generally, the purpose of the present work is to provide a detailed basis 
for future investigation of NS turbulence using NPRG methods.

\section{Acknowledgments}

 The authors acknowledge financial support from the ECOS-Sud France-Uruguay 
program U11E01, and from the PEDECIBA. LC and BD  thank the Universidad de la 
Rep\'ublica (Uruguay) for hospitality during the completion of this work, and NW 
the LPTMC for hospitality during his sabbatical year 2012-2013.  

\begin{appendix}

\renewcommand{\theequation}{A\arabic{equation}}
\section*{Appendix A: General structure of the NPRG propagator}
\setcounter{equation}{0}

In this Appendix, we establish the general structure of the $4\times4$ 
propagator matrix $G_\kappa$ 
 defined as the inverse of  $[\Gamma_\kappa^{(2)} + {\cal R}_\kappa]$.
 The matrix elements of $[\Gamma_\kappa^{(2)} + {\cal R}_\kappa]$ are obtained 
by taking  functional derivatives of  (\ref{anzGk})  and (\ref{deltaSk}) with 
respect to two of the fields $u_\alpha$, $\bar u_\alpha$, $p$ and $\bar p$.  
They are given in Fourier space  and at zero fields by, (omitting the $\kappa$ 
indices to alleviate notation)
\begin{widetext}
\begin{equation}
 \Gamma_\kappa^{(2)}(\omega,\vec p) + {\cal R}_\kappa(\vp)=
 \bordermatrix{~  & u_\beta & \bar u_\beta & p & \bar p\cr
u_\alpha      & 0 & \Gamma_{\alpha\beta}^{(1,1)}(\omega,\vec p) 
+R_{\kappa,\alpha\beta}(\vec p)& 0 & -i p_\alpha \cr
\bar u_\alpha & \Gamma_{\beta\alpha}^{(1,1)}(-\omega,\vec 
p)+R_{\kappa,\beta\alpha}(\vec p) & \Gamma_{\alpha\beta}^{(0,2)}(\omega,\vec p) 
-2 N_{\kappa,\alpha\beta}(\vec p)& i p_\alpha/\rho & 0 \cr
p &  0 & -i p_\beta/\rho & 0 & 0\cr
\bar p & i p_\beta & 0  & 0 & 0}.
\label{gamma2}
\end{equation}
Using rotational invariance and parity, one may infer that the propagator matrix 
is endowed with the following generic structure
\begin{equation}
 G_\kappa(\omega,\vec p)=
 \bordermatrix{~  & u_\beta & \bar u_\beta & p & \bar p\cr
u_\alpha      & G^{u u}_{\alpha\beta}(\omega,\vec p) & G^{u \bar 
u}_{\alpha\beta}(\omega,\vec p) & i p_\alpha G^{up}(\omega,\vec p) & i p_\alpha 
G^{u\bar p}(\omega,\vec p) \cr
\bar u_\alpha & G^{u \bar u}_{\alpha\beta}(-\omega,\vec p) & G^{\bar u\bar 
u}_{\alpha\beta}(\omega,\vec p) & i p_\alpha G^{\bar up}(\omega,\vec p) & i 
p_\alpha G^{\bar u\bar p}(\omega,\vec p) \cr
p & -i p_\beta G^{up}(-\omega,\vec p)& -i p_\beta G^{\bar up}(-\omega,\vec p) &  
G^{pp}(\omega,\vec p) & G^{p\bar p}(\omega,\vec p)\cr
\bar p & -i p_\beta G^{u\bar p}(-\omega,\vec p)& -i p_\beta G^{\bar u\bar 
p}(-\omega,\vec p) &  G^{p\bar p}(-\omega,\vec p) & G^{\bar p\bar p}(\omega,\vec 
p)}
\end{equation}
\end{widetext}
in obvious notation for the two upper indices of the different matrix elements 
of $G_\kappa$.
 The latters are obtained by requiring that the product 
$[\Gamma_\kappa^{(2)}+{\cal R}_\kappa](\omega,\vp)G_\kappa(\omega,\vp)$ is the 
identity matrix. This yields in the pressure sector
\begin{align}
 G^{pp}(\omega,\vec p)&=\frac {2\rho} {p^2}f^\td_\parallel(\vec p)\nonumber\\
G^{p\bar p}(\omega,\vec p)&=\frac{\rho}{p^2}(-i\omega+f^{\nu}_\parallel(\vec 
p))\nonumber\\
 G^{\bar p \bar p}(\omega,\vec p)&=0 .
\end{align}
 In the mixed sector, only two elements are non-vanishing, which are 
\begin{equation}
 G^{u \bar p}(\omega,\vec p) =-\frac 1 {p^2}, \hspace{.4cm}G^{\bar u 
p}(\omega,\vec p)=\frac \rho {p^2}.
\end{equation}
As for the velocity sector, one obtains that all longitudinal components vanish. 
As a consequence, the propagator in this sector is purely transverse and given 
by
\begin{align}
G^{u \bar u}_{\alpha\beta}(\omega, \vec q)&= P_{\alpha\beta}^\perp(\vec q) 
\frac{1}{\Gamma^{(1,1)}_\perp(-\omega, \vec q) + R_\kappa(\vec q)}\nonumber\\
 G^{u u}_{\alpha\beta}(\omega, \vec q)&= -P_{\alpha\beta}^\perp(\vec q) 
\frac{\Gamma^{(0,2)}_\perp(\omega, \vec q) -2 N_\kappa(\vec 
q)}{\left|\Gamma^{(1,1)}_\perp(\omega, \vec q)+R_\kappa(\vec 
q)\right|^2}\nonumber\\
G^{\bar u\bar u}_{\alpha\beta}(\omega, \vec q)&= 0.
\end{align}

\begin{widetext}
\renewcommand{\theequation}{B\arabic{equation}}
\section*{Appendix B: Derivation of the NPRG flow equations at LO}
\setcounter{equation}{0}

In this Appendix, we derive the flow equations for the transverse components of 
the two running functions $f_{\alpha\beta}^\nu(\vp)$ and  
$f_{\alpha\beta}^\td(\vp)$ of the LO \anz. They are related to the flows of
 $\Gamma_{\alpha\beta}^{(1,1)}(\nu=0,\vp)$ and 
$\Gamma_{\alpha\beta}^{(0,2)}(\nu=0,\vp)$, respectively, which we now calculate.
 
\subsection*{Flow equations of $\Gamma_{\alpha\beta}^{(1,1)}$ and 
$\Gamma_{\alpha\beta}^{(0,2)}$ at LO}

  According to Eq.~(\ref{dkgam2}), the flow equation of 
$\Gamma_{\alpha\beta}^{(1,1)}(\nu=0,\vp)$ is given by
\begin{equation}
 \partial_\kappa \Gamma^{(1,1)}_{\alpha\beta}(\nu=0,\vec p) =\mathrm{Tr} 
\int_{\omega,\vec q}\partial_\kappa {\cal R}_\kappa(\vec q)\cdot 
G_\kappa(\bq)\cdot \Big\{\Gamma^{(3)}_{\kappa,u_\alpha}(\bp,\bq)\cdot 
G_\kappa(\bp+\bq)\cdot \Gamma^{(3)}_{\kappa,\bar 
u_\beta}(\bp+\bq,-\bp)\Big\}\cdot G_\kappa(\bq),
 \label{eqdkgam11}
\end{equation}
 omitting the contributions of the 4-point vertices which are vanishing at LO. 
As apparent in (\ref{gamma2}), the regulator matrix ${\cal R}_\kappa$ has only 
three non-vanishing entries: $[{\cal R}_\kappa]_{12}=[{\cal 
R}_\kappa]_{21}=R_{\kappa}(\vp)$ and $[{\cal R}_\kappa]_{12}=-2N_{\kappa}(\vp)$. 
Moreover, at LO, only the vertex
 function $\Gamma^{(2,1)}_{\alpha\beta\gamma}$ is non-zero and contributes in 
the  $\Gamma_{\kappa,i}^{(3)}$ matrices. Performing the 
 matrix product (\ref{eqdkgam11}) and taking the trace, one is left with only 
four  terms, which are
 \begin{align}
 \partial_\kappa \Gamma^{(1,1)}_{\alpha\beta}(\nu=0,\vec p)&=
 \lambda^2 \int_{\omega,\vec q} \partial_\kappa R_\kappa(\vec q) \Bigg\{i ( q_l 
\delta_{i\beta}-(p+q)_i \delta_{l\beta})     i ( (p+q)_\alpha \delta_{kj}-p_k 
\delta_{\alpha j}) \frac{P_{ij}^\perp(\vec q) P_{kl}^\perp(\vec p+\vec q) 
}{\big(-i\omega+\tf_\perp^{\nu}(\vec q)\big)^2} 
\frac{2  \tf_\perp^\td(\vec p+\vec q)}{\big(\omega^2+ (\tf_\perp^{\nu}(\vec 
p+\vec q))^2\big)}\Bigg\}\nonumber\\
&+\lambda^2 \int_{\omega,\vec q}  \Bigg\{ \partial_\kappa R_\kappa(\vec q) 
\frac{ 4 \tf_\perp^\td(\vec q)  \tf_\perp^{\nu}(\vec 
q)}{\big(\omega^2+(\tf_\perp^{\nu}(\vec q))^2\big)^2}- 2 \frac{ \partial_\kappa 
N_\kappa(\vec q)}
{\omega^2+(\tf_\perp^{\nu}(\vec q))^2}\Bigg\}\nonumber \\
 &\times  \Bigg\{i (-q_l \delta_{i\beta}+(q+p)_i \delta_{\beta l}) i ( q_\alpha 
\delta_{jk}+p_j \delta_{\alpha k})
\frac{P_{ij}^\perp(\vec q) P_{kl}^\perp(\vec p+\vec q)}{i\omega+ 
\tf_\perp^{\nu}(\vec p+\vec q)} \Bigg\}
\label{dkgam11-2}
\end{align}
with the notation
\begin{equation}
 \tf_\perp^{\nu}(\vec q)= f_\perp^{\nu}(\vec q) + R_\kappa(\vq)  
\quad\quad\quad\quad\hbox{and}\quad\quad \quad\quad\tf_\perp^{\td}(\vec q)= 
f_\perp^{\td}(\vec q) + N_\kappa(\vq).
\end{equation}
This flow equation is to be projected onto the transverse sector. Hence, all the 
terms proportional 
 to $p_\alpha$ or $p_\beta$ can be discarded since $P_{\alpha\beta}^\perp(\vec 
p) p_\alpha= P_{\alpha\beta}^\perp(\vec p) p_\beta=0$. 
 The two tensor structures in Eq.~(\ref{dkgam11-2}) can  be then simplified in 
the following way
\begin{align}
 T_a^{(1,1)}&\equiv-P_{ij}^\perp(\vec q)( (p+q)_\alpha \delta_{kj}-p_k 
\delta_{\alpha j})P_{kl}^\perp(\vec p+\vec q)( q_l \delta_{i\beta}-(p+q)_i 
\delta_{l\beta})  \nonumber\\
 &=\Big[-\vec p\,^2+\frac{(\vec p \cdot (\vec p+\vec q))^2}{(\vec p+\vec 
q)^2}\Big]\delta_{\alpha\beta}-2 \frac{q_\alpha q_\beta}{\vec q\,^2} \vec p\cdot 
\vec q \quad+\quad\hbox{longitudinal parts}\nonumber\\
 T_b^{(1,1)}&\equiv-P_{ij}^\perp(\vec q)( q_\alpha \delta_{kj}+p_j 
\delta_{\alpha k})P_{kl}^\perp(\vec p+\vec q)( -q_l \delta_{i\beta}+(p+q)_i 
\delta_{l\beta})\nonumber \\
 & =\Big[-\vec p\,^2+\frac{(\vec p \cdot \vec q))^2}{\vec 
q\,^2}\Big]\delta_{\alpha\beta}+2 \frac{q_\alpha q_\beta}{(\vec q+\vec p)^2} 
\vec p\cdot (\vec p+\vec q) \quad +\quad\hbox{longitudinal parts}
 \label{T11}
\end{align}
where only the transverse contributions are explicitly specified.

Similarly, the flow equation of $\Gamma_{\alpha\beta}^{(0,2)}(\nu=0,\vp)$ is 
given by Eq.~(\ref{dkgam2}) 
\begin{equation}
 \partial_\kappa \Gamma^{(0,2)}_{\alpha\beta}(\nu=0,\vec p) =\mathrm{Tr} 
\int_{\omega,\vec q}\partial_\kappa {\cal R}_\kappa(\vec q)\cdot 
G_\kappa(\bq)\cdot \Big\{\Gamma^{(3)}_{\kappa,\bar u_\alpha}(\bp,\bq)\cdot 
G_\kappa(\bp+\bq)\cdot \Gamma^{(3)}_{\kappa,\bar 
u_\beta}(\bp+\bq,-\bp)\Big\}\cdot G_\kappa(\bq)
\label{dkgam02-2}
\end{equation}
omitting the vanishing 4-point vertices (at LO). 
Only three terms are left in the trace of the matrix product (\ref{dkgam02-2}), 
which are
\begin{align}
 \partial_\kappa \Gamma^{(0,2)}_{\alpha\beta}(\nu=0,\vec 
p)&=\lambda^2\int_{\omega,\vec q}  i ((q+p)_i \delta_{j\alpha}-q_j 
\delta_{\alpha i}) i ( q_k \delta_{l\beta}-(p+q)_l \delta_{\beta 
k})P_{il}^\perp(\vec q) P_{kj}^\perp(\vec p+\vec q)
\nonumber\\
&\times\Big\{ \partial_\kappa R_\kappa(\vec q) \frac{ 4  \tf_\perp^\td(\vec 
q)\tf_\perp^\nu(\vec q)}{\big(\omega^2+(\tf_\perp^{\nu}(\vec q))^2\big)^2}- 2 
\frac{ \partial_\kappa N_\kappa(\vec q)}
{\omega^2+ (\tf_\perp^{\nu}(\vec q))^2}\Big\}\frac{2 \tf_\perp^\td(\vec q+\vec 
p)}{\omega^2+ (\tf_\perp^{\nu}(\vec p+\vec q))^2}.
\end{align}
 As previously, ignoring the terms proportional to $p_\alpha$ and $p_\beta$, the 
tensor structure in this equation simplifies to
\begin{align}
 T^{(0,2)}&= -P_{il}^\perp(\vec q)( p_i\delta_{j\alpha}+p_j 
\delta_{i\alpha})P_{jk}^\perp(\vec p+\vec q)(-p_l \delta_{k\beta}-p_k 
\delta_{l\beta})  \quad+\quad\hbox{longitudinal parts}\nonumber\\
 &=\Big[2\vec p\,^2+\frac{(\vec p \cdot (\vec p+\vec q))^2}{(\vec p+\vec 
q)^2}-\frac{(\vec p \cdot \vec q)^2}{\vec q\,^2}\Big]\delta_{\alpha\beta}+2 
\frac{q_\alpha q_\beta}{\vec q\,^2 (\vec p+\vec q)^2}
 \Big(\vec p\,^2 \vec p\cdot \vec q+2 (\vec p\cdot \vec q)^2-\vec p\,^2 \vec 
q\,^2\Big)\quad +\quad\hbox{longitudinal parts}.
 \label{T02}
\end{align}

\subsection*{Flow equations  of $f_\perp^\td$ and $f_\perp^\nu$}

According to the expressions (\ref{gam2LO})  of the 2-point functions at LO in 
the velocity sector, the
 flow equations of the transverse functions $f_\perp^\td$ and $f_\perp^\nu$ may 
be defined as
 \begin{align}
\partial_\kappa f_\perp^{\nu}(\vec p)&  = \frac{1}{(d-1)}\, 
P_{\alpha\beta}^\perp(\vec p) \p_\kappa \Gamma^{(1,1)}_{\alpha\beta}(\nu=0,\vec 
p)  \nonumber\\
\partial_\kappa f_\perp^{\td}(\vec p)&  = -\frac{1}{2(d-1)} 
\,P_{\alpha\beta}^\perp(\vec p)\p_\kappa  
\Gamma^{(0,2)}_{\alpha\beta}(\nu=0,\vec p).
\end{align}
The  flow equations of  $\partial_\kappa \Gamma^{(1,1)}_{\alpha\beta}$ and 
$\partial_\kappa \Gamma^{(0,2)}_{\alpha\beta}$ are proportional to the tensor 
structures  (\ref{T11}) and (\ref{T02}), respectively, which projections 
 onto the transverse sector are straightforward, using
\begin{equation}
P_{\alpha\beta}^\perp(\vec p)\delta_{\alpha\beta}=d-1 
\quad\quad\quad\quad\hbox{and}\quad\quad\quad\quad
P_{\alpha\beta}^\perp(\vec p)q_\alpha q_\beta =q^2 - \frac{(\vec p\cdot \vec 
q)^2}{\vec p\,^2}.
 \end{equation}
 One  obtains
\begin{align}
 \partial_\kappa f_\perp^{\nu}(\vec p)&= 
\frac{\lambda^2}{(d-1)}\int_{\omega,\vec q}\Bigg\{ \frac{2  \tf_\perp^\td(\vec 
p+\vec q)\partial_\kappa R_\kappa(\vec q) }{\big(-i\omega+\tf_\perp^{\nu}(\vec 
q)\big)^2\big(\omega^2+ (\tf_\perp^{\nu}(\vec p+\vec q))^2\big)}
 \Big[\Big(-\vec p\,^2+\frac{(\vec p \cdot (\vec p+\vec q))^2}{(\vec p+\vec 
q)^2}\Big)(d-1)-2 \vec p\cdot \vec q\Big(1-\frac{(\vec p\cdot \vec q)^2}{\vec 
q\,^2 \vec p\,^2}\Big) \Big]\nonumber\\
&+ \Big\{ \partial_\kappa R_\kappa(\vec q) \frac{ 4 \tf_\perp^\td(\vec q)  
\tf_\perp^{\nu}(\vec q)}{\big(\omega^2+(\tf_\perp^{\nu}(\vec q))^2\big)^2}- 2 
\frac{ \partial_\kappa N_\kappa(\vec q)}
{\omega^2+(\tf_\perp^{\nu}(\vec q))^2}\Big\}\frac{1}{i\omega+ 
\tf_\perp^{\nu}(\vec p+\vec q)} 
\nonumber\\
&\times \Big[\Big(-\vec p\,^2+\frac{(\vec p \cdot \vec q)^2}{\vec 
q\,^2}\Big)(d-1)+2 \frac{\vec p\cdot (\vec p+\vec q)}{(\vec q+\vec 
p)^2}\Big(\vec q\,^2-\frac{(\vec p\cdot \vec q)^2}{\vec p\,^2}\Big) 
\Big]\Bigg\}\nonumber\\
 \partial_\kappa f_\perp^\td(\vec p)&
 =-\frac{\lambda^2}{2(d-1)}\int_{\omega,\vec q}  \Bigg\{
\Big[ \partial_\kappa R_\kappa(\vec q) \frac{ 4  \tf_\perp^\td(\vec 
q)\tf_\perp^\nu(\vec q)}{\big(\omega^2+(\tf_\perp^{\nu}(\vec q))^2\big)^2}- 2 
\frac{ \partial_\kappa N_\kappa(\vec q)}
{\omega^2+ (\tf_\perp^{\nu}(\vec q))^2}\Big]\frac{2 \tf_\perp^\td(\vec q+\vec 
p)}{\omega^2+ (\tf_\perp^{\nu}(\vec p+\vec q))^2}
\nonumber\\
&\times\Big[\Big(2\vec p\,^2+\frac{(\vec p \cdot (\vec p+\vec q))^2}{(\vec 
p+\vec q)^2}-\frac{(\vec p \cdot \vec q)^2}{\vec q\,^2}\Big)(d-1)+2 
\frac{1}{\vec q\,^2 (\vec p+\vec q)^2}\Big(\vec q\,^2-\frac{(\vec p\cdot \vec 
q)^2}{\vec p\,^2}\Big) 
 \Big(\vec p\,^2 \vec p\cdot \vec q+2 (\vec p\cdot \vec q)^2-\vec p\,^2 \vec 
q\,^2\Big)\Big]\Bigg\}.
\end{align}
Within the LO approximation, the frequency dependence remains the bare one, and 
the integration over the internal frequency $\omega$ can be carried out 
analytically in the above expressions. Denoting
$A=\tf_\perp^{\nu}(\vec q)^2$ and
 $B=\tf_\perp^{\nu}(\vec p+\vec q)^2$, the different frequency integrals are 
given by 
\begin{align}
 I_1&=\int \frac{d\omega}{2\pi} 
\frac{1}{(-i\omega+A)^2}\frac{1}{\omega^2+B^2}=\frac{1}{2 B (A+B)^2} \nonumber\\
 I_2&=\int \frac{d\omega}{2\pi} 
\frac{1}{i\omega+B^2}\frac{1}{\omega^2+A^2}=\frac{1}{2 A (A+B)} \nonumber\\
 I_3&=\int \frac{d\omega}{2\pi} 
\frac{1}{i\omega+B}\frac{1}{(\omega^2+A^2)^2}=\frac{2A+B}{4 A^3 (A+B)^2} 
\nonumber\\
 I_4&=\int \frac{d\omega}{2\pi} 
\frac{1}{\omega^2+B^2}\frac{1}{\omega^2+A^2}=\frac{1}{2 A B (A+B)} \nonumber\\
 I_5&=\int \frac{d\omega}{2\pi} 
\frac{1}{\omega^2+B^2}\frac{1}{(\omega^2+A^2)^2}=\frac{2 A +B}{4 A^3 B (A+B)^2}
 \end{align}
which yields the two flow equations (\ref{dkfnu}) and (\ref{dkfd}).
\end{widetext}

\renewcommand{\theequation}{C\arabic{equation}}
\section*{Appendix C: Average injected power per unit mass}
\setcounter{equation}{0}

The injected power per unit mass is $f_\alpha(\bx) v_\alpha(\bx)$. The average 
of a quantity linearly depending on the stochastic forcing $\vf$ can be 
calculated using the Janssen-de Dominicis procedure (see Ref. \cite{Canet15a} 
for detail), which yields
\begin{equation}
 \langle f_\alpha(t,\vec x) {\cal O}[\vv] \rangle=2  \int_{\vec x'} 
N_{\kappa,\alpha\beta}(|\vec x-\vec x'|) \langle  \bar v_\beta(t,\vec x')    
{\cal O}[\vv]\rangle. \label{avO}
\end{equation}
denoting $\kappa$ the inverse integral scale.
 Averages of quantities linear in $\vf$ are hence related to response 
functions. 
 In the particular case where $f_\alpha$ and $\mathcal{O}$ are defined at equal 
times, one must carefully
 consider the It$\bar{\rm o}$'s prescription.  
 As a matter of fact, the average of the  injected power per unit mass would 
naively be given, according to Eq.~(\ref{avO}), by
\begin{align}
 \langle f_\alpha(t,\vec x) v_\alpha(t,\vec x)\rangle\stackrel{\rm naive}{=} &2  
\int_{\vec x'} N_{\kappa,\alpha\beta}(|\vec x-\vec x'|) \nonumber\\
&\times \langle v_\alpha(t,\vec x) \bar v_\beta(t,\vec x')\rangle,\label{naive}
\end{align}
but the response fonction  at equal
times is zero because of  It$\bar{\rm o}$'s prescription. However, the precise 
meaning of equal time 
 must be carefully specified in discrete time.  In particular, for the injected 
power, one should examine the kinetic energy theorem and  properly discretize 
it.
The energy at a given space point comes from both  direct injection by the 
external force and  transfer from its neighbouring points. We here only seek the 
variation of
the velocity induced by the external force and thus  omit tranferred power from 
one point to another. Accordingly, in discrete time,
 It$\bar{\rm o}$ forward discretization is
\begin{equation}
f_\alpha(t,\vec x)= \partial_t v_\alpha(t,\vec x) =\frac{1}{\delta 
t}\big(v_\alpha (t+\delta t,\vec x)-v_\alpha (t,\vec x)\big).
\end{equation}
Thus, the discretized kinetic energy theorem is 
\begin{align}
 \partial_t \big(\frac 1 2 v_\alpha v_\alpha\big)(t,\vec x)& = 
\Big[\big(v_\alpha v_\alpha\big) (t+\delta t,\vec x)-\big(v_\alpha v_\alpha\big) 
(t,\vec x)\Big]/{(2 \delta t)}
 \nonumber\\
& =\frac 1 2 f_\alpha(t,\vec x) \big(v_\alpha (t+\delta t,\vec x)+v_\alpha 
(t,\vec x)\big)
\end{align}
 which indicates that half of the quantity to be averaged is not defined at  
coinciding but at successive times and the associated response function has a 
non-zero contribution. In conclusion, the average injected power is precisely 
defined as
\begin{align}
\langle f_\alpha(t,\vec x) v_\alpha(t,\vec x) \rangle  =&\lim_{\delta t\to 
0^+}\int_{\vec x'} N_{\kappa,\alpha\beta}(|\vec x-\vec x'|) \nonumber\\
 &\times \langle v_\alpha(t+\delta t,\vec x) \bar v_\beta(t,\vec x')\rangle,
\end{align}
which, apart from the correct limit process and the  one half factor, coincides 
with the expression (\ref{naive}).

\renewcommand{\theequation}{D\arabic{equation}}
\section*{Appendix D: Numerical integration of the LO flow equations}
\setcounter{equation}{0}

In this Appendix, we expound the detail of the numerical procedure implemented 
to integrate
 the LO flow equations (\ref{flowh}). The running functions $\hat 
h^{\nu,\td}(\hat \vp)$ only depend
 on the modulus $\hat p$ of the wave-vector $\hat \vp$.
 The dimensionless wave-numbers are discretized on a $\sqrt{\hat p}$ grid of 
typical size $\sqrt{\hat p_{\rm max}}\simeq 30$
 and spacing $\Delta \equiv \Delta \sqrt{\hat p} \simeq 1/20$. The integrals 
over the internal wave-vector $\hat \vq$ are calculated numerically using 
Simpson's rule, in cartesian coordinates, chosen with the $\hat q_1$ axis along  
the external wave-vector $\hat \vp$  and the $(d-1)$ other axes $\hat q_i$ 
spanning  the hyperplane perpendicular to $\hat \vp$. With this choice, one 
simply has $\hat \vp \cdot \hat \vq = \hat p\,\hat q_1$
  and $(\hat \vp+\hat \vq)^2 = (\hat p+\hat q_1)^2 + \sum_{i=2}^d \hat q_i^2$. 
  As a square root  grid is used for the wave-numbers, the values of the 
functions $\hat h^{\nu,\td}(\hat \vp+\hat \vq)$ 
   for arguments $|\hat \vp+\hat \vq|^{1/2}$  not falling onto mesh points are 
interpolated using cubic splines. 

 The  presence of the  $\partial_s {\cal R}_\kappa$ terms in (\ref{dkfnu}) and 
(\ref{dkfd}) ensures that the integrands decrease exponentially with $\hat q$, 
such that the internal wave-number integral can be safely cut at  an upper 
finite bound $\hat p_{\rm up} \le\hat p_{\rm max}$. 
For wave-numbers such that $|\hat \vp + \hat \vq|  > \hat p_{\rm max}$, the 
functions $\hat h^{\nu,\td}(\hat \vp+\hat \vq)$ are extended outside the grid 
using power law extrapolations. This corresponds to the expected asymptotics of 
the flowing functions, at least close  to the fixed point. 
 The derivative terms $\hat p \p_{\hat p}$ are computed using 5-point 
differences. 
  For the propagation in renormalization time $s$,  explicit Euler time stepping 
is used with a typical time step $\Delta s =- 1 \times 10^{-4}$.
Starting at $s = 0$ from the bare action ($\hat h^\nu(\hat \vp)=1$ and $\hat 
h^\td(\hat \vp)=0$), we observe that  the two functions $\hat h^{\nu,\td}$ are 
smoothly deformed 
from their flat initial shapes to acquire their fixed point profiles,  typically 
after $|s| \gtrsim 8$.
 The fixed point profiles are  recorded at $s=-30$ ({\it e.g.} in 
Fig.~\ref{fig2}).
 
 \begin{widetext}
 \renewcommand{\theequation}{E\arabic{equation}}
\section*{Appendix E:  Ward identities}
\setcounter{equation}{0}

In this Appendix, we derive the Ward identities for the vertex functions which 
 originate in the time-gauged shift and Galilean symmetries. 

\subsection{Ward identities for the time-gauged shift symmetry}

Let us consider the (functional) Ward identity (\ref{wardshift}) associated with 
the time-gauged shift symmetry, and 
rewrite it using the explicit Ward identity for the pressure sector:
\begin{equation}
\frac{\delta \Gamma_\kappa}{\delta \bar p(\bx)}=\frac{\delta {\cal S}_0}{\delta 
\bar p(\bx)} =\p_\alpha u_\alpha.
\end{equation}
 One obtains  
 \begin{equation}
\label{wardshiftbis}
\int_{\vx} \Big\{\frac{\delta \Gamma_\kappa}{\delta \bar u_\beta(\bx)}
 + u_\beta(\bx) \p_\gamma u_\gamma(\bx) \Big\}= \int_{\vx}  \p_t u_\beta(\bx).
\end{equation}

Differentiating this equation with respect to $\bar u_\alpha(t_y,\vec y)$ and 
evaluating it at vanishing fields, one obtains 
 \begin{equation}
\int_{\vec x}  \Gamma_{\beta\alpha}^{(0,2)}(t,\vec x, t_y,\vec y) =0 
\label{Wshift02r}
\end{equation}
or equivalently in Fourier space
\begin{equation}
 \Gamma_{\alpha\beta}^{(0,2)}(\omega,\vec p=\vec 0)=0.\label{Wshift02}
\end{equation}
Similarly, differentiating Eq.~(\ref{wardshiftbis}) with respect to 
$u_\alpha(t_y,\vec y)$ and evaluating it at vanishing fields yields
\begin{equation}
\int_{\vec x}  \Gamma_{\alpha\beta}^{(1,1)}(t_y,\vec y, t,\vec x) = \int_{\vec 
x} \delta_{\alpha\beta} \p_t \delta(t-t_y)\delta^d(\vec x-\vec y)
\label{eqward11}
\end{equation}
that is  in Fourier space
\begin{equation}
 \Gamma_{\alpha\beta}^{(1,1)}(\omega,\vec p=\vec 
0)=i\omega\delta_{\alpha\beta}.\label{gam11ward}
\end{equation}
Lastly,  taking two derivatives with respect to $u_\alpha(t_y,\vec y)$  and 
$u_\gamma(t_z,\vec z)$ of Eq.~(\ref{wardshiftbis}), one obtains the identity 
 \begin{equation}
  \int_{\vec x}  \Big\{\Gamma_{\alpha\gamma\beta}^{(2,1)}(t_y,\vec y,t_z,\vec z, 
t,\vec x) +  \delta_{\alpha\beta} \p_\gamma^y\delta(t_y-t_z)\delta^d(\vec y-\vec 
z) + \delta_{\beta\gamma} \p_\alpha^z\delta(t_z-t_y)\delta^d(\vec z-\vec y) 
\Big\} =0 \label{wardshift21}
 \end{equation}
which yields in Fourier space, relabeling the indices
\begin{equation}
  \Gamma_{\alpha\beta\gamma}^{(2,1)}(\omega_1,\vec p_1,\omega_2,-\vec p_1) = i 
p_1^\alpha \delta_{\beta\gamma} -ip_1^\beta \delta_{\alpha\gamma}.
\label{gam21ward}
\end{equation}

By taking  additional derivatives of (\ref{Wshift02r}) and (\ref{wardshift21}) 
with respect to either fields $\vu$ or $\bar\vu$ 
 and evaluating the resulting identity at zero external fields, one can infer 
the general property
 \begin{equation}
 \Gamma_{\alpha_1,\cdots,\alpha_{n+m}}^{(m,n)}(\omega_1,\vec p_1, 
\cdots,\omega_m,\vec p_m, \omega_{m+1},\vec p_{m+1}=\vec 0,\cdots)= 0 
\hbox{\hspace{1cm}for all $(m,n)$  but $(1,1)$ and $(2,1)$}\label{WshiftNM}
\end{equation}
which means that any $(m,n)$-point vertex function with a zero wave-vector on a 
$\bar \vu$-leg vanishes, except  the functions $(m,n)=(1,1)$ and   $(m,n)=(2,1)$ 
which  keep their bare forms (\ref{gam11ward}) and (\ref{gam21ward}).
 
 \subsection{Ward identities for the time-gauged Galilean symmetry}

Let us derive the (functional) Ward identities ensuing from the time-gauged 
Galilean symmetry. 
 Retaining only the terms which give a non-zero contribution at vanishing fields 
in the velocity sector ({\it i.e.} dropping the pressure terms), the Ward 
identity (\ref{wardgalilee}) reads
\begin{equation}
\label{wardcut2}
  \int_{\vec x} \Bigg\{\partial_\alpha u_\beta(t,\vec x)\frac{\delta 
\Gamma_\kappa}{\delta u_\beta(t,\vec x)}
  +\partial_t \frac{\delta \Gamma_\kappa}{\delta u_\alpha(t,\vec x)}
+\partial_\alpha \bar u_\beta(t,\vec x)  \frac{\delta \Gamma_\kappa}{\delta \bar 
u_\beta(t,\vec x)} \Bigg\} = -\int_{\vec x} \p_t^2 \bar u_\alpha(t,\vec x).
\end{equation}
Differentiating this equation with respect to $\bar\vu_\beta(t_y,\vec y)$, and 
evaluating the resulting identity at vanishing fields, one obtains
  \begin{equation}
 \delta_{\alpha\beta}\partial_t^2 \delta(t-t')+\int_{\vx} \partial_t 
\Gamma_{\alpha\beta}^{(1,1)}(t,\vec x,t_y,\vec y)=0
\end{equation}
which leads in Fourier space to
\begin{equation}
 \Gamma_{\alpha\beta}^{(1,1)}(\omega,\vec p=\vec 0)=i\omega 
\delta_{\alpha\beta}.\label{Wgal11}
\end{equation}

Then, taking two derivatives of Eq.~(\ref{wardcut2}) with respect to 
$u_\mu(t_y,\vec y)$ and $\bar u_\nu(t_z, \vec z)$ yields
  at vanishing fields
\begin{equation}
\label{der2}
\int_{\vec x} \Bigg\{ \partial_{t_x} \Gamma^{(2,1)}_{\alpha\mu\nu}(t_x,\vec 
x,t_y,\vec y,t_z, \vec z) - \delta(t_x-t_y)\delta(\vec x-\vec y)\partial_\alpha  
\Gamma^{(1,1)}_{\mu\nu}(t_y,\vec y,t_z, \vec z)- \delta(t_x-t_z)\delta(\vec 
x-\vec z) \partial_\alpha  \Gamma^{(1,1)}_{\mu\nu}(t_y,\vec y,t_z,\vec z)  
\Bigg\} = 0.
\end{equation}
This provides in Fourier space an exact identity relating the 3-point vertex 
$\Gamma^{(2,1)}$ with a zero wave-vector on a  $\vu$-leg to the 2-point function 
$\Gamma^{(1,1)}$, which reads
\begin{equation}
\label{WG21}
 \Gamma^{(2,1)}_{\alpha\beta\gamma}(\omega_1,\vec p_1=\vec 0;\omega_2,\vec 
p_2)=-\frac{p_2^\alpha}{\omega_1}\Big(\Gamma^{(1,1)}_{\beta\gamma}
(\omega_1+\omega_2,\vec p_2)-\Gamma^{(1,1)}_{\beta\gamma}(\omega_2,\vec 
p_2)\Big).
\end{equation}
Similarly, taking two derivatives of Eq.~(\ref{wardcut2}) with respect to $\bar 
u$,
 one obtains at vanishing fields
\begin{equation}
\int_{\vec x} \Bigg\{ \partial_{t_x} \Gamma^{(1,2)}_{\alpha\mu\nu}(t_x,\vec 
x,t_y,\vec y,t_z, \vec z) - \delta(t_x-t_y)\delta(\vec x-\vec y)\partial_\alpha  
\Gamma^{(0,2)}_{\mu\nu}(t_y,\vec y,t_z, \vec z)- \delta(t_x-t_z)\delta(\vec 
x-\vec z) \partial_\alpha  \Gamma^{(0,2)}_{\mu\nu}(t_y,\vec y,t_z,\vec z)  
\Bigg\} = 0.
\end{equation}
This yields in Fourier space an exact identity relating the 3-point vertex 
$\Gamma^{(1,2)}$ with a zero wave-vector on its  $\vu$-leg 
to the 2-point function $\Gamma^{(0,2)}$ as
\begin{equation}
\label{WG12}
 \Gamma^{(1,2)}_{\alpha\beta\gamma}(\omega_1,\vec p_1=\vec 0;\omega_2,\vec 
p_2)=-\frac{p_2^\alpha}{\omega_1}\Big(\Gamma^{(0,2)}_{\beta\gamma}
(\omega_1+\omega_2,\vec p_2)-\Gamma^{(0,2)}_{\beta\gamma}(\omega_2,\vec 
p_2)\Big).
\end{equation}

As for the 4-point vertices, taking one additional derivative of 
Eq.~(\ref{der2}) with respect to $\bar u_\rho(t_s,\vec s)$ and evaluating the 
obtained identity at vanishing fields yields
\begin{align}
& \int_{\vec x} \Bigg\{ \partial_{t_x} 
\Gamma^{(2,2)}_{\alpha\mu\rho\nu}(t_x,\vec x,t_y,\vec y,t_s ,\vec s, t_z, \vec 
z) - \delta(t_x-t_y)\delta(\vec x-\vec y)\partial_\alpha  
\Gamma^{(1,2)}_{\mu\rho\nu}(t_y,\vec y, t_s, \vec s, t_z, \vec z)\nonumber \\
 &- \delta(t_x-t_z)\delta(\vec x-\vec z) \partial_\alpha  
\Gamma^{(1,2)}_{\mu\rho\nu}(t_y,\vec y, t_s, \vec s, t_z,\vec z) - 
\delta(t_x-t_s)\delta(\vec x-\vec s) \partial_\alpha  
\Gamma^{(1,2)}_{\mu\rho\nu}(t_y,\vec y, t_s,\vec s, t_z, \vec z) \Bigg\} = 0.
\end{align}
Similarly, taking one additional derivative  of Eq.~(\ref{der2}) with respect to 
 $u_\rho(t_s,\vec s)$ and evaluating the ensuing identity at zero fields leads 
to 
 \begin{align}
& \int_{\vec x} \Bigg\{ \partial_{t_x} 
\Gamma^{(3,1)}_{\alpha\mu\rho\nu}(t_x,\vec x,t_y,\vec y,t_s ,\vec s, t_z, \vec 
z) - \delta(t_x-t_y)\delta(\vec x-\vec y)\partial_\alpha  
\Gamma^{(2,1)}_{\mu\rho\nu}(t_y,\vec y, t_s, \vec s, t_z, \vec z)\nonumber \\
 &- \delta(t_x-t_z)\delta(\vec x-\vec z) \partial_\alpha  
\Gamma^{(2,1)}_{\mu\rho\nu}(t_y,\vec y, t_s, \vec s, t_z,\vec z) - 
\delta(t_x-t_s)\delta(\vec x-\vec s) \partial_\alpha  
\Gamma^{(2,1)}_{\mu\rho\nu}(t_y,\vec y, t_s,\vec s, t_z, \vec z) \Bigg\} = 0.
\end{align}
Fourier transforming the two previous relations, one   deduces  two exact 
identities relating the 4-point vertices $\Gamma^{(2,2)}$ and  $\Gamma^{(3,1)}$ 
with one zero wave-vector on a $\vu$-leg to 3-point functions
 \begin{align}
 \Gamma^{(2,2)}_{\alpha\beta\gamma\delta}(\omega_1,\vec p_1 = \vec 0, \omega_2, 
\vec p_2,\omega_3, \vec p_3)
 &= -\frac{1}{\omega_1}\Bigg[ p_2^\alpha \Gamma^{(1,2)}_{\beta\gamma\delta}( 
\omega_1+\omega_2, \vec p_2,\omega_3, \vec p_3) +  p_3^\alpha 
\Gamma^{(1,2)}_{\beta\gamma\delta}(\omega_2, \vec p_2, \omega_1+\omega_3, \vec 
p_3) \nonumber\\
& +  (- p_2 - p_3)^\alpha \Gamma^{(1,2)}_{\beta\gamma\delta}(\omega_2, \vec 
p_2,\omega_3, \vec p_3)\Bigg]
\label{WG22}\\
 \Gamma^{(3,1)}_{\alpha\beta\gamma\delta}(\omega_1,\vec p_1 = \vec 0, \omega_2, 
\vec p_2,\omega_3, \vec p_3)& = -\frac{1}{\omega_1}\Bigg[ p_2^\alpha 
\Gamma^{(2,1)}_{\beta\gamma\delta}( \omega_1+\omega_2, \vec p_2,\omega_3, \vec 
p_3)+ p_3^\alpha \Gamma^{(2,1)}_{\beta\gamma\delta}(\omega_2, \vec p_2, 
\omega_1+\omega_3, \vec p_3)  \nonumber\\
& +  (-p_2 - p_3)^\alpha \Gamma^{(2,1)}_{\beta\gamma\delta}(\omega_2, \vec 
p_2,\omega_3, \vec p_3)\Bigg]. \label{WG31}
\end{align}

 \renewcommand{\theequation}{F\arabic{equation}}
\section*{Appendix F: Exact flow equations for the 2-point  vertex functions in 
the large external wave-number limit}
\setcounter{equation}{0}

In this Appendix, we derive an expression for the flow equations of 
$\Gamma^{(0,2)}_\perp(\nu,\vp)$ and $\Gamma^{(1,1)}_\perp(\nu,\vp)$ 
which becomes exact in the limit of large  external wave-number $|\vp| \gg 
\kappa$. The diagrams entering these flow equations are
 schematically depicted in Figs.~\ref{fig4} and  \ref{fig5}. Some of them, 
diagrams (a), (b), (d) of Fig.~\ref{fig4} and diagrams (a), (b), (f) of 
Fig.~\ref{fig5} are vanishing (see Sec. \ref{SEC-LP}). We calculate below  the 
contributions of 
    the remaining non-zero diagrams.

\subsection*{Flow equation of $\Gamma^{(0,2)}_\perp(\nu,\vp)$ in the large $\vp$ 
limit}

We separately analyze the three diagrams (c), (d) and (e) of Fig.~\ref{fig5} 
which  give non-vanishing contributions to the flow of $\Gamma^{(0,2)}_\perp$. 
We
begin with determining the expression of diagram (d) in the limit of large 
external wave-number  $|\vp|\gg \kappa$. Introducing the operator
 $\tilde \p_s \equiv \displaystyle \p_s R_\kappa \frac{\p}{\p R_\kappa} + \p_s 
N_\kappa \frac{\p}{\p N_\kappa}$, this contribution may be written as
\begin{equation}
\left[\partial_s \Gamma^{(0,2)}_{\alpha\beta}(\nu, \vec p)\right]_{(d)} = -\frac 
1 2\tilde\partial_s \int_{\omega,\vec q} \Gamma^{(2,1)}_{ij\alpha}(\omega,\vec 
q,-\omega-\nu,-\vec p-\vec q)G^{u u}_{jk}(\omega+\nu,\vec p+\vec 
q)\Gamma^{(2,1)}_{kl\beta}(\omega+\nu,\vec p+\vec q,-\omega,-\vec q)G^{u 
u}_{li}(\omega,\vec q).\label{dsgam02a}
\end{equation}
Either the operator $\tilde \p_s$ acts on $G^{u u}_{li}(\omega,\vec q)$ and the 
internal wave-vector $\vq$ is cut off to $|\vq|\lesssim\kappa$ so that it is  
negligible
 compared to $\vp$ and can be set to zero. Or it acts on $G^{u 
u}_{jk}(\omega+\nu,\vec p+\vec q)$, in which case the combination  $\vp+\vq$ is 
cut off. Changing variables, this last contribution identifies with the first 
one. Hence, in the large $|\vp|$ limit,  the flow equation (\ref{dsgam02a}) 
becomes
\begin{equation}
 \left[\partial_s \Gamma^{(0,2)}_{\alpha\beta}(\nu, \vec p)\right]_{(d)}=- 
\int_{\omega} \Gamma^{(2,1)}_{ij\alpha}(\omega,\vec 0,-\omega-\nu,-\vec p)G^{u 
u}_{jk}(\omega+\nu,\vec p)\Gamma^{(2,1)}_{kl\beta}(\omega+\nu,\vec 
p,-\omega,\vec 0)
\; \tilde\partial_s \int_{\vec q}G^{u u}_{li}(\omega,\vec q).
\end{equation}
 Then, using the Ward identity (\ref{ward-21}) and projecting onto the 
transverse sector, one deduces 
\begin{align}
P_{\alpha\beta}^\perp(\vec p)\left[\partial_s \Gamma^{(0,2)}_{\alpha\beta}(\nu, 
\vec p)\right]_{(d)}&= -(d-1)\left(1-\frac 1 d\right)p^2 \int_{\omega} \frac 1 
{\omega^2}\left(\Gamma^{(1,1)}_\perp(-\nu, \vec p)- 
\Gamma^{(1,1)}_\perp(-\nu-\omega, \vec p)\right)\nonumber \\ 
 &\times \left(\Gamma^{(1,1)}_\perp(\nu, \vec p)- 
\Gamma^{(1,1)}_\perp(\nu+\omega, \vec p)\right) 
G^{(2,0)}_{\perp}(\omega+\nu,\vec p) \; \tilde\partial_s \int_{\vec q} 
G^{(2,0)}_{\perp}(\omega,\vec q) \label{diaA}
\end{align}
where parity in $\vp$ and the identity $\displaystyle\int_{\vec q} (\vec p \cdot 
\vec q)^2 f(q^2) = \frac{p^2}{d} \int_{\vec q} q^2 f(q^2)$
 were used.
The contribution of diagram (c) of Fig.~\ref{fig5} can be written as
\begin{align}
\left[\partial_s \Gamma^{(0,2)}_{\alpha\beta}(\nu, \vec p)\right]_{(c)}&=
-\int_{\omega,\vec q} \Gamma^{(1,2)}_{i\alpha j}(\omega,\vec q,\nu,\vec p)G^{u 
\bar u}_{jk}(-\omega-\nu,\vec p+\vec q)
\Gamma^{(2,1)}_{kl\beta}(\omega+\nu,\vec p+\vec q,-\omega,-\vec 
q)\;\tilde\partial_s G^{u u}_{li}(\omega,\vec q)+c.c.\nonumber\\
 &= -\int_{\vec q} \Gamma^{(1,2)}_{i\alpha j}(\omega,\vec 0,\nu,\vec p)G^{u \bar 
u}_{jk}(-\omega-\nu,\vec p)
 \Gamma^{(2,1)}_{kl\beta}(\omega+\nu,\vec p,-\omega,\vec 0)\;\tilde\partial_s 
\int_{\vec q}G^{u u}_{li}(\omega,\vec q)+c.c.
\end{align}
where  the second equality holds in the large $|\vp|$ limit, when the internal 
wave-vector $\vq$ is negligible compared to $\vp$.
Inserting the Ward identities (\ref{ward-21}) and (\ref{ward-12}) for the 
3-point vertices and projecting onto the transverse sector, 
 one obtains
\begin{align}
P_{\alpha\beta}^\perp(\vec p)\left[\partial_s \Gamma^{(0,2)}_{\alpha\beta}(\nu, 
\vec p)\right]_{(c)}&= -\frac{(d-1)^2}{d} p^2 \int_{\omega}
\left[\frac{\Gamma^{(0,2)}_\perp(\omega+\nu, \vec p)- \Gamma^{(0,2)}_\perp(\nu, 
\vec p)}{\omega}\right] \times
\left[\frac{\Gamma^{(1,1)}_\perp(\omega+\nu, \vec p)- \Gamma^{(1,1)}_\perp(\nu, 
\vec p)}{\omega}\right] \nonumber\\
 &\times G^{u \bar u}_{\perp}(-\omega-\nu,\vec p) \;\tilde\partial_s \int_{\vec 
q} G^{u u}_{\perp}(\omega,\vec q)+c.c..\label{diaB}
\end{align}
Similarly, the contribution of diagram (e) of Fig.~\ref{fig5} simplifies to
\begin{equation}
\left[\partial_s \Gamma^{(0,2)}_{\alpha\beta}(\nu, \vec p) \right]_{(e)}= \frac 
1 2\tilde\partial_s \int_{\omega,\vec q}
\Gamma^{(2,2)}_{ij\alpha\beta}(\omega,\vec q,-\omega,-\vec q,\nu,\vec p)G^{u 
u}_{ij}(\omega,\vec q)
=\frac 1 2 \tilde\partial_s \int_{\omega} 
\Gamma^{(2,2)}_{ij\alpha\beta}(\omega,\vec 0,- \omega, \vec 0,\nu,\vec p)\; 
 \int_{\vec q} G^{u u}_{ij}(\omega,\vec q)
\end{equation}
 which transverse projection reads, using the Ward identity (\ref{ward-22}) for 
the 4-point vertex function 
with two vanishing  wave-vectors on its $\vu$-legs,
\begin{align}
P_{\alpha\beta}^\perp(\vec p) \left[\partial_s\Gamma^{(0,2)}_{\alpha\beta}(\nu, 
\vec p) \right]_{(e)} &= \frac 1 2\frac{(d-1)^2}{d}p^2 \int_\omega \frac 1 
{\omega^2}\left[ \Gamma^{(0,2)}_\perp(\omega+\nu,\vec p) -2 
\Gamma^{(0,2)}_\perp(\nu,\vec p) +\Gamma^{(0,2)}_\perp(-\omega+\nu,\vec p)  
\right]\; \tilde\partial_s \int_{\vec q} G^{u u}_{\perp}(\omega,\vec q). 
\label{diaC}
\end{align}
The exact flow equation of  $\Gamma^{(0,2)}_\perp(\nu,\vp)$ in the large 
external wave-number limit is the sum of the three contributions (\ref{diaA}), 
(\ref{diaB}) and (\ref{diaC}), which yields Eq.~(\ref{dtGam02}).
 
\subsection*{Flow equation of $\Gamma^{(1,1)}_\perp(\nu,\vp)$ in the large $\vp$ 
limit}

Only the two diagrams (c) and (e) of Fig.~\ref{fig4} give a non-vanishing 
contribution to the flow of $\Gamma^{(1,1)}_\perp$ in the large external 
wave-number limit.
The contribution of diagram (c)  can be written as
\begin{align}
\left[\partial_s \Gamma^{(1,1)}_{\alpha\beta}(\nu, \vec p)\right]_{(c)}=& 
-\tilde\partial_s \int_{\omega,\vec q} \Gamma^{(2,1)}_{i\alpha j}(\omega,\vec 
q,\nu,\vec p)
G^{u \bar u}_{jk}(-\omega-\nu,\vec p+\vec 
q)\Gamma^{(2,1)}_{kl\beta}(\omega+\nu,\vec p+\vec q,-\omega,-\vec q)G^{u 
u}_{li}(\omega,\vec q) \nonumber\\
 =&- \int_{\omega,\vec q} \Gamma^{(2,1)}_{i\alpha j}(\omega,\vec 0,\nu,\vec 
p)G^{u \bar u}_{jk}(-\omega-\nu,\vec p)
 \Gamma^{(2,1)}_{kl\beta}(\omega+\nu,\vec p,-\omega,\vec 0)\;\tilde\partial_s 
G^{u u}_{li}(\omega,\vec q)\nonumber\\
 &- \int_{\omega,\vec q} \Gamma^{(2,1)}_{i\alpha j}(-\omega-\nu,-\vec p,\nu,\vec 
p)G^{u u}_{li}(-\nu-\omega,\vec q+\vec p)
 \Gamma^{(2,1)}_{kl\beta}(-\omega,-\vec q,\omega+\nu,\vec p)\;\tilde\partial_s 
G^{u \bar u}_{jk}(\omega,\vec q)
\end{align}
where again, the second equality holds in the large $\vp$ limit, when the 
internal wave-vector $\vq$ can be set to zero.
The transverse projection of this expression,  inserting the Ward identity 
(\ref{ward-21}), is given by
\begin{align}
P_{\alpha\beta}^\perp(\vec p)\left[\partial_s \Gamma^{(1,1)}_{\alpha\beta}(\nu, 
\vec p)\right]_{(c)}&= -\frac{(d-1)^2}{d} p^2 \int_{\omega} 
\left[\frac{\Gamma^{(1,1)}_\perp(\omega+\nu, \vec p)- \Gamma^{(1,1)}_\perp(\nu, 
\vec p)}{\omega}\right]^2  G^{u \bar u}_{\perp}(-\omega-\nu,\vec p) 
\;\tilde\partial_s \int_{\vec q} G^{u u}_{\perp}(\omega,\vec q).\label{diagA}
\end{align}
Lastly, the contribution of diagram (e) of Fig.~\ref{fig4} is very similar to 
the one of diagram (e) in the flow of $ \Gamma^{(0,2)}_\perp$ and its  
transverse projection, using the Ward identity (\ref{ward-31}), is given by
\begin{align}
P_{\alpha\beta}^\perp(\vec p) \left[\partial_s \Gamma^{(1,1)}_{\alpha\beta}(\nu, 
\vec p)\right]_{(e)}&=\frac 1 2 P_{\alpha\beta}^\perp(\vec p)\; \tilde\partial_s 
\int_{\omega,\vec q} \Gamma^{(3,1)}_{ij\alpha\beta}(\omega,\vec 0,-\omega, \vec 
0, \nu,\vec p)G^{u u}_{ij}(\omega,\vec q) \nonumber \\
 &= \frac 1 2 P_{\alpha\beta}^\perp(\vec p)  \int_{\omega} 
\Gamma^{(3,1)}_{ij\alpha\beta}(\omega,\vec 0,-\omega, \vec 0, \nu,\vec p) \; 
\tilde \p_s \int_{\vq}G^{u u}_{ij}(\omega,\vec q) \nonumber\\
&=\frac 1 2\frac{(d-1)^2}{d}p^2 \int_\omega \frac 1 {\omega^2}\Bigg[ 
\Gamma^{(1,1)}_{\perp}(\omega+\nu,\vec p) -2 \Gamma^{(1,1)}_{\perp}(\nu,\vec p) 
+\Gamma^{(1,1)}_{\perp}(-\omega+\nu,\vec p)  \Bigg] \;\tilde \p_s\int_{\vec q} 
G^{u u}_{\perp}(\omega,\vec q).
\label{diagB}
\end{align}
The exact flow equation of  $\Gamma^{(1,1)}_\perp(\nu,\vp)$ in the large 
external wave-number limit is the sum of the two contributions
 (\ref{diagA}) and (\ref{diagB}), which yields Eq.~(\ref{dtGam11}).

\end{widetext}
\end{appendix}

\bibliographystyle{prsty}


\end{document}